%% file: main.tex
\documentclass[a4paper,11pt]{article}
\usepackage[normalem]{ulem}
\usepackage{amsmath,amssymb,color,comment}
\usepackage{slashed, tensor, bm, physics}
\usepackage{caption}
 \usepackage{graphicx}
\usepackage{float}
\usepackage{enumitem}

\usepackage[compat=1.1.0]{tikz-feynhand}
\usepackage{jheppub} 
\usepackage[T1]{fontenc} 
\usepackage{booktabs} 
\usepackage{subcaption}

\title{Dark vector meson effects on composite dark matter}
\author[a]{Tomohiro Abe,}
\author[b]{Ryosuke Sato,}
\author[c]{Ozan Semin}
\author[b]{and Takumu Yamanaka}

\affiliation[a]{Department of Physics and Astronomy, Faculty of Science and Technology, Tokyo University of
Science, Yamazaki, Noda, Chiba 278-8510, Japan}
\affiliation[b]{Department of Physics, The University of Osaka, Machikaneyama-cho, Toyonaka, Osaka 560-0043, Japan}
\affiliation[c]{Institute for Theoretical Physics, University of Tübingen, Auf der Morgenstelle 14, 72076 Tübingen, Germany}

\emailAdd{abe.tomohiro@rs.tus.ac.jp}
\emailAdd{rsato@het.phys.sci.osaka-u.ac.jp}
\emailAdd{ozan.semin@uni-tuebingen.de}
\emailAdd{yamanaka@het.phys.sci.osaka-u.ac.jp}

\abstract{
We extend previous studies of $\mathcal{G}$-parity-odd dark pion dark matter in a QCD-like $\mathrm{SU}(N_c)$ gauge theory with electroweakly interacting dark quarks to include the effects of $\rho$ mesons. We show that vector mesons can substantially affect relic abundance calculations through resonant enhancement, significantly affecting the conditions required to reproduce the observed relic abundance. Electroweak interactions further induce mixing between the dark $\rho$ mesons and Standard Model gauge bosons, opening new decay channels for the dark $\rho$ and modifying the effective interactions of the dark pions. We recast existing ATLAS and CMS searches to constrain the model, and perform a collider phenomenology study targeting the future FCC-hh and muon colliders, focusing on the resonant enhancement of cross sections in the $e^+e^-$, $\mu^+\mu^-$, and $W^+W^-$ final states for the latter. We show that the FCC-hh could probe $\rho$ mesons whose mass is up to 40 TeV.
}
\begin{document}
\begin{flushright}
	OU-HET-1331
\end{flushright}

\maketitle

\flushbottom

\input{sections/section1.tex}
\input{sections/section2.tex}
\input{sections/section3.tex}
\input{sections/section4.tex}
\input{sections/acknowledgments.tex}

\appendix

\input{sections/appendix1.tex}
\input{sections/appendix2.tex}
\input{sections/appendix3.tex}
\clearpage
\bibliography{ref}
\bibliographystyle{JHEP}

\end{document}

%% file: sections/section1.tex
\section{Introduction}\label{sec1}
The Standard Model (SM), while remarkably successful at predicting fundamental particle properties and interactions, leaves several critical questions unanswered. Notably, it cannot account for dark matter (DM), which comprises approximately 27\% of the universe's mass-energy content \cite{Planck:2015fie,Rubin:1980zd,Zwicky:1933gu}. Despite gravitational evidence for the presence of DM, its fundamental nature remains an open question, pointing toward physics Beyond the Standard Model (BSM) \cite{Saikumar:2024ahz,Jungman:1995df,FileviezPerez:2019cyn,Gehrlein:2019iwl}. 

How this abundance of DM was produced in the early universe is also a significant question. The thermal freeze-out mechanism, where DM particles were once in thermal equilibrium with the Standard Model particles and then decoupled as the universe expanded and cooled, is a widely studied scenario \cite{Kolb:1990vq}. 

A well-motivated possibility in this context is very heavy dark matter, with mass in the $\mathrm{TeV}$ regime, where present direct-detection bounds are comparatively weaker. For thermal relic dark matter, the mass scale is still constrained from above by partial-wave unitarity of the annihilation cross section \cite{Griest:1989wd}. On the indirect-detection side, substantial progress has been made by gamma-ray observatories, with sensitivity that now reaches well into the multi-TeV regime \cite{Abeysekara:2017hyn,HESS:2016mib,MAGIC:2016xys,Weekes:2001pd}. Looking ahead, next-generation and complementary facilities such as CTA \cite{CTAConsortium:2017dvg}, LHAASO \cite{LHAASO:2019qtb}, SWGO \cite{Abreu:2019ahw}, HERD \cite{HERD:2014bpk}, and IceCube-Gen2 \cite{IceCube-Gen2:2020qha} are expected to significantly extend the reach for heavy-dark-matter annihilation signals.

In this paper, we will explore the case of heavy composite DM, where the underlying symmetry structure is assumed to be QCD-like. Specifically, we consider $N_f=3$ flavors of dark quarks breaking the dark chiral symmetry $\mathrm{SU}(3)_L \times \mathrm{SU}(3)_R$ down to the diagonal subgroup $\mathrm{SU}(3)_V$, resulting in a spectrum of dark pions as pseudo-Nambu-Goldstone bosons \cite{Bai:2010qg, Antipin:2015xia}. The lightest dark pion, stabilized by $\mathcal{G}$-parity, serves as the DM candidate. Additionally, we assume dark quarks form a triplet under the Standard Model $\mathrm{SU}(2)_W$ gauge group, which is embedded in the unbroken symmetry $\mathrm{SU}(3)_V$. 

This model has been studied in several contexts, under various assumptions. We build on a previous analysis from Ref.~\cite{Abe:2024mwa} that studied the mass spectrum and phenomenology of the dark pions. We extend this analysis by including the effects of heavier composite dark vector mesons, which are expected to be present in the spectrum and can significantly affect the annihilation cross section of the dark pions.

In section~\ref{sec:models} we present the model and the effective Lagrangian describing the interactions of the dark pions and dark $\rho$ mesons within the hidden local symmetry (HLS) framework \cite{Cremmer:1978n,Bando:1984ej,DAdda:1978dle,Breitenlohner:1983pi,Golo:1978de,DAdda:1978vbw,Cremmer:1979up}. We derive mass relations and characterize the mixing of the dark $\rho$ mesons with the SM gauge bosons induced by the electroweak charges of the dark quarks. In section~\ref{sec:pheno} we establish the lower bound on the HLS coupling $g_\rho$, then compute the relic abundance of the dark mesons in the freeze-out scenario, accounting for the new annihilation channels and resonance effects introduced by the $\rho$ mesons, and map out the viable parameter space.
We then turn to collider phenomenology. The signals for this class of models at collider experiments have been studied, for example in ref.~\cite{Asadi:2025vfr}. In this paper, we focus on the $\rho$ meson signal, which mixes with the SM electroweak gauge bosons \cite{Kilic:2009mi, Kilic:2010et,Liu:2023jta,Acanfora:2026aay}.
We derive $95\%$ confidence level exclusion regions from the existing ATLAS and CMS results, and for future FCC-hh and muon colliders.
For the latter, we focus on the pseudorapidity distributions at fixed center-of-mass energy $\sqrt{s}=10$~TeV, identifying the complementary reach of the two analyses. In section~\ref{sec:discussion} we summarize our findings and outline directions for future work.

%% file: sections/section2.tex
\section{Model}\label{sec:models}
We consider a DM scenario where the additional fundamental particles are dark quarks, which are charged under a new confining gauge group $\mathrm{SU}(N_c)$ with $N_c\geq 3$ and the SM $\mathrm{SU}(3)_c \times \mathrm{SU}(2)_W \times \mathrm{U}(1)_Y$ gauge group. We present two separate frameworks for the dark pion and dark $\rho$ meson composite particles, and construct an effective Lagrangian that describes the interactions of these composite states with each other.

\subsection{Dark Pions}\label{subseq:dark_pions}
For the dark QCD-like sector gauged under $\mathrm{SU}(N_c)$ with $N_c\geq 3$, we introduce three flavors of dark quarks and dark anti-quarks, $\mathbf{q}=(q_i,\bar{q}_i^\dagger)$ ($i=1,2,3$), which lie in the fundamental representation of $\mathrm{SU}(N_c)$ group and a vectorial representation of the SM gauge group $\mathrm{SU}(3)_c \times \mathrm{SU}(2)_W \times \mathrm{U}(1)_Y$. The explicit multiplet assignments are given in Table~\ref{tab:dark_quark_charges}. 
\begin{table}[t]
    \centering
    \begin{tabular}{l|cccc}
        \toprule
        \textbf{Field} & \textbf{$\mathrm{SU}(N_c)$} & \textbf{$\mathrm{SU}(3)_c$} & \textbf{$\mathrm{SU}(2)_W$} & \textbf{$U(1)_Y$}\\
        \midrule
        $q_i$ & $\mathbf{N_c}$ & $\mathbf{1}$ & $\mathbf{3}$ & $0$\\
        $\bar{q}_i$ & $\mathbf{\overline{N_c}}$ & $\mathbf{1}$ & $\mathbf{3}$ & $0$\\
        \bottomrule
    \end{tabular}
    \caption{Gauge multiplet assignments for the dark quark fields under the dark $\mathrm{SU}(N_c)$ symmetry and Standard Model gauge groups. $\mathbf{N_c}$ and $\mathbf{\overline{N_c}}$ denote the fundamental and anti-fundamental representations of the $\mathrm{SU}(N_c)$ gauge group, respectively.}
    \label{tab:dark_quark_charges}
\end{table}
They form a triplet under the $\mathrm{SU}(2)_W$ gauge group, and are singlets under the $\mathrm{SU}(3)_c$ color group. The dark quarks are vector-like with respect to the SM gauge group. The dark quarks are also charged under the new confining gauge group $\mathrm{SU}(N_c)$, which becomes strongly coupled at a scale $\Lambda_{\mathrm{d}}$. Below this confinement scale, the dark quarks condense and form bound states. Under these considerations, above the confinement scale, the general Lagrangian for the dark quarks is given by
\begin{equation}\label{eq:lag_quark}
    \mathcal{L} = \mathcal{L}_{\text{SM}}-\frac{1}{4} G_{\mu\nu}^a G^{a\mu\nu} + \sum_{i=1}^3 \bar{\mathbf{q}}_i \left[i\slashed{D} - m_q\mathrm{exp}\left(-\frac{i\theta}{3}\gamma_5\right)\right] \mathbf{q}_i,
\end{equation}
where $G_{\mu\nu}^a$ is the field strength tensor for the $\mathrm{SU}(N_c)$ gauge field, and $D_\mu$ is the covariant derivative that includes both the $\mathrm{SU}(N_c)$ and SM gauge interactions. Here we have also included a CP-violating $\theta$-term, which originally appears as a $\frac{g_d^2\theta}{32\pi^2} G_{\mu\nu}^a \tilde{G}^{a\mu\nu}$ term in the Lagrangian, but can be rotated into the quark mass term by a chiral transformation 
\begin{equation}
    q_i \to \mathrm{exp}\left(-i\frac{\theta}{6}\gamma_5\right) q_i.
\end{equation}

Additionally, the Lagrangian in Eq.~\eqref{eq:lag_quark} has a global chiral symmetry, $\mathrm{SU}(3)_L \times \mathrm{SU}(3)_R$, which is broken by the quark mass and the EW coupling. Under this symmetry, the quark fields transform as
\begin{equation}
    q_i \to L_{ij} q_j, \quad \bar{q}_i^\dagger \to R_{ij} \bar{q}_j^\dagger,
\end{equation}
where $L$ and $R$ are elements of the $\mathrm{SU}(3)_L$ and $\mathrm{SU}(3)_R$ groups, respectively. The explicit breaking of the chiral symmetry constrains the transformation to the diagonal subgroup $\mathrm{SU}(3)_V$, we embed the gauge group $\mathrm{SU}(2)_W$ into this diagonal subgroup, which transforms the quark fields as
\begin{equation}
    \mathbf{q}_i\to\exp\left(i\alpha^a \tilde{T}^a\right)_{ij} \mathbf{q}_j,
\end{equation}
where $\tilde{T}^a$ are the generators of the $\mathrm{SU}(2)_W$ subgroup:
\begin{equation}
    \alpha^a \tilde{T}^a =\begin{pmatrix}
    0 & -i\alpha^3 & i\alpha^2\\
    i\alpha^3 & 0 & -i\alpha^1\\
    -i\alpha^2 & i\alpha^1 & 0
    \end{pmatrix}.
\end{equation} 
Below the confinement scale $\Lambda_{\mathrm{d}}$, the dark quarks acquire a negative vacuum expectation value (VEV):
\begin{equation}
    \langle \bar{q} q \rangle = -v_d^3 \mathbf{1}.
\end{equation}
This spontaneously breaks the approximate chiral symmetry down to the diagonal subgroup $\mathrm{SU}(3)_V$, in which the gauge group $\mathrm{SU}(2)_W$ is embedded. 

In the low energy regime of the model, instead of the Weyl fermions, the relevant degrees of freedom are implemented by the $3\times3$ unitary matrix fields $U$, as well as the spurion mass matrix field $M$, which transform under the chiral symmetry as
\begin{equation}
    U \to L U R^\dagger, \quad M\equiv 
    e^{i\frac{\theta}{3}}
    \begin{pmatrix}
        m & 0 & 0\\
        0 & m & 0\\
        0 & 0 & m
    \end{pmatrix} \to R M L^\dagger.
\end{equation}
The matrix field $U$ is parameterized in terms of the dark pion fields. Constructing an effective Lagrangian that is invariant under the chiral symmetry, we have
\begin{equation}\label{eq:lag_pion}
    \mathcal{L} = \frac{f^2_d}{4}\mathrm{Tr}\left[D_\mu U D^\mu U^\dagger\right] + \underbrace{v_d^3\mathrm{Tr}\left[U M + M^\dagger U^\dagger\right]}_{\equiv -V(U)},
\end{equation}
where $f_d$ is the dark pion decay constant, and
\begin{equation}
    D_\mu U = \partial_\mu U - i g_W W_\mu^a [\tilde{T}^a, U],
\end{equation}
with $g_W$ being the $\mathrm{SU}(2)_W$ gauge coupling and $W_\mu^a$ being the $\mathrm{SU}(2)_W$ gauge fields.
The first term in the Lagrangian describes the kinetic terms and gauge interactions of the dark pions, while the second term gives rise to the dark pion mass. The latter term is viewed as a potential term for the dark pion fields, which is $\theta$-dependent.

Under these assumptions, for all masses $m$ and $\theta\in(-\pi,\pi]$ the potential term is minimized for \cite{Smilga:1998dh, Abe:2024mwa}
\begin{equation}
    \langle U \rangle = \mathbf{1}.
\end{equation}
The fluctuations around this vacuum correspond to the dark pions. In this work, we consider the CP-conserving case, i.e. $\theta=0$. Under the unbroken $\mathrm{SU}(3)_V$ symmetry, the dark pions transform as an octet. In the embedding of the $\mathrm{SU}(2)_W$ gauge group, the dark pion octet is reducible, and thus decomposes into a triplet ($\Pi_3$) and a quintet ($\Pi_5$). Thus, the unitary matrix field $U$ can be parameterized as
\begin{equation}
    U = \exp\left(i\frac{\sqrt{2}}{f_d}(\Pi_3+\Pi_5)\right),
\end{equation}
with 
\begin{equation}\label{eq:pi3pi5}
    \Pi_3 = \frac{1}{\sqrt{2}}\Pi_3^a \tilde{T}^a,\quad\quad \Pi_5 = \frac{1}{\sqrt{2}}\Pi_5^a T^a,
\end{equation}
such that
\begin{equation}\label{eq:su3_decomposition}
    \underbrace{\mathrm{span}\left\{\tilde{T}^1,\tilde{T}^2,\tilde{T}^3\right\}}_{=\mathfrak{su}(2)}\oplus~\mathrm{span}\left\{T^1,T^2,T^3,T^4,T^5\right\} = \mathfrak{su}(3),
\end{equation}
with $\tr(T^a T^b)=\tr(\tilde{T}^a \tilde{T}^b) = 2\delta^{ab}$.
The triplet part is vectorial in the $\mathrm{SU}(2)_W$ gauge group, and thus contains a neutral component $\Pi_3^0$ and two charged components $\Pi_3^\pm$. Meanwhile, the quintet part contains a neutral component $\Pi_5^0$, two singly charged components $\Pi_5^\pm$ and two doubly charged components $\Pi_5^{\pm\pm}$.

As a final part of the Lagrangian, we also include the Wess-Zumino-Witten (WZW) term, which is a topological term that arises from the non-trivial homotopy group $\pi_5(SU(3)) = \mathbb{Z}$ \cite{Witten:1983tw,Wess:1971yu}. This term arises at the order of $p^4$ in the chiral perturbation theory expansion, and describes the anomalous interactions of the dark pions. At the leading order in dark pion fields, the corresponding WZW term can be written as \cite{Abe:2024mwa}:
\begin{align}\label{eq:wzw}
    \mathcal{L}_{\text{WZW}} =& -\frac{g_W^2N_c}{16\sqrt{2}\pi^2 
    f_d
    } \epsilon^{\mu\nu\rho\sigma} \tr\left[\Pi_5W_{\mu\nu}W_{\rho\sigma}\right]\nonumber\\
    &+\frac{i\sqrt{2}g_WN_c}{12\pi^2f_d^3} \epsilon^{\mu\nu\rho\sigma} \tr\left[(\partial_\mu\Pi_3\partial_\nu\Pi_3+\partial_\mu\Pi_5\partial_\nu\Pi_5)\partial_\rho\Pi_5W_\sigma\right],
\end{align}
with $W_{\mu\nu} $ being the field strength tensor for the $\mathrm{SU}(2)_W$ gauge fields. The WZW term makes it possible for the quintet dark pions to decay into SM particles.

\subsubsection{Dark Pion Masses}
Due to the similar nature of our model to the QCD sector in the SM, some of the dark pion mass contributions can be estimated by scaling up the corresponding contributions to the SM pions. Three different categories of contributions are considered in this work, as in the previous work \cite{Abe:2024mwa}: 
\begin{enumerate}[label=\Roman*.]
    \item The contribution from the quark mass term, which is given by the second term in the Lagrangian in Eq.~\eqref{eq:lag_pion}.
    \item Electroweak loop effects on the dark pion self energy.
    \item Electroweak symmetry breaking (EWSB) effects on the charged dark pions, causing a small mass splitting between the charged and neutral dark pions.
\end{enumerate}
Similar to the QCD case, the dark quark mass contribution can be computed directly from the mass term in the Lagrangian in Eq.~\eqref{eq:lag_pion}, which gives
\begin{equation}
    m_\Pi^2 = \frac{4 v_d^3 m}{f_d^2}.
\end{equation}
This relation is identical for both the triplet and quintet dark pions. In our case, the bilinear dark quark condensate can be matched to the QCD case by using the large $N_c$ scaling, which gives
\begin{equation}
    \frac{v_d^3}{v_\text{QCD}^3} \approx \sqrt{\frac{3}{N_c}} \frac{f_d^3}{f_\pi^3}.
\end{equation}
Using the neutral pion mass in the QCD sector, $m_{\pi^0}\approx134.97~\mathrm{MeV}$ \cite{ParticleDataGroup:2024cfk}, and $v_{\text{QCD}}$ is the QCD biquark condensation value given as $v_{\text{QCD}}^3=\frac{m_{\pi,\text{QCD}}^2\cdot f_\pi^2}{2(m_u+m_d)}$. We can estimate the dark pion masses as 
\begin{equation}
    m_\Pi^2 \approx 57.76\sqrt{\frac{3}{N_c}}m f_d.
\end{equation}
This is the dominant contribution for non-zero quark mass $m$, and is the cause of the approximate degeneracy between the triplet and quintet dark pions.

Similarly, the electroweak loop contribution to the dark pion masses can be estimated by scaling up the corresponding contribution to the SM pions. Fundamentally, the interaction terms with the $W$ bosons cause an explicit breaking of the $\mathrm{SU}(3)_V$, removing the degeneracy between the triplet and quintet dark pions. Due to this, the quintet dark pions become heavier than the triplet dark pions. Since both multiplets interact with the $W$ bosons, each multiplet gets a non-zero contribution. Specifically, using similar scaling arguments as before, this contribution can be estimated as \cite{Abe:2024mwa,Das:1967it}:
\begin{equation}
    \Delta m_{\Pi_X}^2 \approx C_X\frac{\alpha_W\left(\Lambda_d\right)\Lambda_d^2}{\alpha\left(\Lambda_\text{QCD}\right)\Lambda_\text{QCD}^2}\left(m_{\pi_\pm}^2-m_{\pi_0}^2\right),
\end{equation}
where $C_X$ with $X=3,5$ are the quadratic Casimir coefficients for the triplet and quintet representations, respectively, and $\alpha_W$ and $\alpha$ are the $\mathrm{SU}(2)_W$ and $\mathrm{U}(1)_\text{EM}$ gauge couplings, respectively. Using the renormalization group equations for the weak gauge coupling, the masses can be estimated as \cite{Martin:1997ns,Abe:2024mwa,ParticleDataGroup:2024cfk}:
\begin{equation}
    \Delta m_{\Pi_3} = 1.1 f_d\sqrt{\frac{3}{N_c}}\quad\text{and}\quad \Delta m_{\Pi_5} = 1.9 f_d\sqrt{\frac{3}{N_c}}.
\end{equation}
Finally, we obtain an additional, comparably smaller mass splitting between the pions inside each multiplet, due to the electroweak one-loop effects. For each integer of charge $Q$, the mass splitting is given as \cite{Cheng:1998hc,Feng:1999fu}:
\begin{equation}
    \Delta m_{Q} = Q^2 \alpha_W m_W\sin[2](\frac{\theta_W}{2})\approx 166~\mathrm{MeV}\times Q^2,
\end{equation}
where $\theta_W$ is the Weinberg angle.
\subsection{Dark $\rho$ Mesons}\label{subseq:dark_rho}
In addition to the dark pions, we also consider the lightest vector mesons in the spectrum, which are the dark $\rho$ mesons. A rigorous way to include vector mesons in the effective theory is through the hidden local symmetry formalism \cite{Bando:1984ej,Bando:1987br}, which treats the vector mesons as gauge bosons of a hidden local symmetry. 

Specifically, a nonlinear sigma model based on the coset space $G/H$ can be interpreted as a gauge theory of the group $G_{\text{global}} \times H_{\text{local}}$, where $G_{\text{global}}$ is the original global symmetry of the nonlinear sigma model, and $H_{\text{local}}$ is the hidden local symmetry, of which the gauge bosons correspond to composite vector particles \cite{Cremmer:1978n,Bando:1984ej,DAdda:1978dle,Breitenlohner:1983pi,Golo:1978de,DAdda:1978vbw,Cremmer:1979up}. In our case, the original global symmetry is the chiral symmetry $\mathrm{SU}(3)_L \times \mathrm{SU}(3)_R$, and the unbroken subgroup is $\mathrm{SU}(3)_V$. Thus, we can introduce a hidden local symmetry $H_{\text{local}} = \mathrm{SU}(3)_V$ systematically into the non-linear sigma model for the dark pions by separating the unitary matrix field $U$ into two parts.

To realize this, we introduce the unitary matrix field $U(x)\in SU(3)_L\times SU(3)_R/\mathrm{SU}(3)_V$, with
\begin{equation}\label{eq:pion_redundant}
    U(x) = \xi_L(x) \xi_R(x)^\dagger,
\end{equation}
where $\xi_L$ and $\xi_R$ are unitary matrix fields that are defined as \begin{equation}
    \xi_{L/R}(x) = \exp\left(\pm i\frac{\Pi_3+\Pi_5}{\sqrt{2} f_d}\right),
\end{equation}
where $\Pi_3$ and $\Pi_5$ are the dark pion fields transforming in the triplet and quintet representations, respectively, as defined in Eqs.~\eqref{eq:pi3pi5} and \eqref{eq:su3_decomposition}. The fields $\xi_L$ and $\xi_R$ transform under the chiral symmetry as
\begin{equation}
    \xi_L \to L \xi_L h^{-1}(x) , \quad \xi_R \to R \xi_R h^{-1}(x),
\end{equation}
where $h(x)$ is an element of the hidden local symmetry group $\mathrm{SU}(3)_V$. This field is dependent on the transformations $L$ and $R$, as well as the pion fields at a given spacetime point. 

Next, we introduce the gauge fields $V_\mu$ for the hidden local symmetry $\mathrm{SU}(3)_V$, which are identified as the dark $\rho$ mesons. The gauge fields transform under the hidden local symmetry as
\begin{equation}
    V_\mu \to h(x) V_\mu h^\dagger(x) - \frac{i}{g_\rho} h(x) \partial_\mu h^\dagger(x).
\end{equation}
with the vector field defined as
\begin{equation}
    V_\mu = \frac{1}{2}\left(\tilde{V}_{3,\mu}^a \tilde{T}^a+V_{5,\mu}^a T^a\right),\label{eq:decomposition}
\end{equation}
with the generators defined in Eq.~\eqref{eq:su3_decomposition}. The exact reason why we chose this decomposition will be clear below, when both the $V$ and $W$ fields get mixed.
Now, also considering the SM gauge interactions assumed in this work, the Lagrangian for the non-linear model under consideration becomes 
\begin{equation}
    \mathcal{L} = - \frac{1}{4} V_{\mu\nu}^a V^{a\mu\nu} + \mathcal{L}_V + \mathcal{L}_A,
\end{equation}
where $V_{\mu\nu}^a$ is the field strength tensor for the hidden local symmetry gauge fields. The terms $\mathcal{L}_V$ and $\mathcal{L}_A$ are given by
\begin{equation}
    \mathcal{L}_V =  f_\rho^2\tr(a_\parallel^2), \quad \mathcal{L}_A = f_d^2\tr(a_\perp^2),
\end{equation} 
where $f_d$ and $f_\rho$ are the decay constants for the dark pions and dark rho mesons, respectively, and the Maurer-Cartan 1-forms $a_\parallel$ and $a_\perp$ are defined as
\begin{equation}
    a_{\parallel\mu} = \frac{\xi_L^\dagger \cdot D_\mu \xi_L + \xi_R^\dagger \cdot D_\mu \xi_R}{2i}, \quad a_{\perp\mu} = \frac{\xi_L^\dagger\cdot D_\mu \xi_L - \xi_R^\dagger \cdot D_\mu \xi_R }{2i},
\end{equation}
where the covariant derivatives are defined as 
\begin{equation}\label{eq:cov_derivative}
    D_\mu \xi_{L,R} = \partial_\mu \xi_{L,R} - i g_\rho  \xi_{L,R} 
    V_\mu + i g_W^{UV} \tilde{W}_\mu \xi_{L,R} ,
\end{equation}
with $g_\rho$ and $g_W^{UV}$ being the UV coupling constants for the hidden local symmetry and the SM $\mathrm{SU}(2)_W$ gauge fields, respectively. It is crucial to note that the hidden local gauge acts from the right.
Typically, the relevant parameter for the $\rho$ mesons is not given as the decay constant $f_\rho$, but rather the dimensionless parameter $a\equiv f_\rho^2/f_d^2$. In the QCD case, this parameter is estimated to be $a= 2$ as it leads to vector meson dominance as well as to the KSRF relation \cite{Bando:1984ej,Kawarabayashi:1966kd,Riazuddin:1966sw}. In our work, we will consider different values of this parameter to explore different phenomenological implications of the model, but focus on the case of $a=2$ for the main results.

In our work, we consider the unitary gauge for the hidden local symmetry, which corresponds to setting $\xi:=\xi_L = \xi_R^\dagger$. In this gauge, the $\mathcal{L}_A$ term reduces to the kinetic term of the dark pions by using the relation in Eq.~\eqref{eq:pion_redundant}.
At the same time, the $\mathcal{L}_V$ term gives rise to the interaction terms between the dark pions and the dark $\rho$ mesons. By integrating out the dark $\rho$ mesons, or using the equations of motion for $V_\mu$, this term vanishes, which is equivalent to removing the hidden local symmetry. 
Finally, we include the mass terms and the WZW terms for the dark pions from Eq.~\eqref{eq:wzw}, giving the complete Lagrangian 
\begin{equation}
    \mathcal{L} = - \frac{1}{4} V_{\mu\nu}^a V^{a\mu\nu} + af_d^2\tr(a_\parallel^2) + f_d^2\tr(a_\perp^2) + v_d^3\mathrm{Tr}\left[U M + M^\dagger U^\dagger\right]+\mathcal{L}_{\text{WZW}}.
\end{equation}
This represents the dark sector of the model, which will be used as an addition to the SM Lagrangian in the phenomenological analysis in the following sections.
Note that, in addition to the WZW term, there are extra anomalous terms with the $\rho$ meson \cite{Fujiwara:1984mp,Harada:2003jx}. However, those anomalous terms are not relevant for our analysis in this paper, and we neglect them for simplicity.

\subsubsection{Mass Eigenstates}\label{sec:mass}
Unlike the dark pions, due to the hidden-gauge nature of the dark $\rho$ mesons, their masses are initially not related to the quark mass term. Instead, the dark $\rho$ meson masses arise from the $\mathcal{L}_V$ term, as it should give a non-zero mass term. 
Using the definition of the covariant derivative given in Eq.~\eqref{eq:cov_derivative}, the $a_{\parallel\mu}$ can be decomposed as
\begin{align}
    a_{\parallel\mu} &= \frac{\xi^\dagger(\partial_\mu\xi+i g_W^{UV} \tilde{W}_\mu \xi - i g_\rho \xi V_\mu)  + \xi(\partial_\mu\xi^\dagger+i g_W^{UV} \tilde{W}_\mu \xi^\dagger- i g_\rho \xi^\dagger V_\mu) }{2i}\nonumber\\
    &= \frac{\xi^\dagger \cdot \tilde{D}_\mu \xi + \xi\cdot \tilde{D}_\mu \xi^\dagger}{2i}-g_\rho V_\mu\nonumber\\
    &= \tilde{a}_{\parallel\mu} - g_\rho V_\mu,
\end{align}
where $\tilde{D}_\mu$ is the covariant derivative that only contains the SM $\mathrm{SU}(2)_W$ gauge fields. Thus, we only need to expand the derivative and the gauge terms. We define $\Pi:=\frac{\Pi_3+\Pi_5}{\sqrt{2}}$ and obtain
\begin{equation}
    \xi^\dagger \cdot \partial_\mu \xi = i\frac{\partial_\mu\Pi}{f_d} + \frac{1}{2f_d^2}\left[\Pi,\partial_\mu\Pi\right]-\frac{i}{6f_d^3}\left[\Pi,\left[\Pi,\partial_\mu\Pi\right]\right]+\cdots,\nonumber
\end{equation}
and
\begin{equation}
    \xi^\dagger \cdot \tilde{W}_\mu \xi = \tilde{W}_\mu - i\frac{1}{f_d}\left[\Pi,\tilde{W}_\mu\right] - \frac{1}{2f_d^2}\left[\Pi,\left[\Pi,\tilde{W}_\mu\right]\right]+\cdots.
\end{equation}
Using these two expansions, we can expand the $\mathcal{L}_V$ term in the Lagrangian as
\begin{align}
    \mathcal{L}_V =& af_d^2\tr((g_W^{UV} \tilde{W}_\mu-g_\rho V_\mu)(g_W^{UV} \tilde{W}^\mu-g_\rho V^\mu))-\frac{a}{4f_d^2}\tr([\Pi,\partial_\mu\Pi][\Pi,\partial^\mu\Pi])\nonumber\\
    &-ia\tr([\Pi,\partial_\mu\Pi](g_W^{UV} \tilde{W}^\mu-g_\rho V^\mu))-ag_W^{UV}\tr(\left[\Pi,\left[\Pi,\tilde{W}_\mu\right]\right](g_W^{UV} \tilde{W}^\mu-g_\rho V^\mu))+\cdots.
\end{align} 
Here, $V_\mu$ is the full HLS field of Eq.~\eqref{eq:decomposition}, i.e.\ it contains both the triplet $\tilde{V}_{3,\mu}$, which mixes with $\tilde{W}_\mu$ and is therefore not a mass eigenstate, and the quintet $V_{5,\mu}$, which does not mix.
We see that the first term gives rise to the mass terms for the dark $\rho$ mesons. However, due to the mixing with the $W$ bosons, the total mass matrix is not diagonal anymore. Since before the EWSB, the $W$ bosons are massless, we could simply diagonalize by a simple rotation of the $V$ and $\tilde{W}$ fields. However, the trade-off is that we would get a non-diagonal mass matrix again after the EWSB, whose diagonalization is more involved. Thus, we choose to first perform the EWSB, and then diagonalize the mass matrix. After the EWSB, the weak gauge bosons acquire a mass the same way as in the SM, which we can substitute back into our model. Now, the mass terms in the Lagrangians can be nicely written as
\begin{eqnarray*}
    \mathcal{L}_{M}=&&4a f_d^2
    \begin{pmatrix}
          \tilde{W}^{-,\mu}, &   \tilde{V}_{3}^{-,\mu}
    \end{pmatrix}
    \begin{pmatrix}
     {g_W^{UV}}^2 +\frac{{m_{W}^{\text{SM}}}^2}{4a f_d^2} & -\frac{g_W^{UV} g_\rho}{2} \\
    - \frac{g_W^{UV} g_\rho}{2}  &  \frac{g_\rho^2}{4} 
    \end{pmatrix}
    \begin{pmatrix}
    \tilde{W}_\mu^+\\
    \tilde{V}_{3,\mu}^+
    \end{pmatrix}+ \frac{a f_d^2 g_\rho^2}{2}V_5^\mu V_{5,\mu}\\
    &&+2a f_d^2
    \begin{pmatrix}
    \tilde{Z}^\mu,&
    \tilde{A}^{\mu},&
    \tilde{V}_{3}^{0,\mu}
    \end{pmatrix}
    \begin{pmatrix}
     {g_W^{UV}}^2c_W^2 +\frac{{m_{Z}^{\text{SM}}}^2}{4a f_d^2} & c_W {g_W^{UV}}^2 s_W & -\frac{c_W g_{\rho } g_W^{UV}}{2} \\
 c_W {g_W^{UV}}^2 s_W & {g_W^{UV}}^2 s_W^2 & -\frac{g_{\rho } g_W^{UV} s_W}{2} \\
 -\frac{c_W g_{\rho } g_W^{UV}}{2} & - \frac{g_{\rho } g_W^{UV} s_W}{2} & \frac{g_\rho^2}{4} 
    \end{pmatrix}
    \begin{pmatrix}
    \tilde{Z}_\mu\\
    \tilde{A}_{\mu}\\
    \tilde{V}_{3,\mu}^0
    \end{pmatrix},
\end{eqnarray*}
where the superscripts denote the charge of the corresponding particles, and
\begin{align}
    m_W^\text{SM} &= \frac{g_W^{UV}v}{2},\\
    m_Z^\text{SM} &= \frac{v\sqrt{{g_W^{UV}}^2 + g^{\prime2}}}{2},
\end{align}
with $v$ being the SM Higgs VEV, and $g^\prime$ being the hyperweak $\mathrm{U}(1)_Y$ gauge coupling constant.
Furthermore, $c_W$ and $s_W$ are the cosine and sine of the weak mixing angle, and $\tilde{W}^\pm_\mu$, $\tilde{Z}_\mu$ and $\tilde{A}_\mu$ are the SM charged and neutral weak gauge bosons after EWSB, i.e.\ the mass eigenstates of the SM before the mixing with the dark $\rho$ mesons is taken into account. Likewise, $\tilde{V}^\pm_{3,\mu}=(\tilde{V}_3^1\mp i\tilde{V}_3^2)/\sqrt{2}$ and $\tilde{V}^0_{3,\mu}$ are the charge eigenstates of the triplet part of the dark $\rho$ mesons. The advantage of this route is that the mass matrix is block diagonal, and thus can be diagonalized by three separate rotations for the different charge sectors. 
These are written in the charge eigenstate basis to make the mass matrix block diagonal. Note that the quintet part of the dark $\rho$ mesons does not mix with the $W$ bosons, and thus their masses are simply given by 
\begin{equation}
    m_{V_5}^2 = a  f_d^2 g_\rho^2.
\end{equation}
The mass eigenstates can be obtained by diagonalizing the mass matrices by orthogonal transformations. We see immediately that the neutral sector has a zero determinant, ensuring the existence of a massless eigenstate, which is identified as the photon. The total mass eigenvalues are given by
\begin{align}
    m_{W^\pm}^2 &= \frac{\mathcal{C}_{{m_{W}^{\text{SM}}}}}{2}- \frac{1}{2}\sqrt{\mathcal{C}_{{m_{W}^{\text{SM}}}}^2-4a f_d^2 g_{\rho }^2 {m_{W}^{\text{SM}}}^2},\label{eq:mass_Wpm}\\
    m_{Z}^2 &= \frac{\mathcal{C}_{{m_{Z}^{\text{SM}}}}}{2} -\frac{1}{2}\sqrt{\mathcal{C}_{{m_{Z}^{\text{SM}}}}^2-4 a f_d^2
   {m_{Z}^{\text{SM}}}^2 \left(g_{\rho }^2+4{g_W^{UV}}^2 s_W^2\right)},\label{eq:mass_Z}\\
   m_{A}^2 &= 0,\\    
    m_{V_3^\pm}^2 &= \frac{\mathcal{C}_{{m_{W}^{\text{SM}}}}}{2}+\frac{1}{2}\sqrt{\mathcal{C}_{{m_{W}^{\text{SM}}}}^2-4a f_d^2 g_{\rho }^2 {m_{W}^{\text{SM}}}^2},\label{eq:mass_V3pm}\\
    m_{V_3^0}^2 &= \frac{\mathcal{C}_{{m_{Z}^{\text{SM}}}}}{2}+\frac{1}{2}\sqrt{\mathcal{C}_{{m_{Z}^{\text{SM}}}}^2-4 a f_d^2{m_{Z}^{\text{SM}}}^2 \left(g_{\rho }^2+4{g_W^{UV}}^2 s_W^2\right)},
\end{align}
where
\begin{equation*}
    \mathcal{C}_{m} := a f_d^2 (g_\rho^2 + 4{g_W^{UV}}^2) + m^2.
\end{equation*}
In the limit of $g_\rho \gg g_W^{UV}$, or equivalently $m_\rho \gg m_{W,Z}$, which corresponds to decoupling the dark $\rho$ mesons from the SM gauge bosons, we recover 
\begin{eqnarray}
    m_{W^\pm}^2 \approx {m_{W}^{\text{SM}}}^2, \quad m_Z^2 \approx {m_{Z}^{\text{SM}}}^2, \quad m_A^2 = 0,\\
    m_{V_3^\pm}^2 = a f_d^2 g_\rho^2+{g_W^{UV}}^2\left(4af_d^2+\frac{4{m_W^{\mathrm{SM}}}^2}{g_\rho^2}\right)+\mathcal{O}({g_W^{UV}}^4), \\
    m_{V_3^0}^2 = a f_d^2 g_\rho^2+{g_W^{UV}}^2\left(4af_d^2+\frac{4{m_W^{\mathrm{SM}}}^2}{g_\rho^2}\right)+\mathcal{O}({g_W^{UV}}^4),
\end{eqnarray}
and the degeneracy between the triplet and quintet dark $\rho$ mesons is restored. In the opposite limit of $g_\rho \ll g_W^{UV}$, the masses for the gauge bosons decrease, giving us a natural lower bound for the coupling $g_\rho$ by using the experimental measurements of the SM $W$ and $Z$ boson masses, which will be discussed in the following sections.
Similarly, the diagonalization of the mass matrices gives us the mass eigenstates as linear combinations of the original fields. For the charged sector, the mass eigenstates are given by
\begin{equation*}
    \begin{pmatrix}
    W_\mu^\pm\\[4pt]
    V_{3,\mu}^\pm
    \end{pmatrix}
    =
    \begin{pmatrix}
    \cos\alpha_\pm & -\sin\alpha_\pm\\[4pt]
    \sin\alpha_\pm & \cos\alpha_\pm
    \end{pmatrix}
    \begin{pmatrix}
    \tilde{W}_\mu^\pm\\[4pt]
    \tilde{V}_{3,\mu}^\pm
    \end{pmatrix}.
\end{equation*}
where the mixing angle $\alpha_\pm$ is given by
\begin{equation*}
    \tan(\alpha_\pm) = -\frac{4 a f_d^2 g_{\rho } g_W^{UV}}{\sqrt{\mathcal{D}_{{m_{W}^{\text{SM}}}}+8 a f_d^2 {g_W^{UV}}^2 m_{W,\text{SM}}^2}-m_{W,\text{SM}}^2+a f_d^2 \left( g_{\rho }^2-4{g_W^{UV}}^2\right)},
\end{equation*}
with
\begin{equation*}
    \mathcal{D}_{m} := m^4-2 a f_d^2 g_{\rho }^2 m^2+a^2 f_d^4\left(g_\rho^2+4{g_W^{UV}}^2\right)^2.
\end{equation*}
For the neutral sector, the mass eigenstates are given by two successive rotations. 
\begin{equation*}
    \begin{pmatrix}
        A_\mu\\[4pt]
        Z_\mu\\[4pt]
        V_{3,\mu}^0
    \end{pmatrix}
    =
    \begin{pmatrix}
        1 & 0 & 0\\[4pt]
        0 & \cos\beta_0 & \sin\beta_0\\[4pt]
        0 & -\sin\beta_0 & \cos\beta_0
    \end{pmatrix}
    \begin{pmatrix}
        \cos\alpha_0 & 0 & \sin\alpha_0\\[4pt]
        0 & 1 & 0\\[4pt]
        -\sin\alpha_0 & 0 & \cos\alpha_0
    \end{pmatrix}
    \begin{pmatrix}
        \tilde{A}_\mu\\[4pt]
        \tilde{Z}_\mu\\[4pt]
        \tilde{V}_{3,\mu}^0
    \end{pmatrix},
\end{equation*}
where the mixing angles $\alpha_0$ and $\beta_0$ are given by
\begin{equation*}
    \tan(\alpha_0) = \frac{2g_W^{UV} s_W}{g_{\rho }}
\end{equation*}
and
\begin{equation*}
    \tan(\beta_0) = \frac{8 a c_W f_d^2 g_W^{UV} \sqrt{\frac{g_{\rho }^2}{4}+{g_W^{UV}}^2 s_W^2}}{\sqrt{\mathcal{D}_{{m_{Z}^{\text{SM}}}}+8 a f_d^2 m_{Z,\text{SM}}^2 {g_W^{UV}}^2 \left(c_W^2-s_W^2\right)}-m_{Z,\text{SM}}^2+a f_d^2 \left(4{g_W^{UV}}^2
   \left(s_W^2-c_W^2\right)+g_{\rho }^2\right)},
\end{equation*}
which, in the limit of $m_\rho\gg m_{W,Z}$ becomes
\begin{equation*}
    \tan(\beta_0)= \frac{2g_W^{UV}c_W}{g_\rho}+\mathcal{O}\left(g_\rho^{-3}\right).
\end{equation*}
Since the transformations are orthogonal, the mixings do not cause any additional kinetic mixing terms, removing the need for further field redefinitions. As expected, in the limit of $g_\rho \gg g_W$, the mixing angles $\alpha_\pm$, $\alpha_0$ and $\beta_0$ all approach zero, and thus the mass eigenstates approach the original fields, consistent with the masses.

Furthermore, it is important to note that due to the structure of the covariant derivative, weak coupling in the UV theory is not the same as weak coupling in the IR theory. Since the experimentally acquired value for the weak gauge coupling is the IR value, one can determine the UV value of the weak gauge coupling by using the mixing angles obtained from the diagonalization of the mass matrix. Specifically, we obtain
\begin{equation}
     g_{W }^{UV}= \frac{g_W^{IR}g_\rho}{\sqrt{g_{\rho }^2-4{g_W^{IR}}^2}}.
\end{equation}
This requires $g_\rho>2g_W^{IR}$. The UV value of the $\rho$ meson coupling remains a free parameter.
The resulting interaction structure is particularly important for the phenomenology of the model. After rotating to the mass eigenstate basis, the gauge interactions inherited from the covariant derivatives induce couplings between the physical $W/Z/\gamma$ bosons, the dark $\rho$ mesons, and the dark pions. In particular, the triplet vector mesons acquire suppressed but non-vanishing couplings to the SM sector through their admixture with $\tilde{W}_\mu$ and $\tilde{Z}_\mu$, while the quintet vectors do not mix with the SM gauge bosons. In general, for any SM lepton, the coupling of the charged $\rho$ triplets to the SM fermions is given by
\begin{equation}
    g_{V_3^\pm}^{\ell \nu} = \frac{g_W^{UV}}{2\sqrt{2}}\sin\alpha_\pm.\label{eq:glnu}
\end{equation}
Likewise, defining the overlap parameters
\begin{equation}
    \kappa_\gamma \equiv -\cos\beta_0\sin\alpha_0, \qquad \kappa_Z \equiv -\sin\beta_0,\label{eq:kappa}
\end{equation}
the effective vector and axial couplings of $V_3^0$ to two same-flavor SM fermions are given as
\begin{align}
    g^{fV}_{V_3^0} &= \kappa_\gamma\, eQ_f
        + \kappa_Z\frac{g_W^{UV}}{2c_W}\!\left(T^3_f - 2Q_f\sin^2\!\theta_W\right),\label{eq:gfV} \\
    g^{fA}_{V_3^0} &= \kappa_Z\frac{g_W^{UV}}{2c_W}\,T^3_f\label{eq:gfA},
\end{align}
where $c_W = \cos\theta_W$. This implies that the triplet states can mediate both production and decay channels involving SM gauge bosons and dark pions, while the quintet states can only participate in processes involving dark pions. More specific interaction vertices can be derived by substituting the mass eigenstate expressions back into the original Lagrangian. 

\subsection{Accidental Symmetries}\label{sec:accidental_symmetry}
The model under consideration has several accidental symmetries that play an important role in the phenomenology. Specifically, for a given global symmetry, the model may exhibit additional conserved quantities or selection rules that constrain the allowed interactions and decay channels, as the lightest \textit{odd} dark particle is protected by the symmetry and thus stable, given the SM particles are \textit{even} under the very same symmetry. 
One immediate accidental symmetry, similar to the QCD case, is the baryon number symmetry, which is a global $U(1)$ symmetry under which the dark quarks transform as
\begin{equation}
    \mathbf{q}_i \to e^{i\alpha} \mathbf{q}_i.
\end{equation}
This symmetry is not explicitly broken by any of the terms in the Lagrangian, and thus is an accidental symmetry of the model. The lightest baryon in the dark sector is stable due to this symmetry, and thus its contribution to the dark matter density will be considered in the following sections \cite{nussinov1985technocosmology}. We will assume its mass to be approximately given by $m_B\approx 4\pi f_d$.

Another accidental symmetry of the model is \textit{$\mathcal{G}$-parity}, a discrete symmetry
that constrains the interactions among dark mesons and their decays into the SM particles. Following Ref.~\cite{Bai:2010qg}, $\mathcal{G}$-parity is defined as the
composition of dark charge conjugation and a global $\pi$-rotation in the
$\mathrm{SU}(2)_W$ flavor space,
\begin{equation}
    \mathcal{G}\;\equiv\; \mathcal{C}_{\mathrm{dark}} \otimes S, \qquad
    S = \exp\!\left(i\pi \tilde{T}_2\right),
\end{equation}
where $\mathcal{C}_{\mathrm{dark}}$ exchanges dark quarks with their antiquarks,
\begin{equation}
    \mathcal{C}_{\mathrm{dark}}: \quad \mathbf{q}_i \;\longleftrightarrow\; \bar{\mathbf{q}}_i,
\end{equation}
and $S$ implements a $\pi$-rotation about the second axis of the $\mathrm{SU}(2)_W$ flavor
space. In the triplet representation of $\mathrm{SU}(2)_W$, the matrix form of $S$ is
\begin{equation}
    S = \exp\!\left(i\pi \tilde{T}_2\right) \;=\;
    \begin{pmatrix} 0 & 0 & 1 \\ 0 & -1 & 0 \\ 1 & 0 & 0 \end{pmatrix},
\end{equation}
which maps $(q_{+1},\,q_0,\,q_{-1})^T \to (q_{-1},\,-q_0,\,q_{+1})^T$, where
subscripts denote the $T_3$ eigenvalue. The combination $\mathcal{G}$ is an exact symmetry of
the Lagrangian in Eq.~\eqref{eq:lag_quark}: $\mathcal{C}_{\mathrm{dark}}$
exchanges $\mathbf{q}$ and $\bar{\mathbf{q}}$, which enter symmetrically due to the
vector-like representation, while $S$ compensates the sign change that
$\mathcal{C}_{\mathrm{dark}}$ introduces in the $\mathrm{SU}(2)_W$ covariant derivative.

For a composite dark meson in the weak-isospin $J$ representation of $\mathrm{SU}(2)_W$,
the $\mathcal{G}$-parity eigenvalue of its neutral ($T_3 = 0$) component is \cite{Bai:2010qg}
\begin{equation}
    \mathcal{G} = (-1)^J \times \mathcal{C}_{\mathrm{dark}}(\text{neutral}),
\end{equation}
where $\mathcal{C}_{\mathrm{dark}}(\text{neutral}) = +1$ for pseudoscalar mesons and $-1$
for vector mesons, by direct analogy with the charge conjugation of their QCD
counterparts. Evaluating this for each species in the spectrum:
\begin{itemize}
    \item \textbf{Triplet dark pions $\Pi_3$} ($J=1$, pseudoscalar):
          $\mathcal{G}(\Pi_3) = (-1)^1 \times (+1) = -1$. $\mathcal{G}$-parity odd.
    \item \textbf{Quintet dark pions $\Pi_5$} ($J=2$, pseudoscalar):
          $\mathcal{G}(\Pi_5) = (-1)^2 \times (+1) = +1$. $\mathcal{G}$-parity even.
    \item \textbf{Triplet dark $\rho$ mesons $V_3$} ($J=1$, vector):
          $\mathcal{G}(V_3) = (-1)^1 \times (-1) = +1$. $\mathcal{G}$-parity even.
    \item \textbf{Quintet dark $\rho$ mesons $V_5$} ($J=2$, vector):
          $\mathcal{G}(V_5) = (-1)^2 \times (-1) = -1$. $\mathcal{G}$-parity odd.
\end{itemize}
Since the SM particles are not charged under the dark $\mathrm{SU}(N_c)$ gauge group
or the dark flavor symmetry, they carry $\mathcal{G} = +1$ by convention. A $\mathcal{G}$-parity-odd dark
particle therefore cannot decay into a purely SM final state, as this would require
a net change in $\mathcal{G}$. The lightest $\mathcal{G}$-parity-odd state is thus absolutely stable. In
our model, the triplet dark pions $\Pi_3$ are lighter than the quintet pions $\Pi_5$
(see section~\ref{sec:mass}) and lighter than the dark $\rho$ mesons. They are
therefore the lightest $\mathcal{G}$-parity-odd states and constitute a cosmologically stable
DM candidate.

The decay channels of the heavier species follow directly. The quintet dark pions
$\Pi_5$ are $\mathcal{G}$-parity even and can decay into the SM particles through the WZW term in
Eq.~\eqref{eq:wzw}. The quintet dark $\rho$ mesons $V_5$ are $\mathcal{G}$-parity odd but
heavier than $\Pi_3$; the dominant decay channel within the dark sector is
$V_5 \to \Pi_3 + \Pi_5$, which is $\mathcal{G}$-parity allowed,
provided it is kinematically accessible. If this two-body channel is closed, $V_5$ instead decays through an
off-shell $\Pi_5$, giving the three-body final state
$V_5 \to \Pi_3\,\Pi_5^{*} \to \Pi_3\,WW$. The triplet dark $\rho$ mesons $V_3$ are $\mathcal{G}$-parity even and can in principle decay to both dark pions and SM particles. Their allowed channels are further constrained by a second discrete symmetry which we now discuss.

\medskip

A further accidental discrete symmetry of the model is \textit{$\mathcal{H}$-parity}, introduced recently in Ref.~\cite{Asadi:2024bbq} in the context of vector-like dark sectors. Since this symmetry becomes relevant for specific collider observables concerning dark $\rho$ decays into dark pions that are not handled in this work, we briefly discuss its effects in the appendix~\ref{app:1}

The symmetry quantum numbers and resulting stability of each species are summarised
in Table~\ref{tab:symmetries}.

\begin{table}[tbp]
    \centering
    \begin{tabular}{l|ccc|l}
        \toprule
        \textbf{Particle} & $\mathcal{G}$ & $\mathcal{H}$ (neutral) & $U(1)$
            & \textbf{Status} \\
        \midrule
        $\Pi_3$ & $-1$ & $-1$ & $0$
            & Stable ($\mathcal{G}$-parity) \\
        $\Pi_5$ & $+1$ & $+1$ & $0$
            & Unstable (WZW $\to$ SM gauge bosons) \\
        $V_3$ & $+1$ & $-1$ & $0$
            & Unstable (EW mixing) \\
        $V_5$ & $-1$ & $+1$ & $0$
            & Unstable \\
        Dark baryon $B$ & $-$ & $\pm 1$ & $\pm 1$
            & Stable (dark baryon number) \\
        \bottomrule
    \end{tabular}
    \caption{Accidental symmetry quantum numbers and stability of each dark-sector
    species. The $\mathcal{H}$-parity eigenvalue is given for the neutral ($m=0$) component;
    charged components are not eigenstates but map to their charge conjugates under
    $\mathcal{H}$, with a sign that depends on the representation. The dark baryon
    number charge is $1/N_c$ per dark quark.}
    \label{tab:symmetries}
\end{table}

%% file: sections/section3.tex
\section{Phenomenology}\label{sec:pheno}
In this section, we discuss the phenomenology of the model, especially the relic abundance of the DM candidates, and include the effects of the dark $\rho$ mesons on the relevant processes. We also comment on other phenomenological aspects such as direct and indirect detection, and collider signatures, although a detailed analysis of these is left for future work.

As mentioned in the previous sections, the lightest $\mathcal{G}$-parity odd dark pion $\Pi_3$ is stable and can serve as a DM candidate. Furthermore, the lightest dark baryon is also stable due to the conservation of the dark baryon number. Considering the QCD-like dynamics with $N_c=3$, the dark baryons form representations of the flavor symmetry $SU(3)_V$, decomposing as $\mathbf{3} \otimes \mathbf{3} \otimes \mathbf{3} = \mathbf{10} \oplus \mathbf{8} \oplus \mathbf{8} \oplus \mathbf{1}$. Since each dark quark carries $T_3 = Q \in \{+1, 0, -1\}$ under the embedded $SU(2)_W$, three-quark combinations can sum to a net $T_3 = 0$, so these representations contain baryon states that can be neutral on top of being stable.

We determine the available parameters of the model, i.e., the dark pion decay constant $f_d$, the dark quark mass $m$, and the HLS coupling $g_\rho$ by requiring that the relic abundance of the dark matter candidates matches the observed value. Furthermore, we also consider different values for the ratio $a=f_\rho^2/f_d^2$, but focus on the case $a=2$, which is motivated by the QCD analogy. 

\paragraph{Electroweak constraints.}
The HLS coupling $g_\rho$ sets the dark $\rho$ mass through the dark-KSRF relation and controls the mixing with the SM gauge bosons, which is suppressed by inverse powers of $g_\rho$. Electroweak precision data therefore translate into a \emph{lower} bound on $g_\rho$ at fixed $\sqrt{a}f_d$. Since we work at tree level, we constrain the model through the universal oblique parameters $\hat S$, $\hat T$, $\hat U$, $W$ and $Y$~\cite{Barbieri:2004qk}. Integrating out the dark $\rho$ triplet generates only the parameter $W$, with $\hat S=\hat T=\hat U=Y=0$, so that all bounds take the form $\sqrt{a}f_d \gtrsim C/g_\rho^{2}$. Using the fits of LEP~\cite{Barbieri:2004qk}, the LHC Drell--Yan data~\cite{CMS:2026dilepton} and the projected FCC-ee Scenario~S1~\cite{Greljo:2025ggc}, we find $C_{\rm LEP}=2.754~\mathrm{TeV}$, $C_{\rm DY}=8.30~\mathrm{TeV}$ and $C_{\rm S1}=28.6~\mathrm{TeV}$ at $95\%$~CL. The details and the resulting exclusion plot are given in appendix~\ref{app:2}. These limits are not very strong: for the QCD-like benchmark $g_\rho=6$ even the FCC-ee projection only requires $\sqrt{a}f_d \gtrsim 795~\mathrm{GeV}$, well below the multi-TeV decay constants favored by the relic density analysis of section~\ref{sec:results}. Only for $g_\rho\sim\mathcal{O}(1)$ do current data push $\sqrt{a}f_d$ into the multi-TeV range.

\subsection{Relic Density}
In the following, we calculate the relic abundance of the dark matter candidates in the freeze-out mechanism, taking into account the effects of the dark $\rho$ mesons and the lightest stable baryon on the relevant annihilation processes. The model was implemented in \texttt{FeynRules}~\cite{Alloul:2013bka}, and the resulting model files was used to compute the relevant annihilation cross sections and the dark matter relic abundance with \texttt{micrOMEGAs}~\cite{Alguero:2023zol}. We then determine the parameters for which the relic abundance matches $\Omega_{\rm DM}h^2=0.12$. 
For the triplet dark pions, we consider annihilation into SM gauge bosons ($\Pi_3\Pi_3 \to WW$), dark pions ($\Pi_3\Pi_3 \to \Pi_5\Pi_5$), and mixed channels ($\Pi_3\Pi_3 \to \Pi_5 W,~VV,~WV$, and $\Pi_5 V$), where $V$ denotes $V_3$ and $V_5$.
\begin{figure}[!tb]
    \centering
    \begin{tabular}{cccc}
        $\vcenter{\hbox{\includegraphics[height=1.75cm]{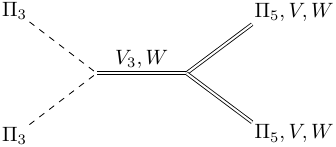}}}$ &
        $\vcenter{\hbox{\includegraphics[height=2.45cm]{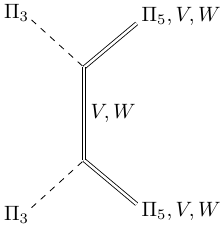}}}$ &
        $\vcenter{\hbox{\includegraphics[height=2.45cm]{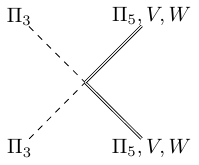}}}$ &
        $\vcenter{\hbox{\includegraphics[height=1.75cm]{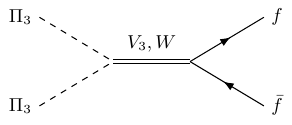}}}$ \\
        (a) & (b) & (c) & (d)
    \end{tabular}
    \caption{Feynman diagrams for the annihilation processes of the triplet dark pions $\Pi_3\Pi_3\to ij$. The labels refer to (a) $s$-channel, (b) $t$-channel, (c) contact diagrams, and (d) $s$-channel annihilation into SM fermions. The double lines indicate both vector ($V_3, W$) and scalar ($\Pi_5$) fields.}
    \label{fig:feynman_p3p3}
\end{figure}
 \subsubsection{Boltzmann Equations}
 
Let us first discuss the relic abundance of the lightest stable dark pion $\Pi_3$. Here we show the qualitative behavior of the DM annihilation cross section and the relic abundance of all components of $\Pi_3$ since the masses for neutral and charged components of $\Pi_3$ are approximately degenerate.
Coannihilation among the components of $\Pi_3$ plays an important role because of the finite mass splitting. We denote $n_{\Pi_3}$ as the total number density of $\Pi_3^0$ and $\Pi_3^\pm$.
 
The time evolution of $n_{\Pi_3}$ is then described by the single Boltzmann equation
\begin{equation}\label{eq:boltzmann_unstable}
    \dot{n}_{\Pi_3} + 3H n_{\Pi_3} \simeq
    -\langle\sigma v\rangle_{\Pi_3\Pi_3}
    \!\left(n_{\Pi_3}^2 - n_{\Pi_3,\rm eq}^2\right),
\end{equation}
where $\langle\sigma v\rangle_{\Pi_3\Pi_3 }$ is the total thermally averaged annihilation cross section into final states with $i,j\in\{W,V,\Pi_5,f\}$, with $W$ representing the $\mathrm{SU}(2)_W$ gauge bosons. $V$ represents both the triplet and quintet dark $\rho$, and $f$ the SM fermions. Tree-level diagrams which contribute to each of these cross sections are shown in Figs.~\ref{fig:feynman_p3p3}. We do not include the Sommerfeld effect.

\subsubsection{Dark Baryon Contribution}
 
Let us estimate the annihilation cross section relevant to the relic abundance of the dark baryon from an analogy with baryon annihilation in real QCD. For this, we use the approximation done in \cite{Abe:2024mwa}. In the
non-relativistic limit, the $s$-wave annihilation cross section is parametrized as
\begin{equation}
    \sigma_B \simeq \frac{4\pi}{k}\,\mathrm{Im}\,b,
\end{equation}
where $b$ is the scattering length and $k$ is the center-of-mass momentum. The annihilation cross section for baryons in real QCD has been measured in the OBELIX experiment \cite{OBELIX:1997nkw}, which shows $\mathrm{Im}\,b \simeq 0.5~\mathrm{fm} \simeq 2.4/m_N$ for $\bar{n}p$ annihilation, where $m_N$ is the nucleon mass. Scaling to the dark sector, the thermally averaged annihilation cross section for the dark baryon is estimated as
\begin{equation}\label{eq:sigma_baryon}
    \langle \sigma_B v \rangle \simeq c\,\frac{4\pi}{m_B^2},
\end{equation}
where $c$ is an $\mathcal{O}(1)$ parameter. Then, by using
$\Omega h^2 \simeq 0.12 \times (2\times10^{-26}~\mathrm{cm}^3/\mathrm{s})/\langle\sigma v\rangle$
\cite{Kolb:1990vq}, the relic abundance of the dark baryon is estimated as
\begin{equation}\label{eq:omega_baryon}
    \Omega_B h^2 \simeq \frac{0.12}{c}\left(\frac{f_d}{6.66~\mathrm{TeV}}\right)^2.
\end{equation}
Here we take a naive estimate of the baryon mass as $m_B \sim \Lambda_d \sim 4\pi f_d$.
The total relic abundance is the sum of the two contributions,
\begin{equation}
    \Omega_{\rm DM} h^2 = \Omega_{\chi} h^2 + \Omega_B h^2,
\end{equation}
where $\Omega_\chi h^2$ consists of dark meson, namely the neutral component of $\Pi_3$.
 
The relative contributions of the dark pions and dark baryons to the total relic abundance depend sensitively on the value of $c$ and on the dark sector parameters.
\subsubsection{Relic Abundance Contours}\label{sec:results}
We now present the main results of the relic density calculation. For each value of $a \in \{2, 3, 7\}$ and each of the two representative HLS coupling values $g_\rho = 6$, $g_\rho = \sqrt{4\pi}$ as well as the high mass limit $m_{V_3^0}\gg f_d$, we determine the curve in the $(m, f_d)$ plane satisfying $\Omega_{\rm DM}h^2 = 0.12$ for three values of the baryon parameter $c \in \{0.3, 1, 3\}$, and identify the $(a, m)$-plane colormaps of the required $f_d$ for values of $a$ in the interval $a\in[0,10]$. To this end, we compute $\Omega_\chi h^2$ with \texttt{micrOMEGAs} on a logarithmic grid with $m\in[10^{-3},10^{2}]~\mathrm{TeV}$ and $f_d\in[0.1,50]~\mathrm{TeV}$, and extract the relic density contours as level sets of $\Omega_{\rm DM}h^2$ on this grid. Since $\Omega_B h^2$ is known analytically, one scan serves all values of $c$. This two-dimensional treatment is necessary because, close to the $\rho$ resonance, the contours are not single-valued in $m$, as shown below.
 
\begin{figure}[!tb]
    \centering

    \begin{subfigure}[t]{0.48\textwidth}
        \centering
        \includegraphics[width=\textwidth]{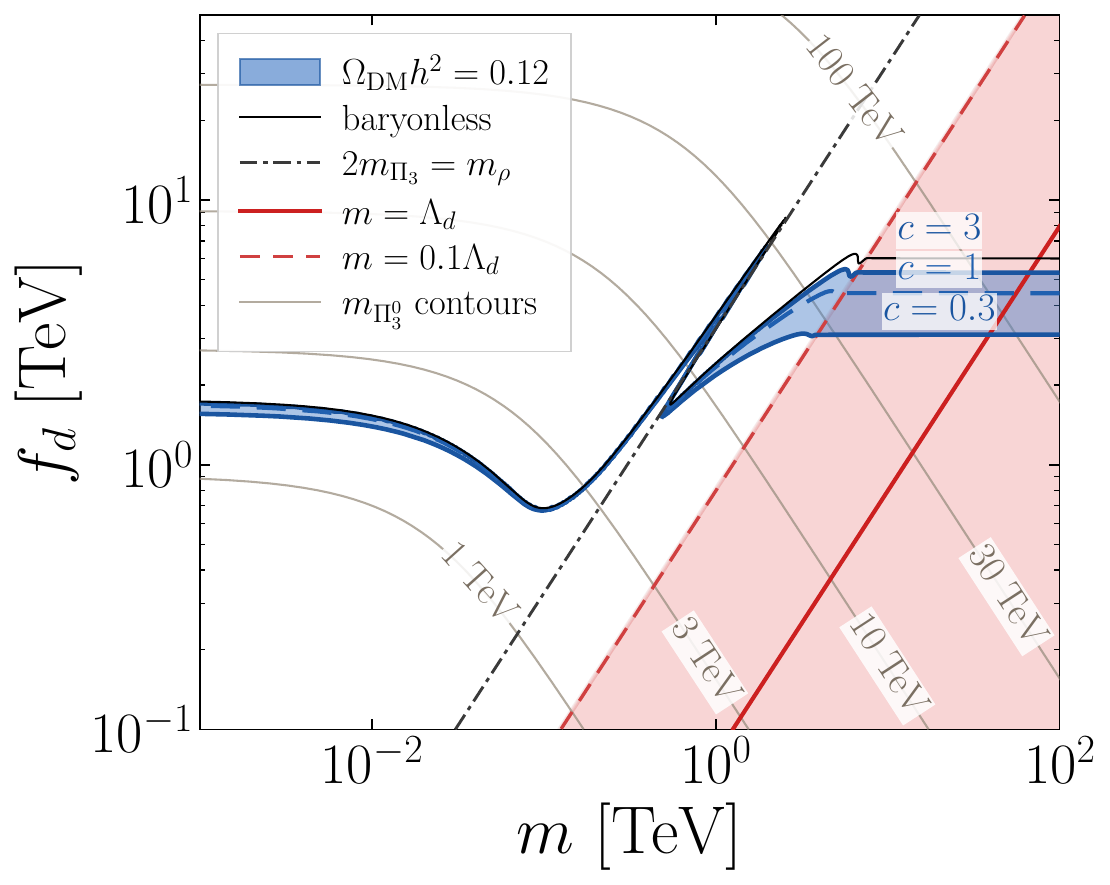}
        \caption{$g_\rho = 6$, $a = 2$}
    \end{subfigure}
    \hfill
    \begin{subfigure}[t]{0.48\textwidth}
        \centering
        \includegraphics[width=\textwidth]{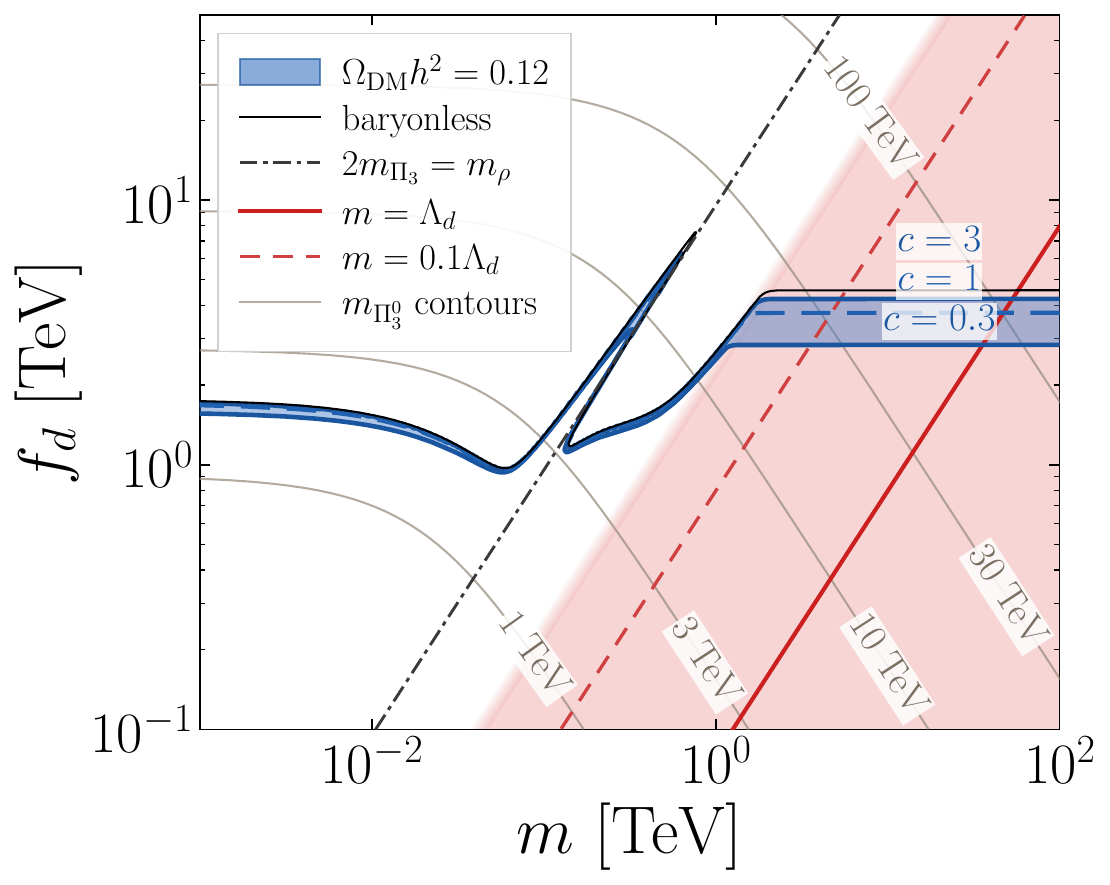}
        \caption{$g_\rho = \sqrt{4\pi}$, $a = 2$}
    \end{subfigure}

    \vspace{0.8em}

    \begin{subfigure}[t]{0.48\textwidth}
        \centering
        \includegraphics[width=\textwidth]{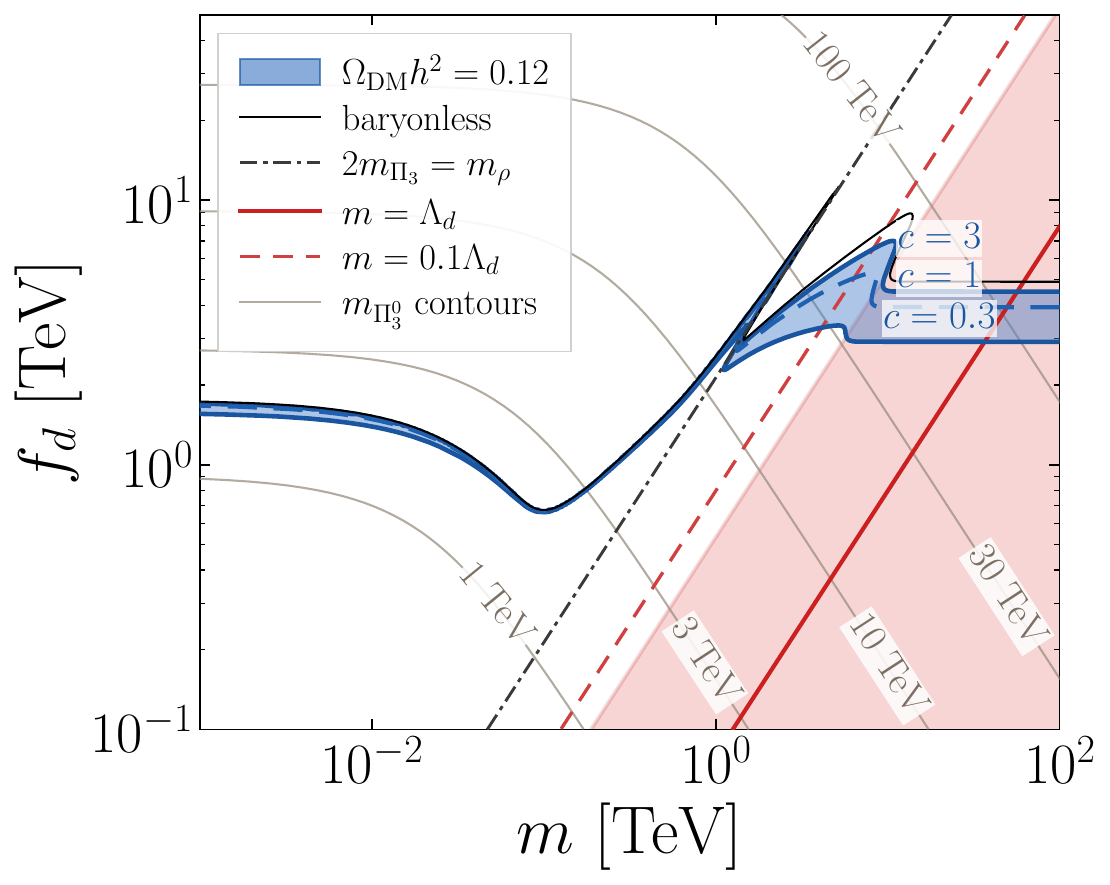}
        \caption{$g_\rho = 6$, $a = 3$}
    \end{subfigure}
    \hfill
    \begin{subfigure}[t]{0.48\textwidth}
        \centering
        \includegraphics[width=\textwidth]{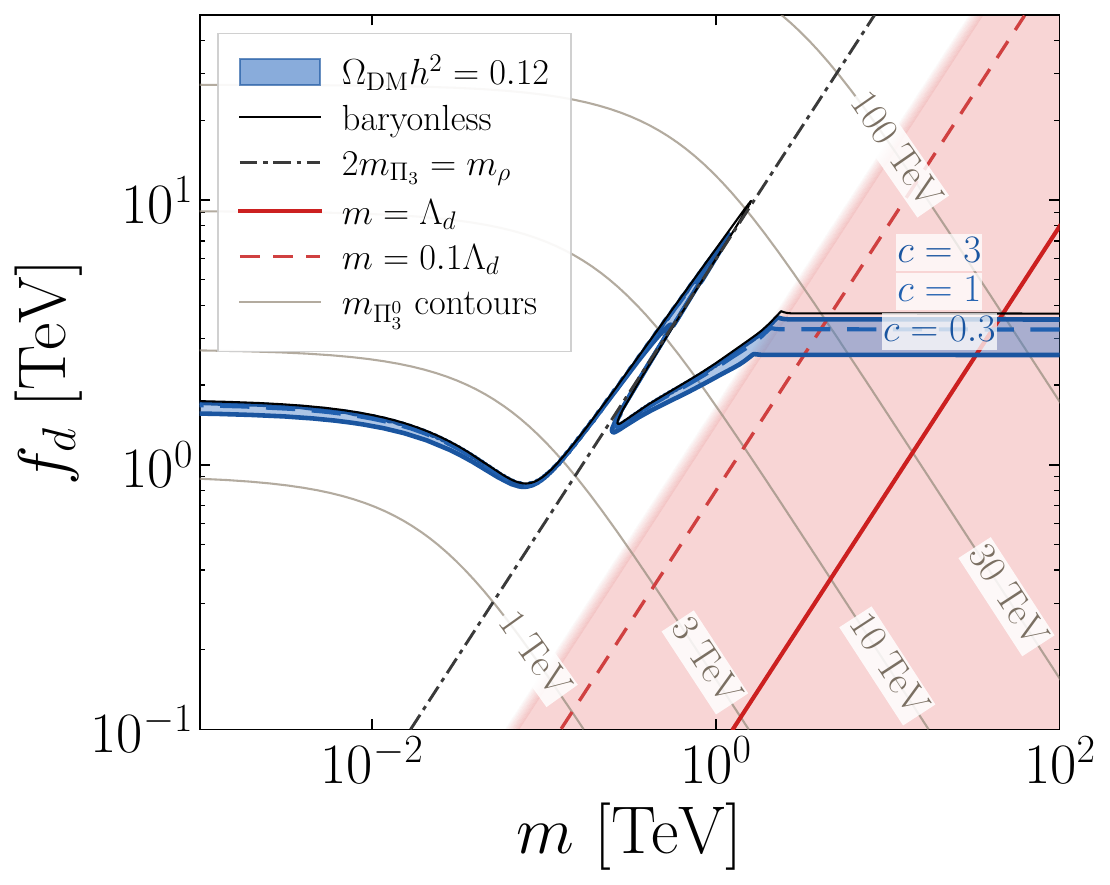}
        \caption{$g_\rho = \sqrt{4\pi}$, $a = 3$}
    \end{subfigure}

    \vspace{0.8em}

    \begin{subfigure}[t]{0.48\textwidth}
        \centering
        \includegraphics[width=\textwidth]{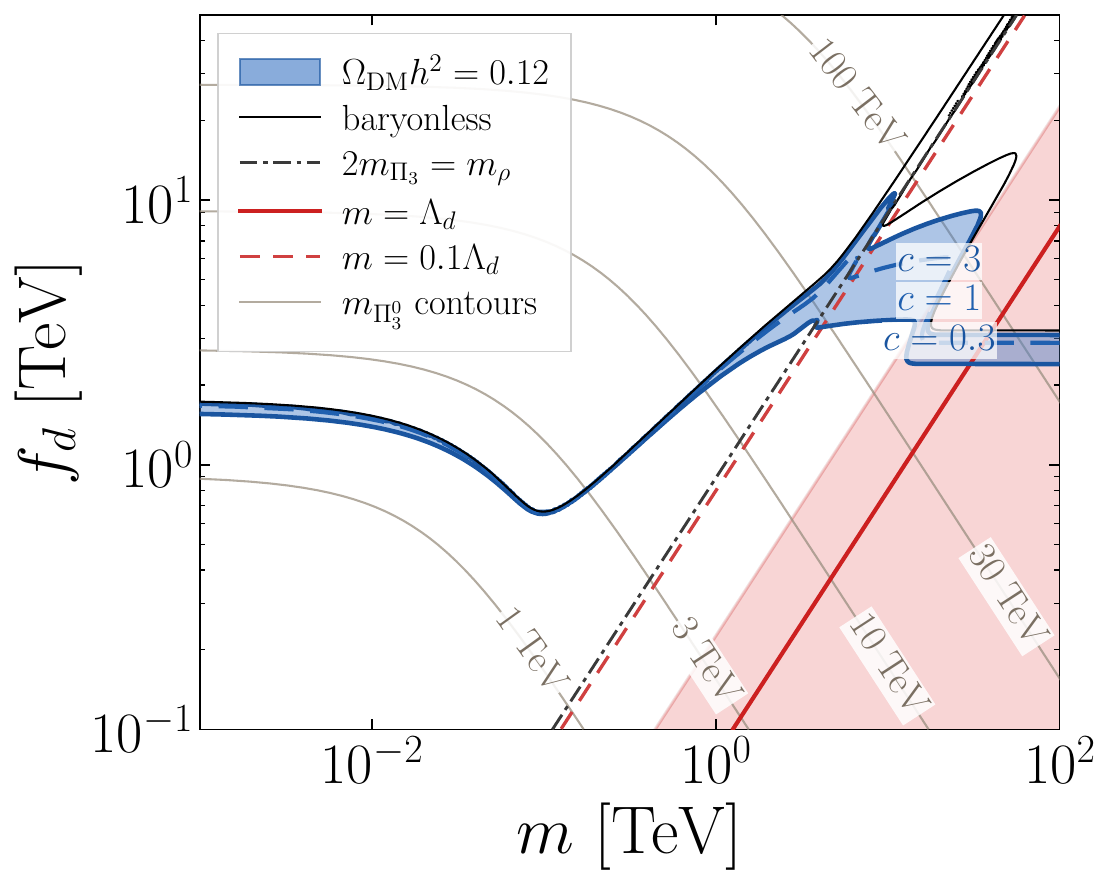}
        \caption{$g_\rho = 6$, $a = 7$}
    \end{subfigure}
    \hfill
    \begin{subfigure}[t]{0.48\textwidth}
        \centering
        \includegraphics[width=\textwidth]{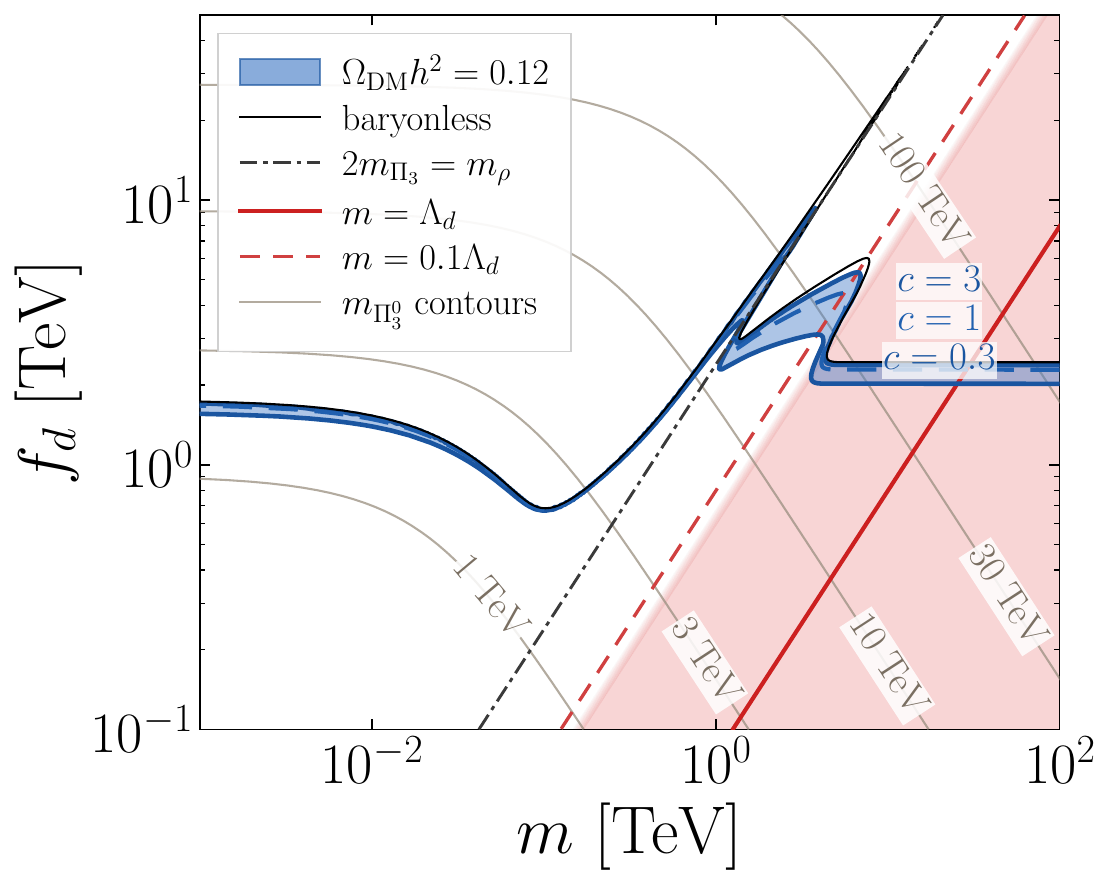}
        \caption{$g_\rho = \sqrt{4\pi}$, $a = 7$}
    \end{subfigure}

    \caption{%
        Relic density contours $\Omega_{\rm DM}h^2 = 0.12$ in the $(m, f_d)$ plane.
        Left column: $g_\rho = 6$; right column: $g_\rho = \sqrt{4\pi}$.
        Rows correspond to $a = 2$, $3$, $7$ (top to bottom).
        Within each panel the three curves correspond to $c = 0.3$, $1$, and $3$
        (bottom to top; $c=1$ dashed); the blue-shaded region marks the parameter
        space for intermediate $c$ values.
        Gray contours are lines of constant $m_{\Pi_3^0}$.
        The red solid (dashed) line marks $m = \Lambda_d$ ($m = 0.1\Lambda_d$).
        The red-shaded region indicates where the chiral EFT is unreliable, either due to the mass degeneracy between the multiplets or the quark mass approaching the confinement scale. The baryonless case is shown as the black line. The dash-dotted line marks the resonance condition $2m_{\Pi_3}=m_\rho$, along which the contours fold back.    
    }
    \label{fig:relic_combined}
\end{figure}

The results for the $(m, f_d)$ parameter space are shown in the figure~\ref{fig:relic_combined} for $g_\rho = 6$ and $g_\rho = \sqrt{4\pi}$. Each figure contains three panels corresponding to $a = 2$, $3$, and $7$. Within each panel, the three curves show the relic density contour for $c = 0.3$, $1$, and $3$, and the blue-shaded region between them indicates the variation in between. The gray lines are contours of constant dark matter mass $m_{\Pi_3^0}$. The thin black line corresponds to the baryonless case, for comparison. The dash-dotted line marks the resonance condition $2m_{\Pi_3}=m_\rho$. The red solid and dashed lines mark $m = \Lambda_d$ and $m = 0.1\,\Lambda_d$ respectively, indicating where the chiral EFT becomes unreliable. Furthermore, we indicate the limit for $m$, after which the $\rho$ meson masses become equal to the pion mass, where we consider the theory to be also unreliable. Combining those cases, we summarize this unreliable region as a red shaded area, where the border at which $\rho$ mesons and pions become equal is illustrated as a very sharp gradient, due to the mass differences between the multiplets.

Several features are common to all panels. For small quark masses
($m \lesssim 10^{-2}~\mathrm{TeV}$), the contours are approximately flat in $f_d$: mesons dominate the dark matter abundance, where the degeneracy between the dark pion multiplets is lifted. As a consequence of this, the annihilation channels containing quintet pions are either forbidden or strongly suppressed. Since the $c$ band is quite narrow, the baryonic contribution is quite small. At intermediate masses up to $m\lesssim 10~\mathrm{TeV}$, the mass difference between the pion multiplets becomes smaller, lifting the suppression of the quintet final product from the annihilation. This significantly increases the total annihilation cross section, reducing the amount of pionic dark matter. To compensate for this, corresponding cross sections, and consequently the scale needs to be reduced, causing the graphs to have a major dip in this region. This dip is most pronounced for larger $g_\rho$, and becomes shallower as $a$ decreases. For $g_\rho=6$, the minimum of the dip lies at $m\simeq0.1~\mathrm{TeV}$ and $f_d\simeq0.66~\mathrm{TeV}$ for all values of $a$. For $g_\rho=\sqrt{4\pi}$ and small $a$, the resonance discussed below already sets in near the bottom of the dip, which therefore remains shallower. The separation between the $c = 0.3$ and $c = 3$ contours reflects the varying baryon contribution: a larger $c$ requires the pion component $\Omega_\chi h^2$ to be larger, which in turn demands a smaller annihilation cross section and hence a larger $f_d$ at fixed $m$.

In addition, as seen from the $s$-channel diagram shown in figure~\ref{fig:feynman_p3p3}, we expect a resonant enhancement of the annihilation cross section when the center of mass energy approaches the dark $\rho$ meson mass. Considering the freeze-out temperatures, the thermally averaged cross section is expected to be dominated by the center of mass energies equal to the sum of the masses of the annihilating particles. Thus, the resonance is expected to occur when $2m_{\Pi_3} \sim m_{V_3}$. Indeed, close to the pole, the enhanced annihilation has to be compensated by a larger $f_d$, so one branch of the contour climbs along the resonance line up to $f_d$ of a few TeV and then folds back onto the non-resonant branch. Within this bifurcation, $\Omega_{\rm DM}h^2=0.12$ is satisfied for three different values of $f_d$ at the same $m$. The bifurcation stays narrow, since the resonant enhancement becomes weaker for larger $\sqrt{a}f_d$. Furthermore, since the mass of the dark $\rho$ mesons is proportional to $g_\rho$, the required pion mass for the resonance is higher for larger $g_\rho$, shifting the resonant peak to the higher end of the dark quark mass range. For the same reason, larger values of $a$, which make the dark $\rho$ mesons heavier at fixed $f_d$, move the bifurcation to larger $m$. For $g_\rho=6$ and $a=7$, the resonance line therefore runs close to the line $m=0.1\Lambda_d$. For even higher values, however, $\rho$ meson mass and pion mass become degenerate, and the validity of our model breaks down. As pions become heavier after this regime, we observe very steep changes in the graph, and it becomes horizontal in the unreliable regime. 
 \begin{figure}[!tbp]
    \centering
    \includegraphics[width=0.9\textwidth]{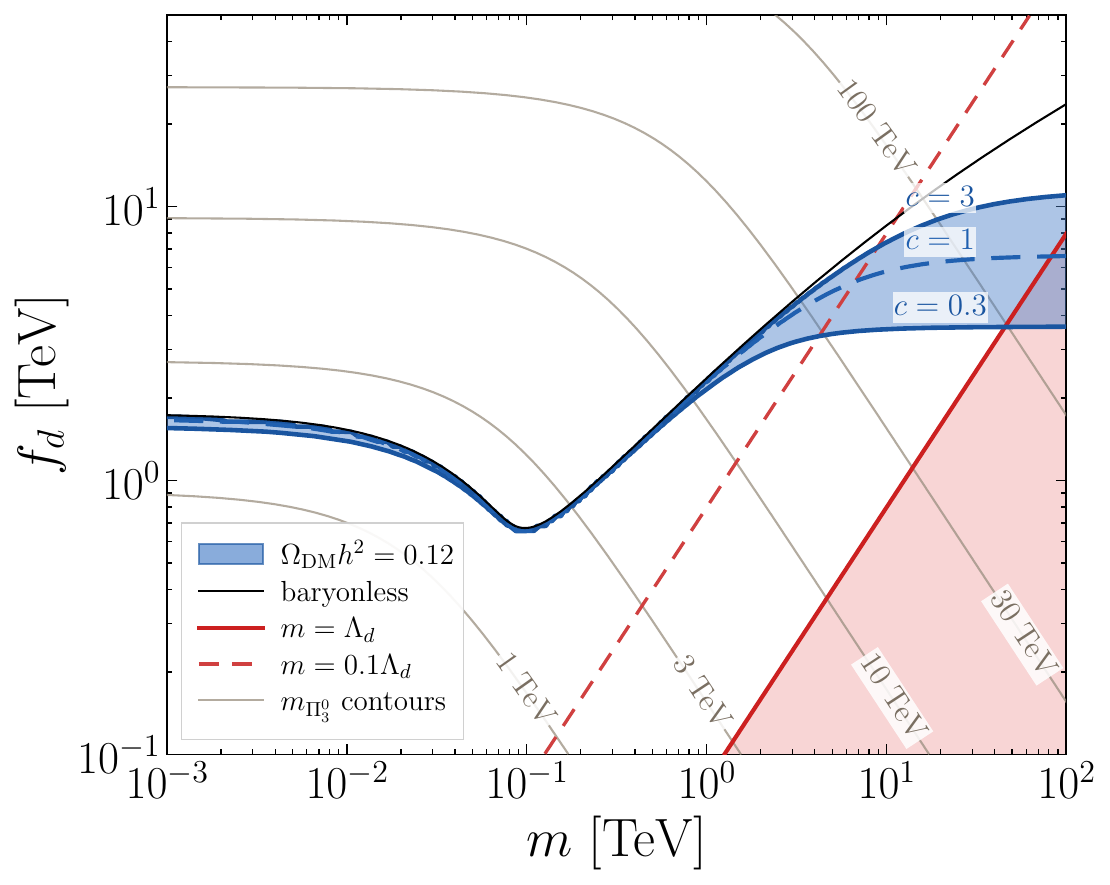}

    \caption{
    Relic density contours $\Omega_{\rm DM}h^2 = 0.12$ in the $(m, f_d)$ plane, for very heavy $\rho$ meson masses. In this limit, the contours are independent of $a$. The three curves correspond to $c = 0.3$, $1$, and $3$ (bottom to top; $c=1$ dashed); the blue-shaded region marks the parameter space for intermediate $c$ values. Gray contours are lines of constant $m_{\Pi_3^0}$. The red solid (dashed) line marks $m = \Lambda_d$ ($m = 0.1\Lambda_d$). The red-shaded region indicates where the chiral EFT is unreliable due to the quark mass approaching the confinement scale. The baryonless case is shown as the black line.}
    \label{fig:heavy_combined}
\end{figure}
\clearpage
Finally, we look at the $m_\rho \to \infty$ limit for a consistency check, where we expect the $\rho$ meson contribution to vanish. The corresponding relic abundance contours are shown in figure~\ref{fig:heavy_combined}. We obtain them with the same grid scan as above, with all $\rho$ masses far beyond the scanned range. Compared to the previous cases, pions and $\rho$ mesons cannot become degenerate, and the resonant enhancement of the annihilation cross section is absent. Thus, the graphs show a much smoother behavior without the resonant bifurcation, the contours are single-valued in $m$, and the validity boundary is determined solely by the chiral EFT breakdown at $m\approx \Lambda_d$. The pictorial interpretation of this limit is that we obtain additional four-point interactions between pions, stemming from diagrams with $\rho$ meson exchange. These, however, cancel the genuine four-point diagrams from $\mathcal{L}_V$, which leaves the theory independent of $a$. As such, we obtain the exact same graph as in the $\rho$-less case from \cite{Abe:2024mwa} in the heavy-mass limit.

\subsubsection{Scan over $a$}
 
\begin{figure}[!tb]
    \centering

    \begin{subfigure}[t]{0.48\textwidth}
        \centering
        \includegraphics[width=\textwidth]{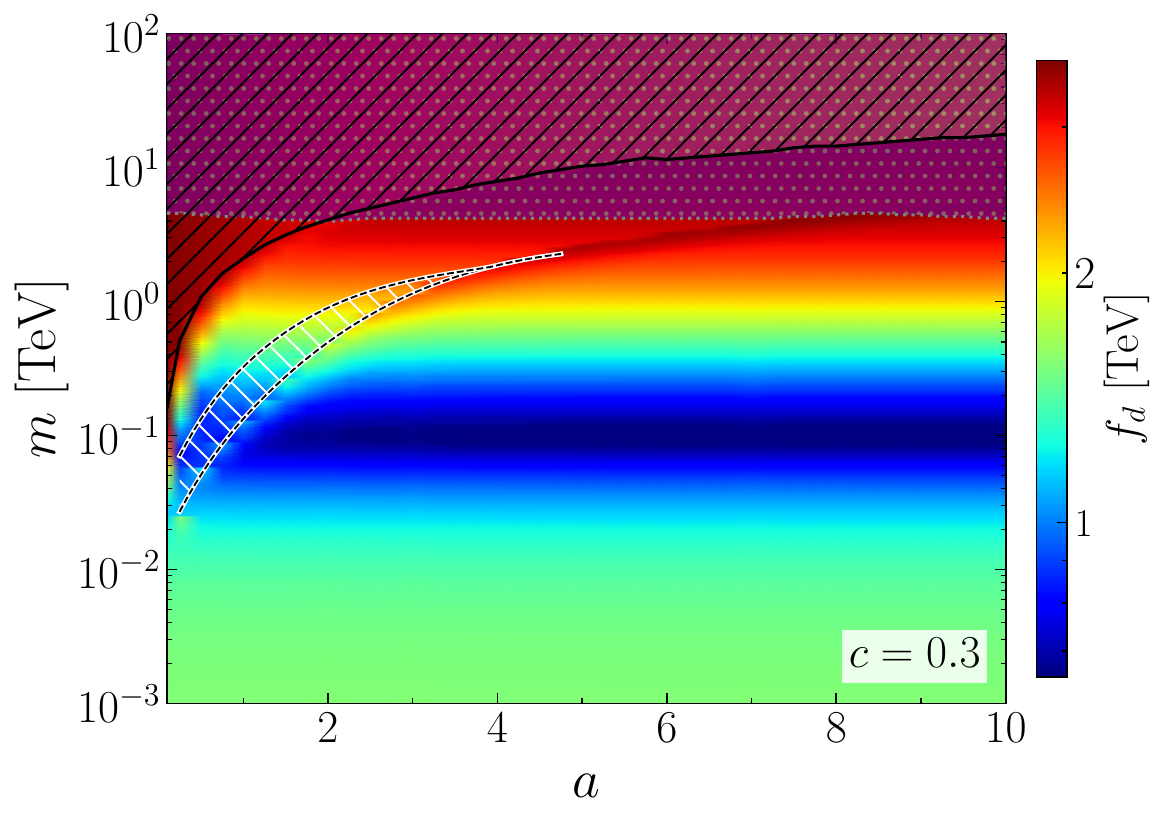}
        \caption{$g_\rho = 6$, $c = 0.3$}
    \end{subfigure}
    \hfill
    \begin{subfigure}[t]{0.48\textwidth}
        \centering
        \includegraphics[width=\textwidth]{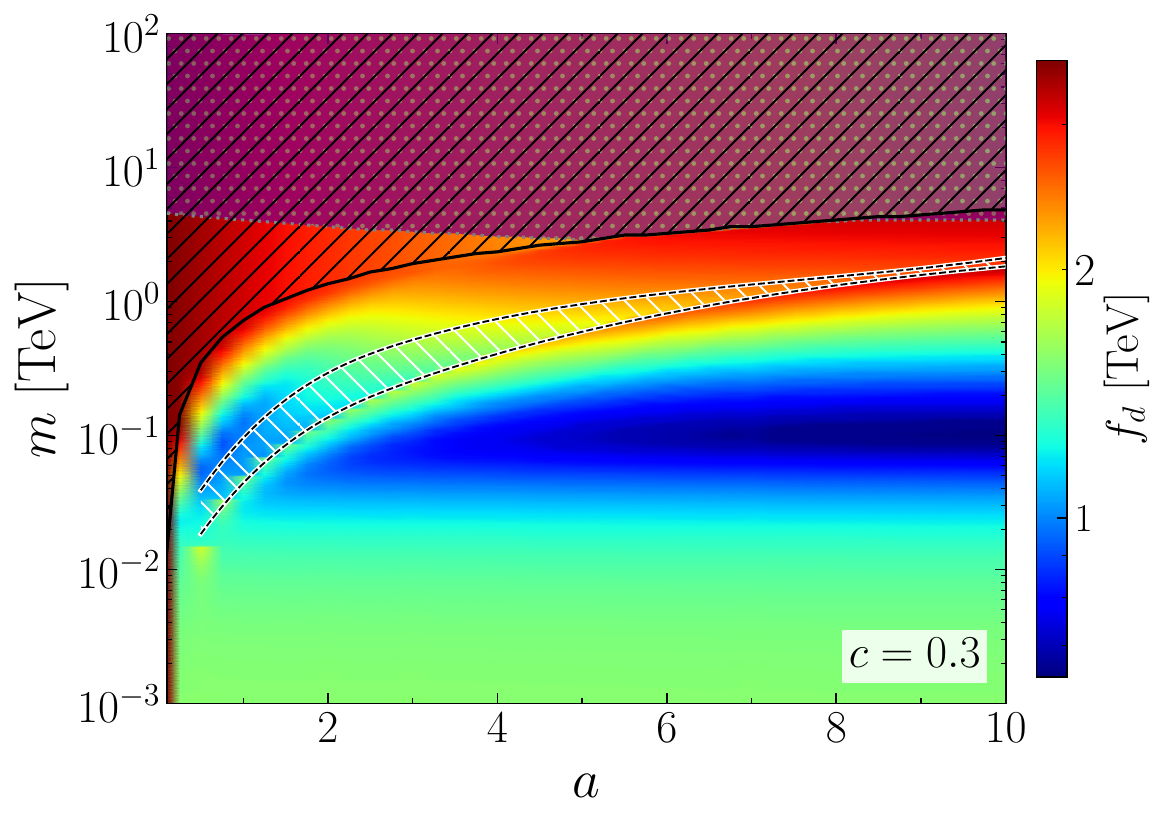}
        \caption{$g_\rho = \sqrt{4\pi}$, $c = 0.3$}
    \end{subfigure}

    \vspace{0.8em}

    \begin{subfigure}[t]{0.48\textwidth}
        \centering
        \includegraphics[width=\textwidth]{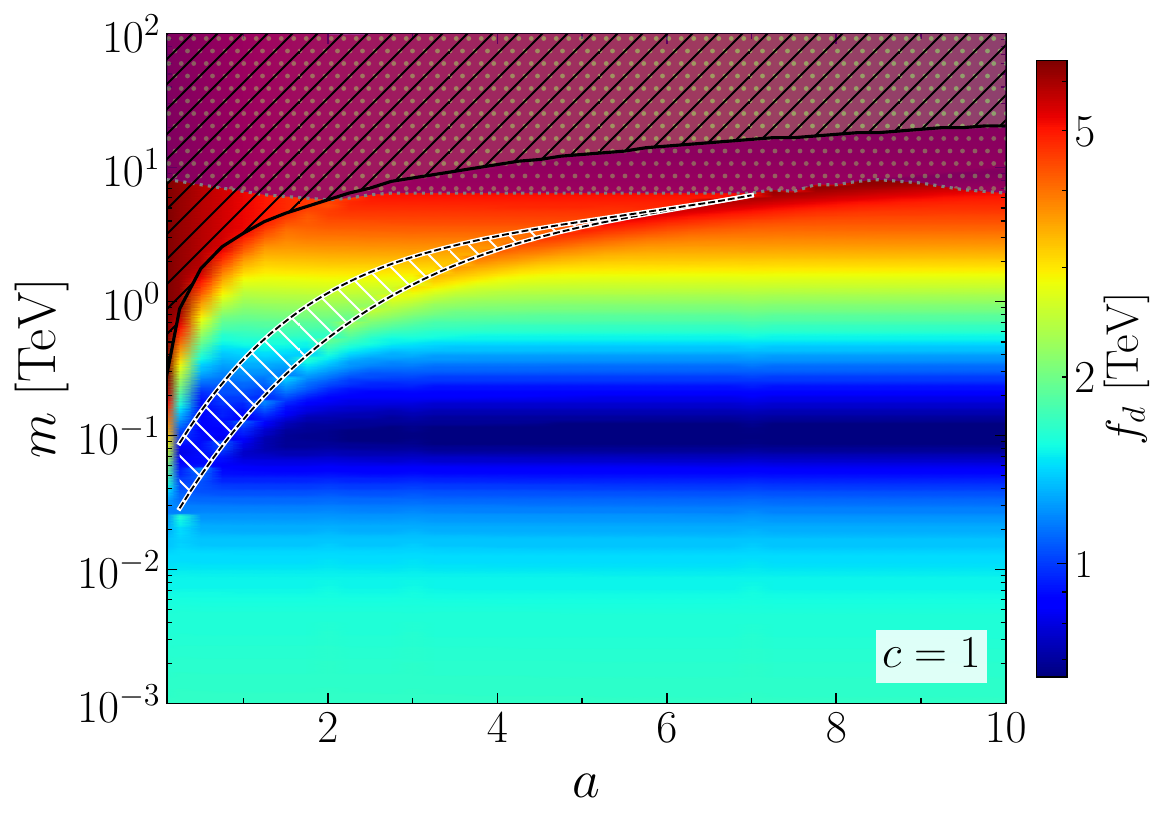}
        \caption{$g_\rho = 6$, $c = 1$}
    \end{subfigure}
    \hfill
    \begin{subfigure}[t]{0.48\textwidth}
        \centering
        \includegraphics[width=\textwidth]{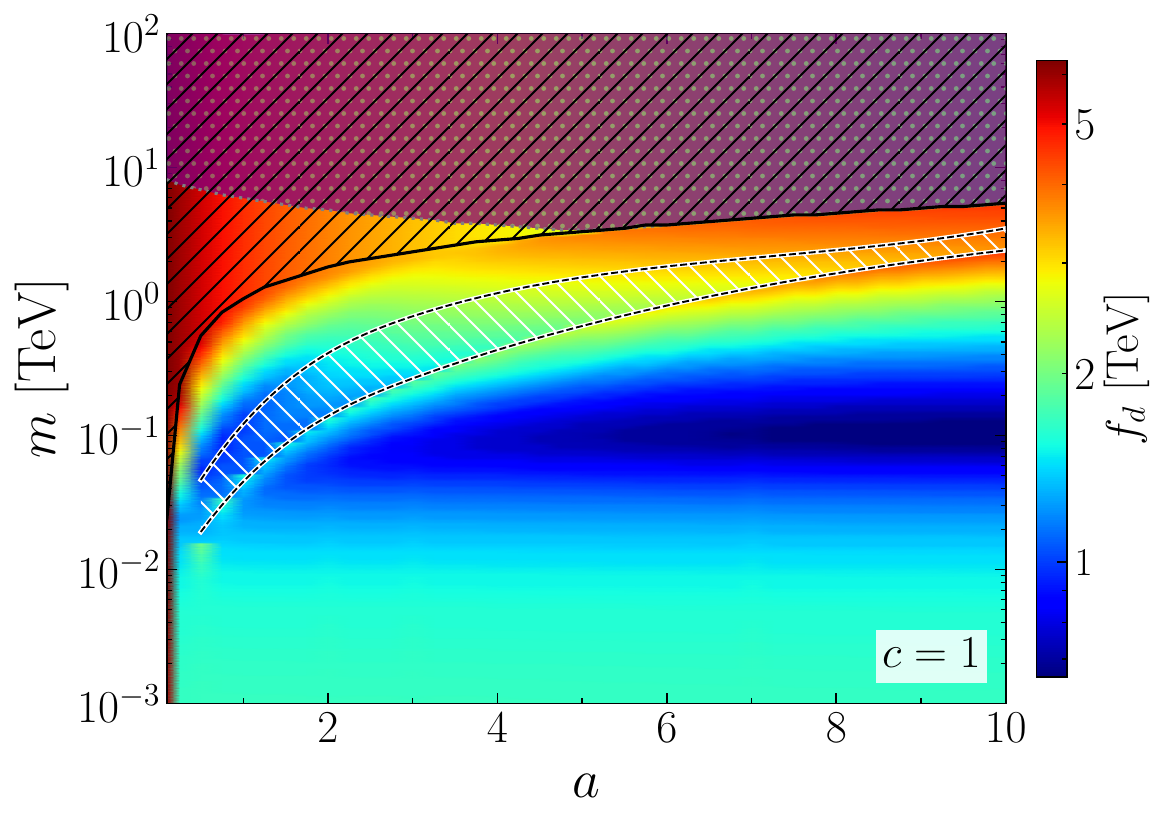}
        \caption{$g_\rho = \sqrt{4\pi}$, $c = 1$}
    \end{subfigure}

    \vspace{0.8em}

    \begin{subfigure}[t]{0.48\textwidth}
        \centering
        \includegraphics[width=\textwidth]{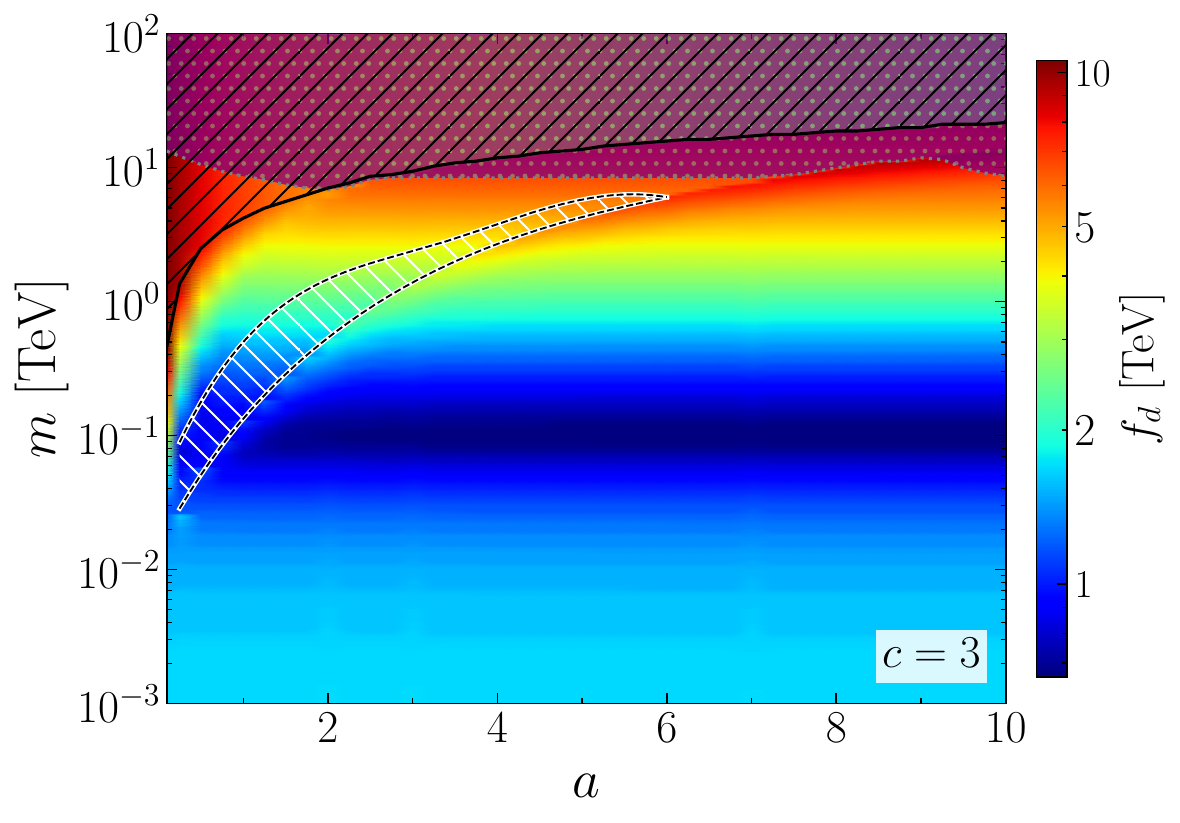}
        \caption{$g_\rho = 6$, $c = 3$}
    \end{subfigure}
    \hfill
    \begin{subfigure}[t]{0.48\textwidth}
        \centering
        \includegraphics[width=\textwidth]{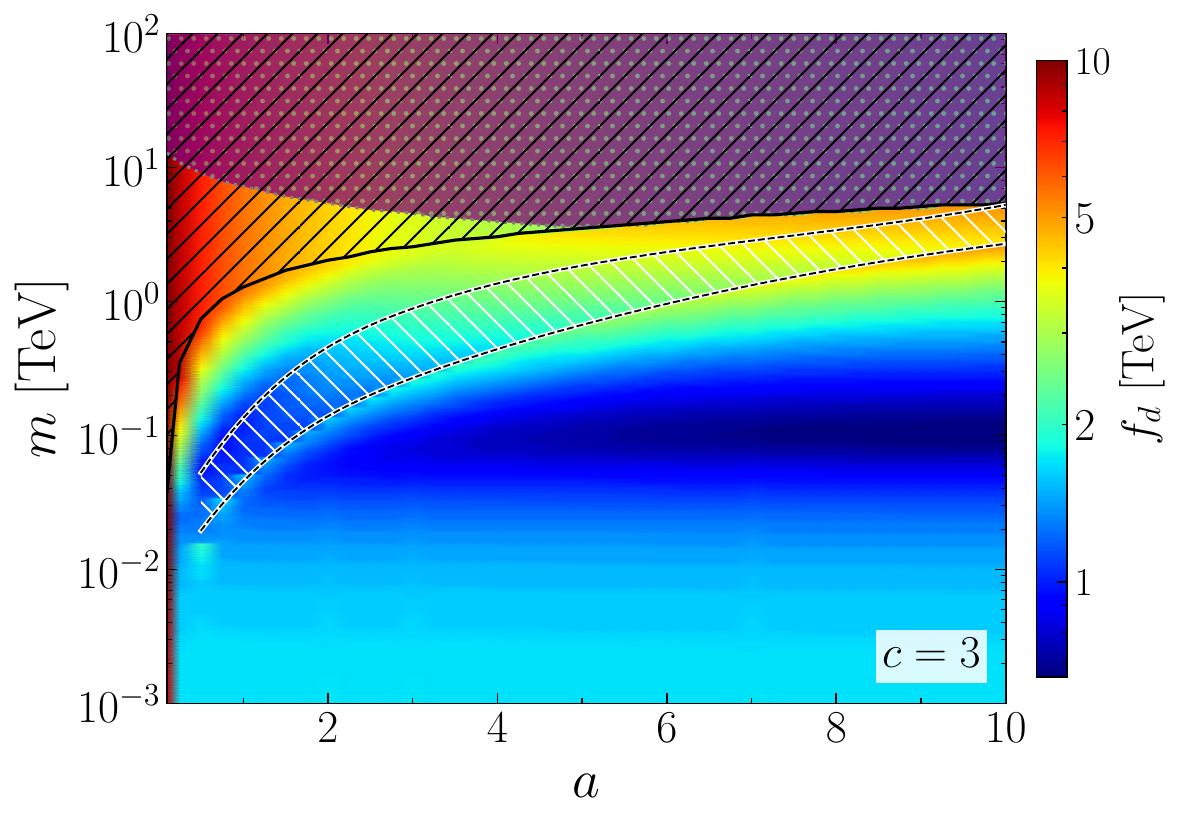}
        \caption{$g_\rho = \sqrt{4\pi}$, $c = 3$}
    \end{subfigure}

    \caption{%
        Colormap of the required $f_d$ to achieve $\Omega_{\rm DM}h^2 = 0.12$ in the $(a, m)$ plane, for $g_\rho = 6$ (left) and $g_\rho = \sqrt{4\pi}$ (right) and for $c = 0.3$ (top), $1$ (middle), and $3$ (bottom). The colours show the smallest $f_d$ on the contour (the non-resonant branch), and the hatched band bounded by dashed lines marks the bifurcation, where the contour has several solutions for $f_d$. The purple dotted area corresponds to the region where $m > 0.1\Lambda_d$, and the black hatched area to the unreliable region of figure~\ref{fig:relic_combined}.}
    \label{fig:color_relic}
\end{figure}

To present a global view of the required confinement scale, we show in figure~\ref{fig:color_relic} the value of $f_d$ satisfying $\Omega_{\rm DM}h^2 = 0.12$ as a function of both $a$ and $m$, for $c \in \{0.3, 1, 3\}$ and the two HLS coupling values considered above, $g_\rho\in\{\sqrt{4\pi},6\}$. We repeat the grid scan of section~\ref{sec:results} for more values of $a$. Since the contours bifurcate at the resonance, $f_d$ is not a single-valued function of $(a,m)$ there. The colours therefore show the smallest solution for $f_d$, i.e.\ the non-resonant branch, while the hatched band bounded by dashed lines marks the bifurcation, i.e.\ the range of $m$ between the two folds of the contour, in which it has several solutions for $f_d$. Its edges are given by the folds of the contours, i.e.\ the ends of the resonant branches shown in figures~\ref{fig:contribution_relic} and~\ref{fig:contribution_relic_sq}. The purple dotted area corresponds to $m>0.1\Lambda_d$, and the black hatched area to the unreliable region of figure~\ref{fig:relic_combined}, both evaluated on the colour-coded solution.
 
The colormaps show certain common regions across both coupling values. Most notably, the dip of the contours appears as a band of small $f_d\simeq0.66~\mathrm{TeV}$ around $m\simeq0.1~\mathrm{TeV}$, which hardly depends on $a$ for $a\gtrsim1.5$ ($g_\rho=6$) and $a\gtrsim7$ ($g_\rho=\sqrt{4\pi}$). For smaller $a$, the resonance approaches the dip, which therefore becomes shallower, with $f_d$ up to $0.85~\mathrm{TeV}$ for $g_\rho=6$ and $1.05~\mathrm{TeV}$ for $g_\rho=\sqrt{4\pi}$, and, for $a\lesssim0.25$ ($g_\rho=\sqrt{4\pi}$) or $a\lesssim0.1$ ($g_\rho=6$), moves to smaller $m$ and larger $f_d$.

Furthermore, we observe a bifurcation of this region, which is more pronounced for smaller $g_\rho$ values. This is due to the resonant enhancement of the annihilation cross section when the center of mass energy approaches the dark $\rho$ meson mass, as discussed above. Following $2m_{\Pi_3}=m_\rho$, the bifurcation runs diagonally through the $(a,m)$ plane, from $m\simeq0.03~\mathrm{TeV}$ at small $a$ to larger $m$ as $a$ increases. For $g_\rho=6$, it ends at $a\simeq5-7$ and $m\simeq2$--$8~\mathrm{TeV}$, in agreement with figure~\ref{fig:relic_combined}, where the $a=7$ contour lies close to this line. For $g_\rho=\sqrt{4\pi}$, it sets in at $a\simeq0.5$ and extends over the whole scanned range, up to $m\simeq2~\mathrm{TeV}$ at $a=10$. For very low values of $g_\rho$, which are excluded due to the constraints from the oblique parameters, the bifurcation is shifted to very small quark mass values, and the resonance is not visible in the scanned region. For smaller values of $c$, the colors become more uniform, highlighting the flattening of the contours. The boundary of the black hatched area, at which the pions become as heavy as the $\rho$ mesons or $m$ reaches $\Lambda_d$, rises with $a$. Above it, the results are not reliable.
 \begin{figure}[!tb]
    \centering
    \begin{subfigure}[t]{0.48\textwidth}
        \centering
        \includegraphics[width=\textwidth]{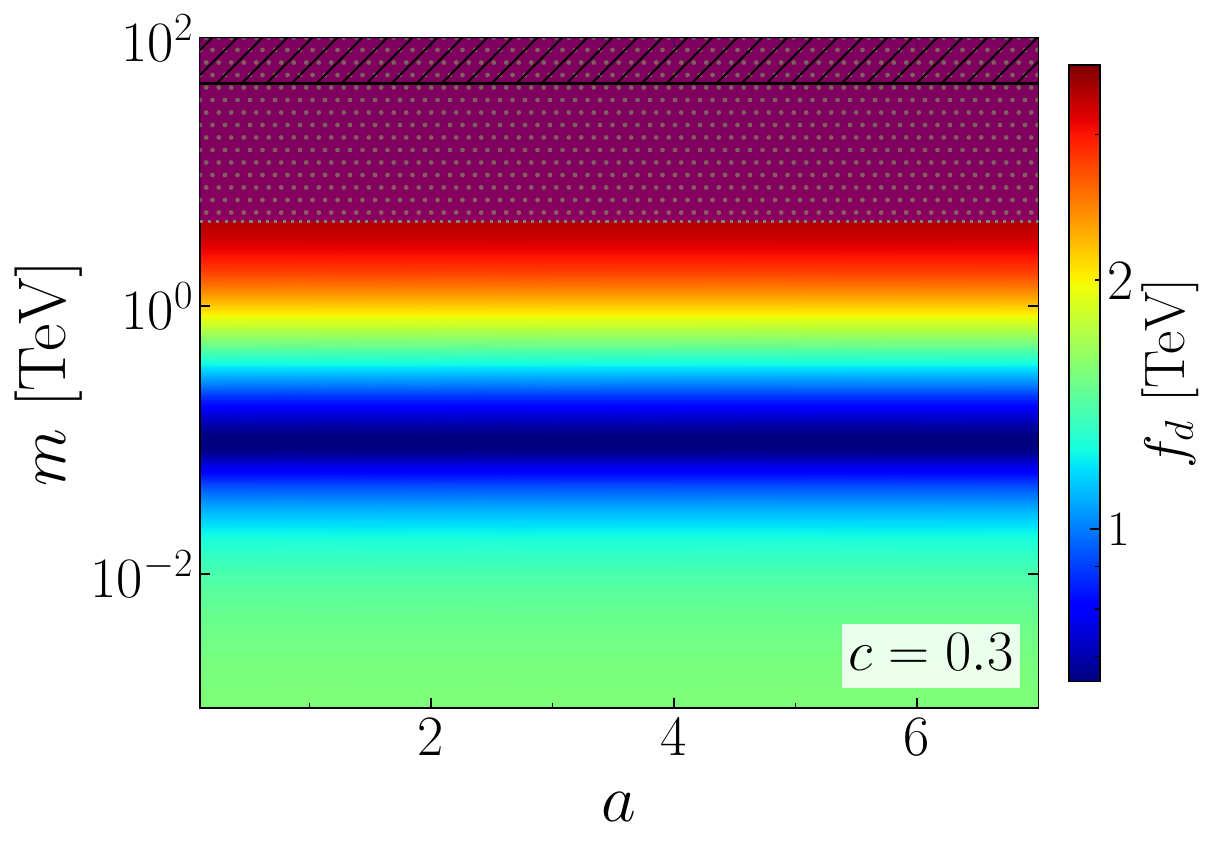}
        \caption{$c = 0.3$}
        \label{fig:sub1}
    \end{subfigure}
    \hfill
    \begin{subfigure}[t]{0.48\textwidth}
        \centering
        \includegraphics[width=\textwidth]{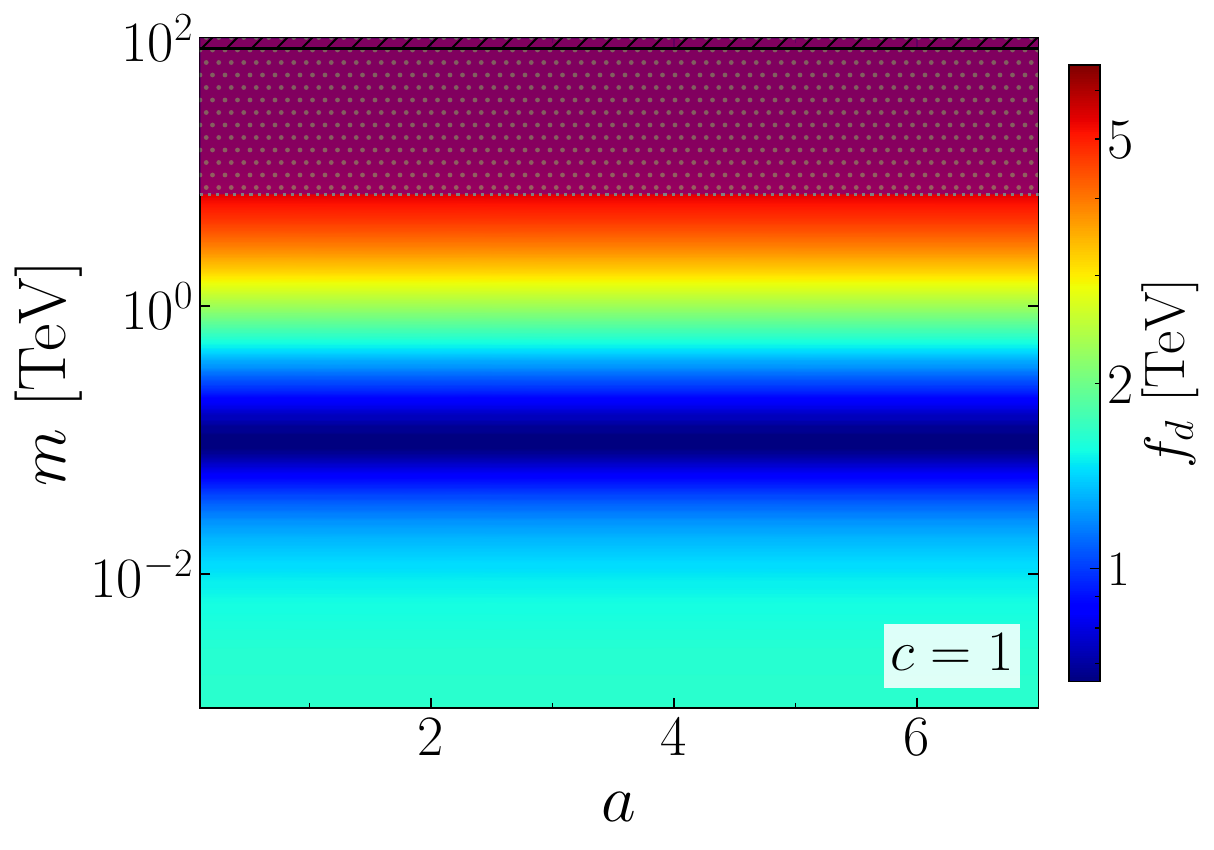}
        \caption{$c = 1$}
        \label{fig:sub2}
    \end{subfigure}

    \vspace{0.8em}

    \begin{subfigure}[t]{0.48\textwidth}
        \centering
        \includegraphics[width=\textwidth]{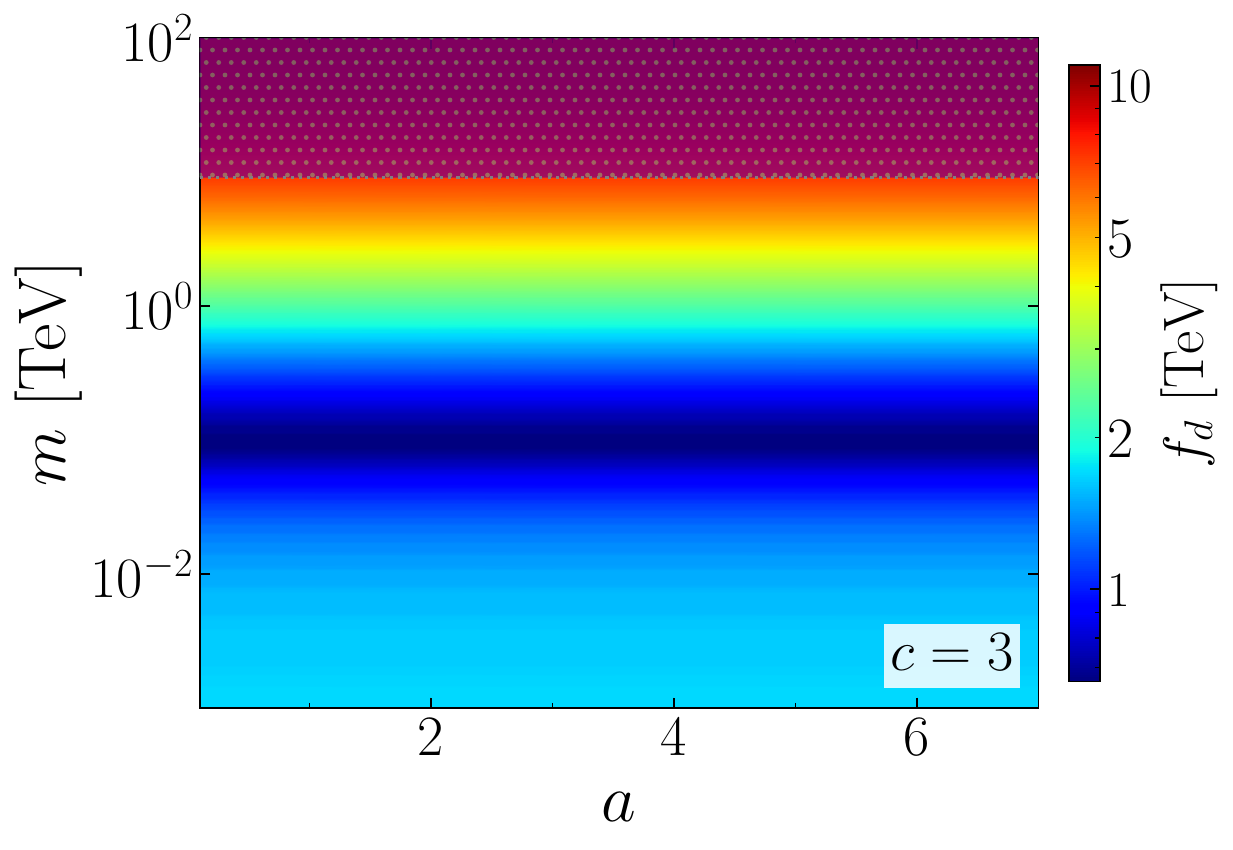}
        \caption{$c = 3$}
        \label{fig:sub3}
    \end{subfigure}

    \caption{Colormap of the required $f_d$ to achieve $\Omega_{\rm DM}h^2 = 0.12$ in the $(a, m)$ plane, for very heavy $\rho$ meson masses. Each panel corresponds to a different value of $c$. The purple dotted area corresponds to the region where $m > 0.1\Lambda_d$, and the black hashed area corresponds to the region where $m > \Lambda_d$, in which the chiral EFT is unreliable.
    }
    \label{fig:heavy_color_relic}
\end{figure}

Similar to the previous subsection, we once again present the colormap for the case where the $\rho$ meson mass is much heavier than the confinement scale, which allows us to integrate out the $\rho$ mesons. The corresponding colormap is shown in figure~\ref{fig:heavy_color_relic}. Compared to the previous cases, the bifurcation is absent, as the resonant enhancement of the annihilation cross section is absent. Furthermore, the plot is constant in $a$ variable and the validity boundary is determined solely by the chiral EFT breakdown at $m\approx \Lambda_d$. No stark saturation boundaries are present in the colormaps.
 
\subsubsection{Dark matter composition}
 
\begin{figure}[!tbp]
    \centering

      \includegraphics[width=\textwidth]{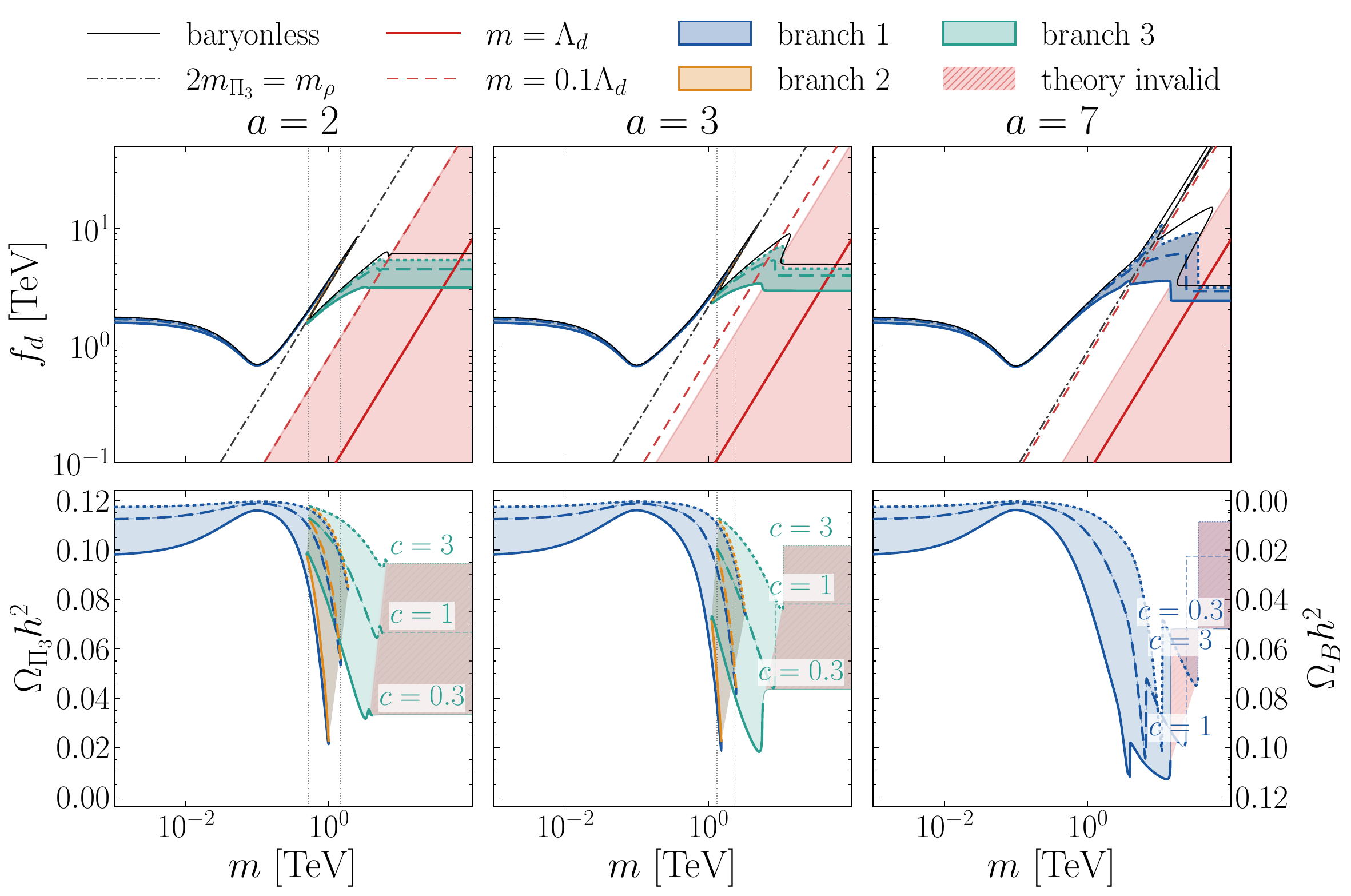}

    \caption{%
        Composition of the dark matter along the relic density contours for $g_\rho=6$ and $a=2$ (left), $3$ (middle) and $7$ (right). Upper panels: the contours $\Omega_{\rm DM}h^2=0.12$ split at their folds into branches that are single-valued in $m$ (colours), together with the baryonless contour (black), the resonance line $2m_{\Pi_3}=m_\rho$ (dash-dotted) and the unreliable region (red). Lower panels: $\Omega_{\Pi_3}h^2$ (left axis) and $\Omega_Bh^2$ (right axis) on the same branches for $c=0.3$ (solid), $1$ (dashed) and $3$ (dotted). Vertical dotted lines mark the folds of the $c=1$ contour, and the hatched red area lies in the unreliable region.
    }
    \label{fig:contribution_relic}
\end{figure}

\begin{figure}[!tbp]
    \centering
    \includegraphics[width=\textwidth]{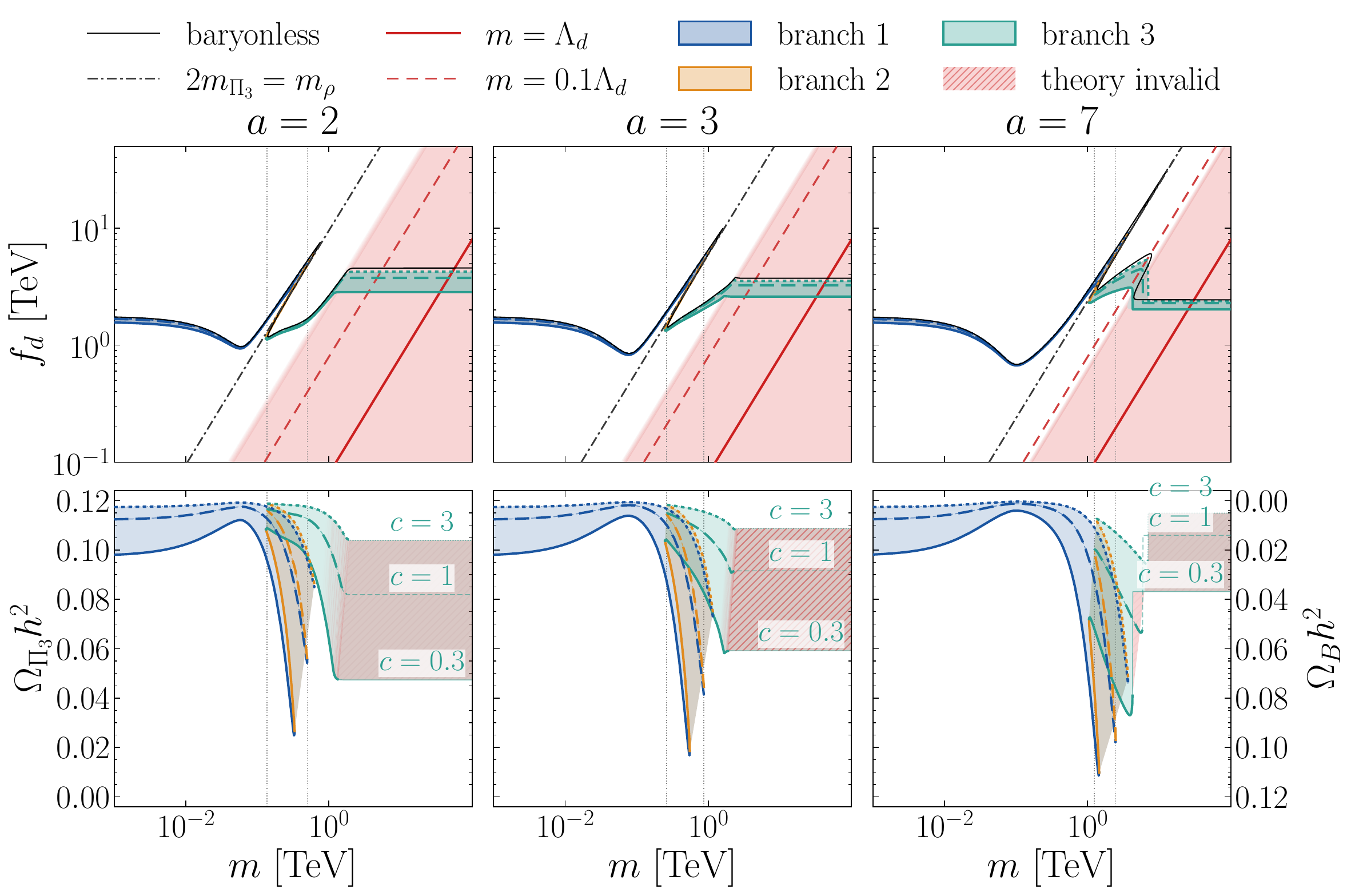}
    \caption{%
        Same as figure~\ref{fig:contribution_relic}, for $g_\rho=\sqrt{4\pi}$.
    }
    \label{fig:contribution_relic_sq}
\end{figure}

Finally, we discuss the individual contributions of the dark meson and dark baryon components to the total relic abundance.
On a contour $\Omega_{\rm DM}h^2=0.12$, the baryonic part follows directly from Eq.~\eqref{eq:omega_baryon},
\begin{equation}\label{eq:baryon_fraction}
    \frac{\Omega_B}{\Omega_{\rm DM}} = \frac{1}{c}\left(\frac{f_d}{6.66~\mathrm{TeV}}\right)^2 ,
\end{equation}
so the composition is fixed by the value of $f_d$ on the contour. Since the contours fold back at the resonance, the composition is not a single-valued function of $m$. We therefore split each contour at its folds into branches on which $f_d$ is single-valued in $m$. Figs.~\ref{fig:contribution_relic} and~\ref{fig:contribution_relic_sq} show these branches in the $(m,f_d)$ plane (upper panels) and the corresponding $\Omega_{\Pi_3}h^2$ (left axis) and $\Omega_Bh^2=0.12-\Omega_{\Pi_3}h^2$ (right axis) as a function of $m$ (lower panels), with the same colour for each branch. The solid, dashed and dotted lines correspond to $c=0.3$, $1$ and $3$, the shaded bands to intermediate values of $c$, and the vertical dotted lines mark the folds of the $c=1$ contour. The hatched red area indicates the part of the bands that lies in the unreliable region of figure~\ref{fig:relic_combined}.

For small quark masses, the dark matter is dominantly mesonic. At $m\simeq1~\mathrm{GeV}$, the baryons contribute only $2$--$18\%$ of $\Omega_{\rm DM}$ (decreasing with $c$), independently of $g_\rho$ and $a$, and in the dip around $m\simeq0.1~\mathrm{TeV}$ their share drops below $5\%$. Within the bifurcation, $f_d$ and hence the baryon fraction grow quickly along the resonant branch, so that the baryons can make up the majority of the relic abundance at its tip for $c\leq1$. On the non-resonant branch, the baryon fraction again rises with $m$. For $g_\rho=6$, the baryons become the dominant component inside the reliable region above $m\simeq1.1$--$1.6~\mathrm{TeV}$ for $c=0.3$ and, for $a=3$ and $7$, above $m\simeq4.9$ and $4.3~\mathrm{TeV}$ for $c=1$; for $a=2$ and $c=1$, and for $c=3$, they stay subdominant on the non-resonant branch. For $g_\rho=\sqrt{4\pi}$, the reliable region ends at smaller $m$, and the baryons become dominant there only for $c=0.3$, above $m\simeq0.9$--$1.6~\mathrm{TeV}$. Deep in the unreliable region, $\Omega_{\Pi_3}\to0$ and the contours approach $f_d\simeq6.66\sqrt{c}~\mathrm{TeV}$, where the relic abundance is purely baryonic.

 \begin{figure}[!tbp]
    \centering
    \includegraphics[width=0.7\textwidth]{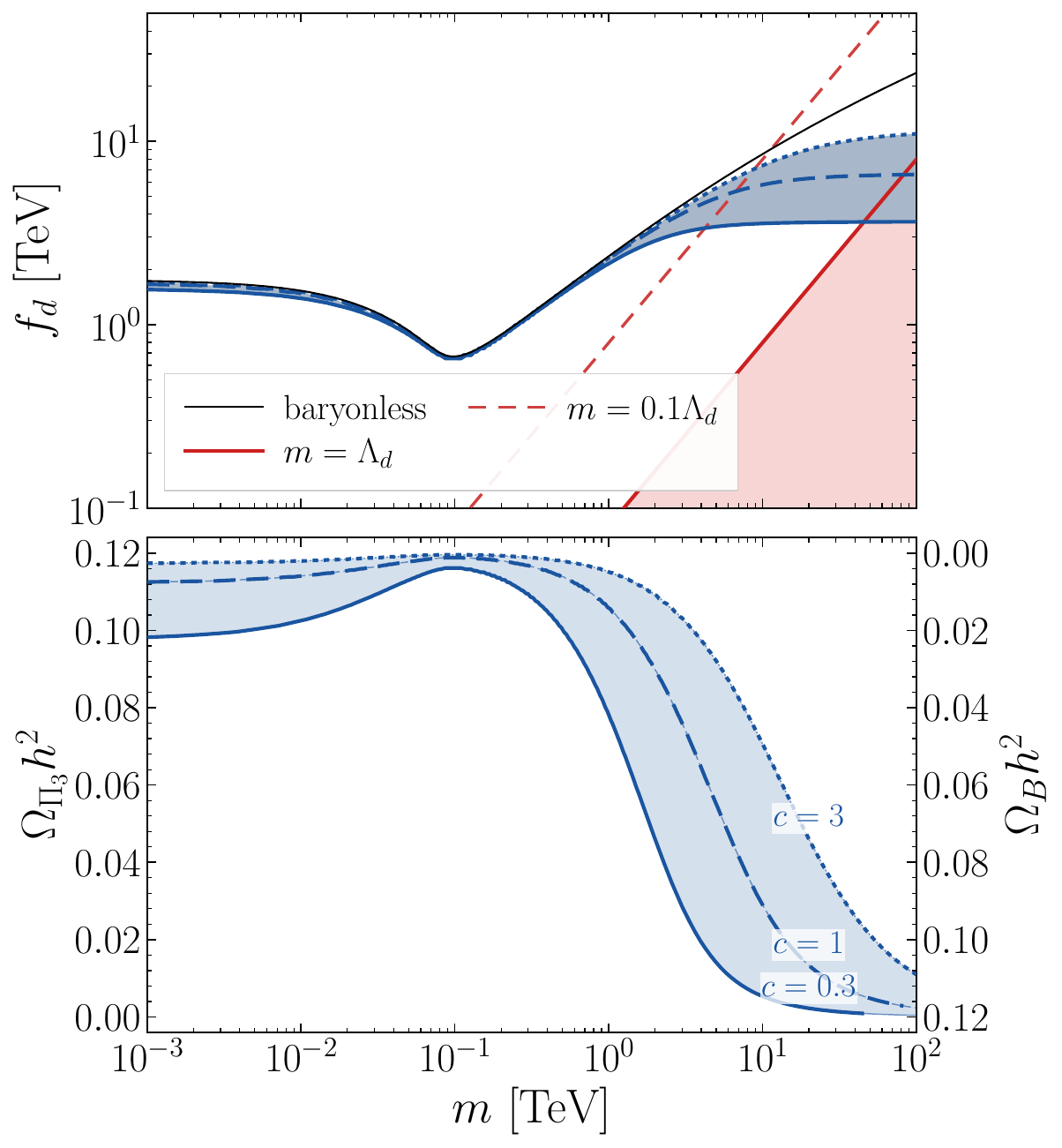}

    \caption{Composition of the dark matter along the relic density contours for very heavy $\rho$ meson masses, shown as in figure~\ref{fig:contribution_relic}. In this limit, the composition is independent of $a$, and without the resonance each contour consists of a single branch.}
    
    \label{fig:heavy_contribution_relic}
\end{figure}

As shown in figure~\ref{fig:heavy_contribution_relic}, in the case where the $\rho$ meson mass is much heavier than the confinement scale, the mesonic contribution is also consistently dominant for the lower quark mass region, for all values of $a$. This HLS parameter region, similar to  figure~\ref{fig:heavy_combined}, converges to the case without $\rho$ mesons. Likewise, no pion-$\rho$ degeneracy region is present, and the baryonic contribution becomes dominant once again for higher quark masses, namely for $m\gtrsim1.5$, $4.6$ and $14~\mathrm{TeV}$ for $c=0.3$, $1$ and $3$, respectively. These values are close to those found for $g_\rho=6$ on the non-resonant branch, and since the reliable region now extends up to $m\simeq\Lambda_d$, the baryon fraction reaches $91$--$99\%$ within the reliable part of the scanned range.

\subsection{Collider Constraints}
We now consider a short discussion of the collider constraints on the model. There are many ways in which the dark sector can be probed at colliders, including direct production of the dark mesons, as well as the effective couplings of the dark pions to the SM gauge bosons and fermions. These can then cause SM/dark hadronic showers, which would also greatly modify signatures. The possibility of many different models and their phenomenologies is an active area of research, see, e.g., \cite{Cheng:2021kjg,Butterworth:2021jto,ATLAS:2025bsz,Asadi:2025vfr,Kribs:2018ilo,Kilic:2009mi, Kilic:2010et,Liu:2023jta}.

In our work, we focus on the collider phenomenology of the $\rho$ meson. The signature depends strongly on the mass spectrum. Here we assume $ m_\Pi> m_\rho/2$ due to the dark quark mass, and the $\rho$ meson dominantly decays into a pair of SM fermions.
We will consider the existing constraints from the LHC as well as the future reach of a muon collider and FCC-hh. For the hadronic collider constraints, we will use a narrow width approximation and probe the resonant production of the dark $\rho$ mesons, which then decay to SM final states. A discussion on the validity region of this approximation and relevant decay channels is provided in appendix~\ref{app:3}. For the muon collider, we will not use the narrow width approximation and will consider the full $2\to2$ processes with SM final states, which are mediated by the dark $\rho$ mesons. 
\subsubsection{Existing Constraints}\label{sec:existing_constraints}
Due to the similarity of the dark $\rho$ triplets to the $Z^\prime$ and $W^\prime$ bosons, we can use the existing constraints from the LHC on these particles to constrain the dark $\rho$ mesons. We will use the results from the ATLAS and CMS collaborations, which have searched for $W^\prime$ bosons via $\ell\nu$ final states. Although there are $Z^\prime$ searches as well, their constraints do not directly translate to our model, as the typically assumed $Z^\prime$ couplings to the SM fermions have the identical structure to the SM $Z$ boson, e.g., the Sequential Standard Model (SSM) $Z^\prime$ boson. In contrast, the neutral $\rho$ triplet couples to the SM fermions only through its $\tilde{W}^3$ admixture, i.e.\ essentially to the left-handed $\mathrm{SU}(2)_W$ current (see section~\ref{sec:mass}). As for the $W^\prime$, we can express the fermionic couplings of the charged $\rho$ triplets in terms of the charged mixing angle $\alpha_\pm$, as explained in section~\ref{sec:mass}, and given in Eq.~\eqref{eq:glnu}. Similarly, for the charged SM vector bosons, leptonic couplings have a very similar structure and are given by 
\begin{equation}
    g_{W^\pm}^{\ell\nu} = \frac{g_W^{UV}}{2\sqrt{2}}\cos\alpha_\pm.
\end{equation}
This relation allows us to recast the existing LHC constraints on $W^\prime$ bosons, which are usually presented in terms of coupling ratio and mass, as constraints on the dark $\rho$ triplets in terms of their mass and the mixing angle $\alpha_\pm$, since ${g_{V_3^\pm}^{\ell \nu}}/{g_{W^\pm}^{\ell \nu}}=\tan\alpha_\pm$. 

\begin{figure}[!tb]
    \centering
    \includegraphics[width=0.95\textwidth]{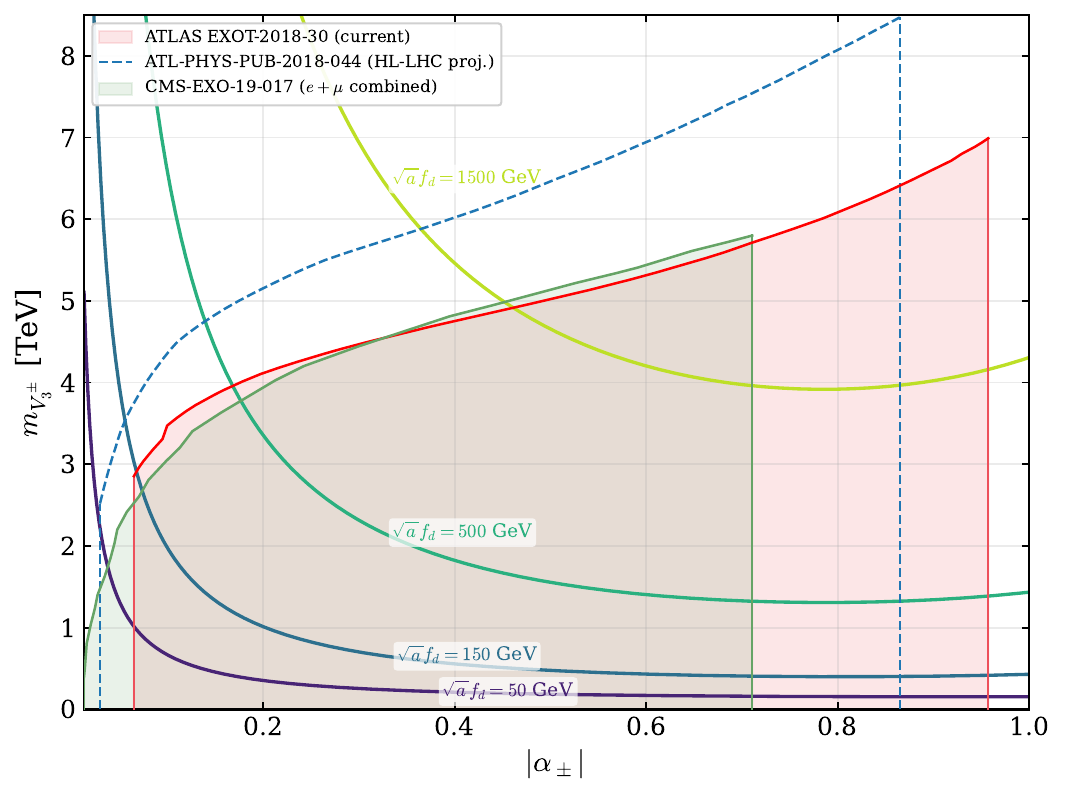}
    \caption{
        Recast of the existing LHC constraints on $W^\prime$ bosons in the $\ell\nu$ final state to the $V_3^\pm$ bosons in terms of their mass and the mixing angle $\alpha_\pm$. The solid lines correspond to the existing constraints from ATLAS and CMS, while the dashed line corresponds to the future reach of the HL-LHC at ATLAS with $3~\mathrm{ab}^{-1}$ of integrated luminosity and $\sqrt{s} = 14~\mathrm{TeV}$. The colored regions correspond to the parameter space that is excluded by the constraints. Colored lines correspond to the mass-angle plots for a given value of $\sqrt{a}f_d$.
    }
    \label{fig:lhc_recast}
\end{figure}

We use the ATLAS \cite{ATLAS:2019lsy} and CMS \cite{CMS:2022krd} results on $W^\prime$ searches in the $\ell\nu$ final state to recast the constraints on the dark $\rho$ triplets. Furthermore, we recast the future reach of the HL-LHC at the ATLAS experiment with $3~\mathrm{ab}^{-1}$ of integrated luminosity and $\sqrt{s} = 14~\mathrm{TeV}$ \cite{ATLAS:2018tvr}. The results in terms of the dark $\rho$ mass and the mixing angle $\alpha_\pm$ are shown in figure~\ref{fig:lhc_recast}.  

\begin{figure}[!tb]
    \centering
    \includegraphics[width=1\textwidth]{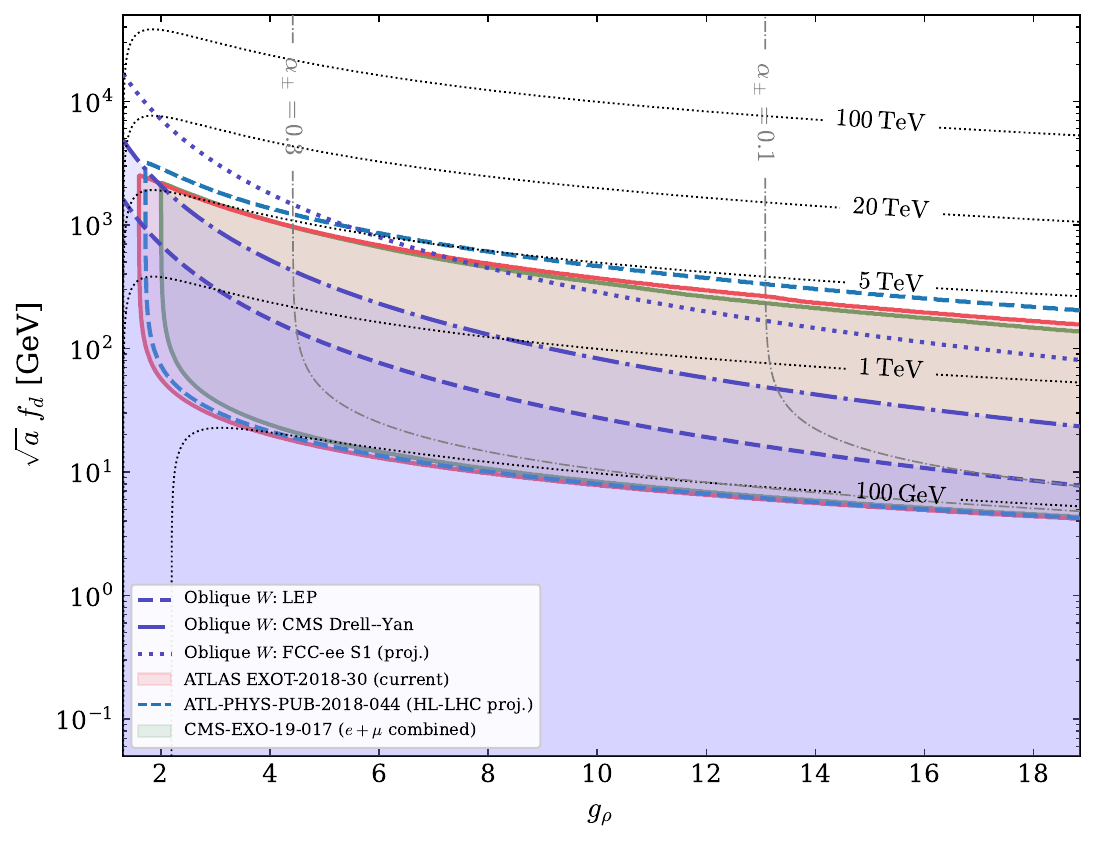}
    \caption{
        Recast of the existing LHC constraints on $W^\prime$ bosons in the $\ell\nu$ final state to the $V_3^\pm$ bosons in terms of $g_\rho$ and $\sqrt{a}f_d$. The red and green areas correspond to the existing constraints from ATLAS and CMS, while the light blue line corresponds to the future reach of the HL-LHC at ATLAS with $3~\mathrm{ab}^{-1}$ of integrated luminosity and $\sqrt{s} = 14~\mathrm{TeV}$. Furthermore, the purple region corresponds to the parameter space excluded due to the $W$ oblique constraints discussed in the appendix \ref{app:2}. The diagonal lines correspond to the $V_3^\pm$ mass contours, and the vertical dash-dotted lines are contours of the charged mixing angle $\alpha_\pm$}.
        
    \label{fig:lhc_recast_grho_f}
\end{figure}

The remarkable feature of the recast is that, for a very broad range of mixing angles, $V_3$ masses below $\mathcal{O}(5)~\mathrm{TeV}$ are excluded. Specifically, the ATLAS analysis has more exclusion power for higher mixing angles, while the CMS analysis has more exclusion power for lower mixing angles. Since we do not expect a strong mixing between the sectors, the CMS analysis is more relevant for our model. The dependence of the angle on the model parameters gives us the opportunity to transform these results to the $(\sqrt{a}f_d,g_\rho)$ plane, which makes it possible to combine with the electroweak precision constraints presented in appendix \ref{app:2} and figure~\ref{fig:ew_constraints_combined}. The results are shown in figure~\ref{fig:lhc_recast_grho_f}. 

The transformed figure shows that the small mixing angle translates to a larger $g_\rho$ value, whereas the ATLAS results are more relevant for smaller $g_\rho$ values. The vertical lines correspond to the mixing angle contours, and the diagonal lines correspond to the $V_3^\pm$ mass contours. Note that none of these are dependent on the quark mass, and thus the recast is valid for any quark mass value, as long as the narrow width approximation is valid. 

\subsection{Future Colliders}
There are many proposals for future colliders, including the Future Circular Collider (FCC) at CERN, which would be a $100~\mathrm{TeV}$ proton-proton collider \cite{FCC:2018vvp}, and the muon collider \cite{Accettura:2023ked}, which would be a high-energy lepton collider. The FCC-hh would have a much higher centre-of-mass energy than the LHC, and would be able to probe much higher mass scales. In this subsection, we will therefore briefly discuss the future reach of the exclusion constraints with the help of these two colliders.

\subsubsection{FCC-hh}
\begin{figure}[!tbp]
    \centering
        \includegraphics[width=0.95\textwidth]{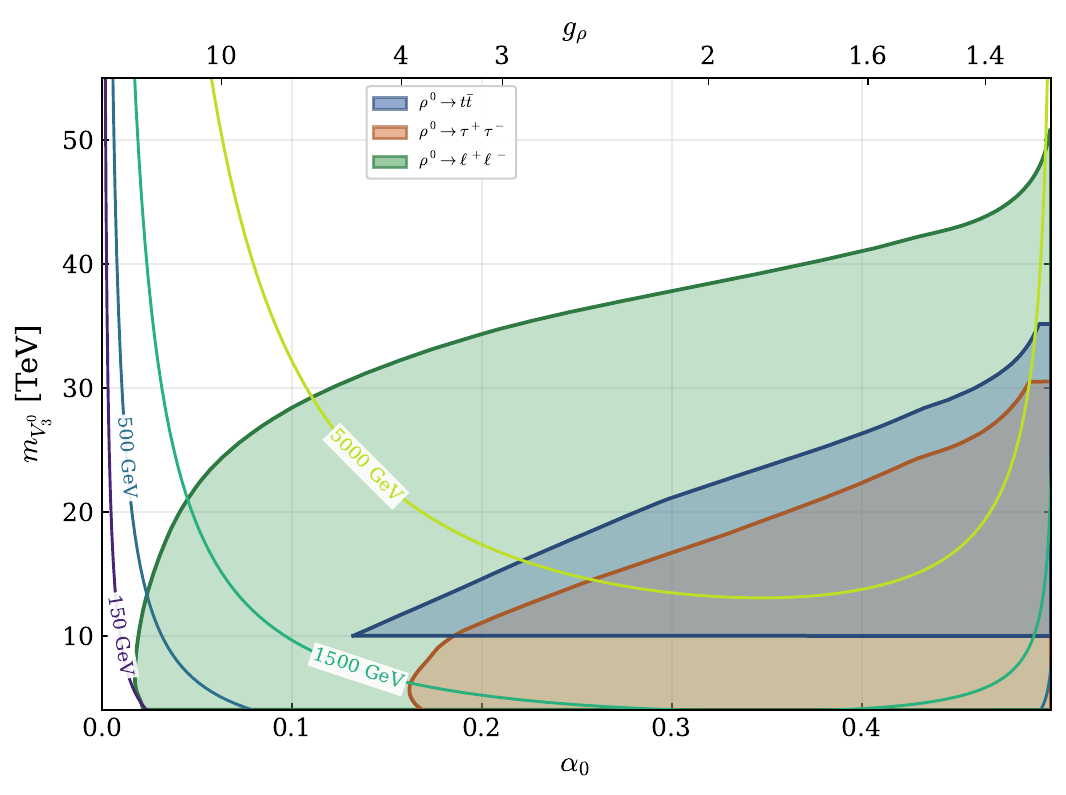}
    \caption{
        Recast of the projected FCC-hh constraints on vector boson resonances in the $\ell\ell$, $\tau\tau$ and $t\bar{t}$ final states~\cite{Helsens:2019bfw} to the $V_3^0$ bosons in terms of their mass and the neutral mixing angle $\alpha_0$, with the corresponding value of $g_\rho$ on the upper axis. The green, orange and blue regions are excluded by the $\ell^+\ell^-$ ($\ell=e,\mu$), $\tau^+\tau^-$ and $t\bar{t}$ channels at FCC-hh with $\sqrt{s} = 100~\mathrm{TeV}$ and an integrated luminosity of $30~\mathrm{ab}^{-1}$, assuming that the decays into dark pions are closed. Colored lines correspond to the mass-angle curves for a given value of $\sqrt{a}f_d$.}
    \label{fig:FCClep}
\end{figure}

For the FCC-hh, we follow a similar strategy to the LHC recast, again working within the narrow width approximation. However, since the existing FCC-hh projections in the literature are formulated for a neutral heavy vector resonance and typically presented as a limit on the mixing angle that is not directly applicable to our model, we instead perform the recast at the level of simulated events. We generate the relevant signal processes in \texttt{MadGraph5}~\cite{Alwall:2011uj} with decays handled by \texttt{MadSpin}~\cite{Artoisenet:2013}, allowing us to directly compare the predicted cross sections of the $V_3$ production against the projected FCC-hh sensitivities, without relying on an intermediate coupling-ratio parametrization. We use the UFO output of our \texttt{FeynRules} implementation and scan $1.31\le g_\rho\le20$ and $3.9~\mathrm{TeV}\le m_{V_3^0}\le51.4~\mathrm{TeV}$ at $\sqrt{s}=100~\mathrm{TeV}$. In the narrow width approximation, the signal factorizes into the production cross section $\sigma(pp\to V_3^0)$ and the branching ratios of $V_3^0$ (see appendix~\ref{app:3}). At fixed $g_\rho$ and $m_{V_3^0}$, the production cross section does not depend on $f_d$ and $m$, whereas the branching ratios do depend on $m$ through the decays into dark pions. We show the most optimistic case, in which the branching ratios are maximized over $f_d$ and $m$. This corresponds to kinematically closed decays into dark pions, $2m_{\Pi_3}>m_{V_3^0}$, as assumed above, for which $\mathrm{BR}(V_3^0\to\ell^+\ell^-)\simeq8\%$ ($\ell=e,\mu$). We draw on two studies targeting different final states. The first is a projection that considers the $\ell\ell,\tau\tau$ and $t\bar{t}$ final states~\cite{Helsens:2019bfw}, and provides the dominant sensitivity in our recast. The second is the observed ATLAS search at $\sqrt{s}=13~\mathrm{TeV}$ that targets diboson final states, $WW$ and $ZW$~\cite{ATLAS:2019nat}; however, we find that the resulting constraints are extremely weak over the relevant parameter space, to the extent that they do not meaningfully restrict the model and can be neglected in the combined exclusion. The results for the former are shown in figure~\ref{fig:FCClep}, where we show the recast of the projected FCC-hh constraints on vector boson resonances in the $\ell\ell,\tau\tau$ and $t\bar{t}$ final states to the $V_3^0$ bosons in terms of their mass and the neutral mixing angle $\alpha_0$, in analogy to figure~\ref{fig:lhc_recast}. Since $\tan\alpha_0=2g_W^{UV}s_W/g_\rho$ depends on $g_\rho$ only, each value of $\alpha_0$ corresponds to one value of $g_\rho$. A center of mass energy of $\sqrt{s} = 100~\mathrm{TeV}$ and an integrated luminosity of $30~\mathrm{ab}^{-1}$ is assumed.  

We observe that almost all of the obtained constraints stem from the leptonic final states, shown as the green area. The $\tau^+\tau^-$ (orange) and $t\bar{t}$ (blue) channels only exclude parameter space for $\alpha_0\gtrsim0.16$ and $\alpha_0\gtrsim0.135$, respectively, which is already covered by the leptonic channel. Since the production cross section grows with the mixing, the mass reach increases with $\alpha_0$. Nothing is excluded for $g_\rho\gtrsim36$ ($\alpha_0\lesssim0.017$), while the $\ell^+\ell^-$ channel excludes $V_3^0$ masses from $4~\mathrm{TeV}$ up to $24$, $29$ and $33~\mathrm{TeV}$ for $g_\rho=10$, $6$ and $\sqrt{4\pi}$, and up to $39~\mathrm{TeV}$ for $g_\rho=2\sqrt{2}\,g_W^{IR}\simeq1.85$ ($\alpha_0\simeq0.35$). In total, we obtain potential constraints for $\rho$ meson masses up to $\sim40~\mathrm{TeV}$. For even larger $\alpha_0$, the excluded region extends up to the $51~\mathrm{TeV}$ end of the projected limit. There, however, the $V_3^0$ becomes broad: with closed decays into dark pions, $\Gamma/m_{V_3^0}$ is independent of $m_{V_3^0}$ and grows from $3.5\%$ at $\alpha_0\simeq0.35$ to $7\%$ at $\alpha_0\simeq0.4$ and above $10\%$ for $\alpha_0\gtrsim0.43$, so that the narrow width approximation, and thus the recast, is no longer reliable in this part of figure~\ref{fig:FCClep}. If the decays into dark pions are open, the leptonic branching ratio, and with it the reach, is reduced; for the benchmark point P1 of Table~\ref{tab:benchmarks}, for instance, $\mathrm{BR}(V_3^0\to\ell^+\ell^-)$ drops to $1.0\%$.

As for the LHC constraints in figure~\ref{fig:lhc_recast_grho_f}, we translate these results to the $(g_\rho,\sqrt{a}f_d)$ plane, which allows a direct comparison with the electroweak precision constraints on the oblique parameter $W$ derived in appendix~\ref{app:2}. The result is shown in figure~\ref{fig:FCC_grho_f}. The $\ell^+\ell^-$ channel excludes a band between the $m_{V_3^0}=4~\mathrm{TeV}$ contour and an upper edge that decreases from $\sqrt{a}f_d\simeq{14.5}~\mathrm{TeV}$ at $g_\rho={2}$ to ${4.7}$ and ${2.4}~\mathrm{TeV}$ at $g_\rho=6$ and $10$. For $g_\rho\gtrsim{7.2}$, this band lies entirely above the bounds on $W$, both the current ones from the LEP and LHC Drell--Yan data and the projected FCC-ee one, so that FCC-hh would probe a region that is inaccessible to electroweak precision measurements. For smaller $g_\rho$, the projected FCC-ee bound on $W$ covers the lower part of the band (up to $\sqrt{a}f_d\simeq{7.1}~\mathrm{TeV}$ at $g_\rho={2}$), and for $g_\rho\lesssim{2.4}$ the current bounds on $W$ also reach into its lower edge.
\begin{figure}[!tbp]
    \centering
    \includegraphics[width=1\textwidth]{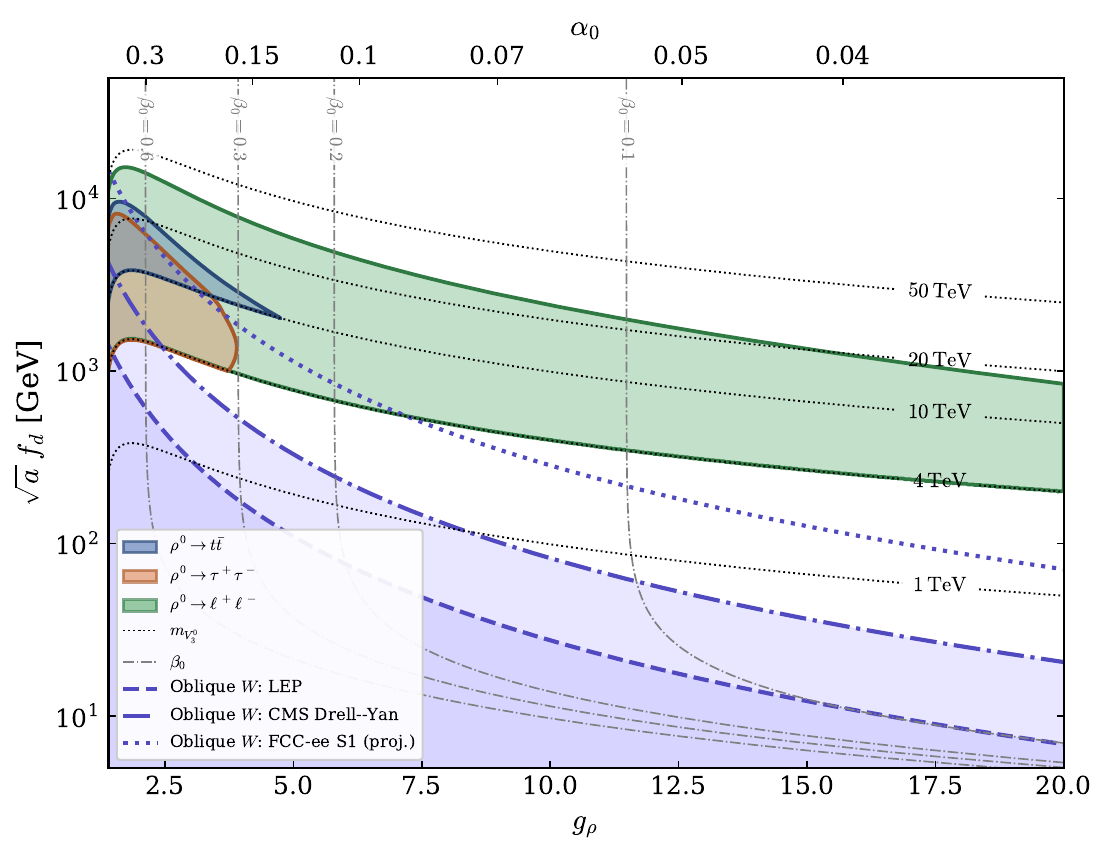}
    \caption{
        Recast of the projected FCC-hh constraints on vector boson resonances in the $\ell\ell$, $\tau\tau$ and $t\bar{t}$ final states to the $V_3^0$ bosons in terms of $g_\rho$ and $\sqrt{a}f_d$. The green, orange, and blue areas correspond to the projected constraints in the $\ell^+\ell^-$, $\tau^+\tau^-$, and $t\bar{t}$ channels at FCC-hh with $\sqrt{s}=100~\mathrm{TeV}$ and $30~\mathrm{ab}^{-1}$, assuming that the decays into dark pions are closed. The purple regions are excluded by the electroweak precision constraints on the oblique parameter $W$ from LEP (dashed) and LHC Drell--Yan data (dash-dotted), see appendix~\ref{app:2}, and the dotted purple line shows the projected FCC-ee (S1) constraint on $W$. The black dotted lines are $V_3^0$ mass contours, the gray dash-dotted lines are contours of the neutral mixing angle $\beta_0$, and the upper axis shows $\alpha_0$.}    
    \label{fig:FCC_grho_f}
\end{figure}

\subsubsection{Muon Collider}
The primary motivation for considering a muon collider is the combination of high centre-of-mass energy with a clean electroweak environment~\cite{Accettura:2023ked, InternationalMuonCollider:2024jyv,Long:2020wfp}. We will probe three possible $2\to2$ processes that are enhanced by $\rho$ meson mediators, and estimate the $95\%$ confidence intervals. Specifically, we consider $\mu^+\mu^-\to \mu^+\mu^-$, $\mu^+\mu^-\to e^+e^-$ and $\mu^+\mu^-\to W^+W^-$. Similar processes have been quite popular for probing $Z'/W'$ bosons at future muon colliders \cite{Lu:2023jlr,Korshynska:2024suh,Lu:2024qrf}. At high energies, alternatively, vector boson fusion (VBF) also becomes a good probe as it is comparable to, or exceeds, Drell-Yan pair production, as the VBF cross section falls more slowly with the $V_3$ mass~\cite{Han:2020uak}. This is especially relevant given that the relic density contours in section~\ref{sec:results} favour dark pion masses in the multi-TeV range, which falls within or near the reach of a $\sqrt{s} = 10\,\mathrm{TeV}$ muon collider~\cite{Long:2020wfp}. For the processes we consider, this target energy is a good probe for vector mediator resonances. We assume an integrated luminosity of $1~\mathrm{ab}^{-1}$.

We will analytically compute the cross sections for the processes listed above, as well as the SM background, and then use the expected luminosity up to $1~\mathrm{ab}^{-1}$ of a future muon collider to estimate the $95\%$ confidence intervals. We will not perform a detailed analysis of the detector effects, but we will assume that the detector can efficiently identify the final states we consider. We will also assume a systematic uncertainty of $10\%$ to be on the conservative side, which is assumed to be achievable at a future muon collider~\cite{InternationalMuonCollider:2024jyv}. Specifically, we take a relative uncertainty $\epsilon_i=10\%$ on the expected number of events in every bin, uncorrelated between the bins (see Eq.~\eqref{eq:q_asimov} below). 

From a given set of Poisson distributed numbers of events, the $95\%$ confidence interval can be computed using the likelihood function 
\begin{equation}
    \mathcal{L}(\mu) = \prod_{i=1}^{n_{\rm bins}} \frac{\mu_i^{N_i
} e^{-\mu_i}}{N_i!},
\end{equation}
where $\mu_i$ is the expected number of events in the $i$-th bin, and $N_i$ is the observed number of events in the $i$-th bin, either from the signal or background. The test statistic is log likelihood ratio, defined as
\begin{equation}
    q = -2\ln\frac{\mathcal{L}(\mu_{\rm BSM})}{\mathcal{L}(\mu_{\rm SM})},
\end{equation}
where the subscript on $\mu$ indicates whether the expected number of events is computed with the BSM signal or the SM background. Due to the lack of data, we will use the Asimov dataset as the observed number of events in a bin $N_i$, which is equal to the expected number of events from the SM background. Assuming a Gaussian distributed systematic uncertainty of $10\%$ in each bin, with an independent nuisance parameter $C_i$ per bin, as mentioned above, the test statistic can be written as 
\begin{equation}
    q_{\bar{\epsilon}}^{\rm Asimov} = -2\sum_{i=1}^{n_{\rm bins}} \left(\mu_i^{\rm SM}-C_i\mu_i^{\rm BSM}+\mu_i^{\rm SM}\ln\frac{C_i\mu_i^{\rm BSM}}{\mu_i^{\rm SM}}-\frac{(C_i-1)^2}{2\epsilon_i^2}\right),\label{eq:q_asimov}
\end{equation}
with
\begin{equation}
    C_i = \frac{1}{2}\left(1-\epsilon_i^2\mu_i^{\rm BSM}+\sqrt{(1-\epsilon_i^2\mu_i^{\rm BSM})^2+4\epsilon_i^2\mu_i^{\rm SM}}\right),
\end{equation}
where $\epsilon_i$ is the systematic uncertainty in the $i$-th bin \cite{Conway:2011in,Acanfora:2026aay}. Finally, the $95\%$ confidence interval can be computed by finding the value of the BSM signal strength $\mu_{\rm BSM}$ such that the $95\%$ confidence level satisfies
\begin{equation}
    {\rm CL}_s=\frac{1-p(\sqrt{{q_{\bar{\epsilon}}^{\rm Asimov}}})}{p(0)}<0.05,
\end{equation}
where $p(z)$ is the cumulative distribution function of the standard normal distribution.

\paragraph{$\mu^+\mu^-\to\mu^+\mu^-$ channel}\mbox{}\\
For this process, the SM background receives $s$- and $t$-channel contributions from $\gamma$ and $Z$ exchange. The dark rho meson $V_3^0$ enters as an additional mediator in both channels, with couplings to SM fermions induced by the mass mixing with $\tilde{Z}$ and $\tilde{A}$ derived in section~\ref{sec:mass}. For muons ($T^3_\mu = -1/2$, $Q_\mu = -1$) the couplings given in 
Eqs.~\eqref{eq:gfV} and \eqref{eq:gfA}
reduce to
\begin{equation}
    g^{\mu V}_{V_3^0} = -\kappa_\gamma e + \kappa_Z\frac{g_W^{UV}}{2c_W}\!\left(-\tfrac{1}{2}+2\sin^2\!\theta_W\right), \qquad
    g^{\mu A}_{V_3^0} = -\frac{\kappa_Z g_W^{UV}}{4c_W}.
\end{equation}
In the limit $g_\rho \gg g_W$, both mixing angles are suppressed, making $V_3^0$ coupling to muons also negligible. Here, $\kappa_\gamma$ and $\kappa_Z$ are the overlap parameters defined in Eq.~\eqref{eq:kappa} of section~\ref{sec:mass}.

\begin{figure}[!tb]
    \centering
    \begin{subfigure}[!tb]{0.5\textwidth}
        \centering
        \includegraphics[width=\textwidth]{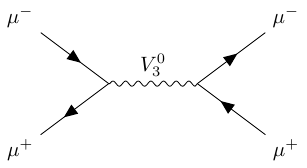}
        \caption{$s$-channel}
    \end{subfigure}
    \hfill
    \begin{subfigure}[!tb]{0.3\textwidth}
        \centering
        \includegraphics[width=\textwidth]{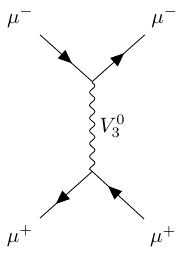}
        \caption{$t$-channel}
    \end{subfigure}
    \caption{Feynman diagrams for the process $\mu^+\mu^-\to\mu^+\mu^-$, with $s$- and $t$-channel contributions from $V_3^0$ exchange.}
    \label{fig:feynman_mumu}
\end{figure}
\begin{figure}[!tb]
    \centering
    \begin{subfigure}[!tb]{0.48\textwidth}
        \centering
        \includegraphics[width=\textwidth]{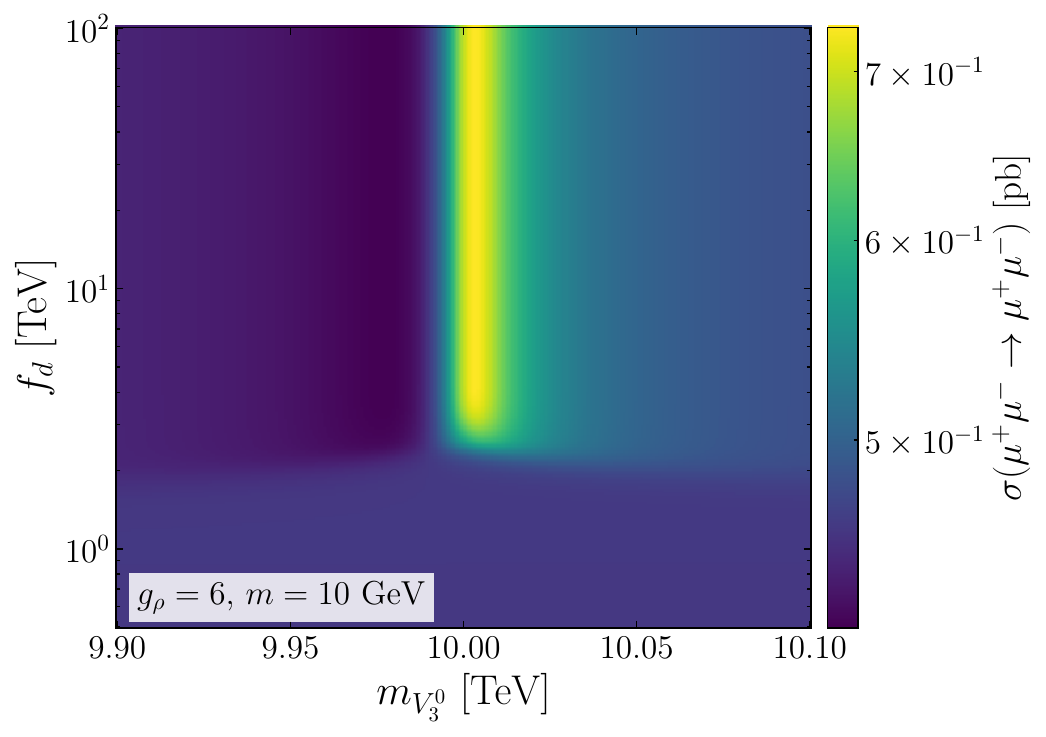}
        \caption{$m = 10$ GeV, $g_\rho = 6$}
    \end{subfigure}
    \hfill
    \begin{subfigure}[!tb]{0.48\textwidth}
        \centering
        \includegraphics[width=\textwidth]{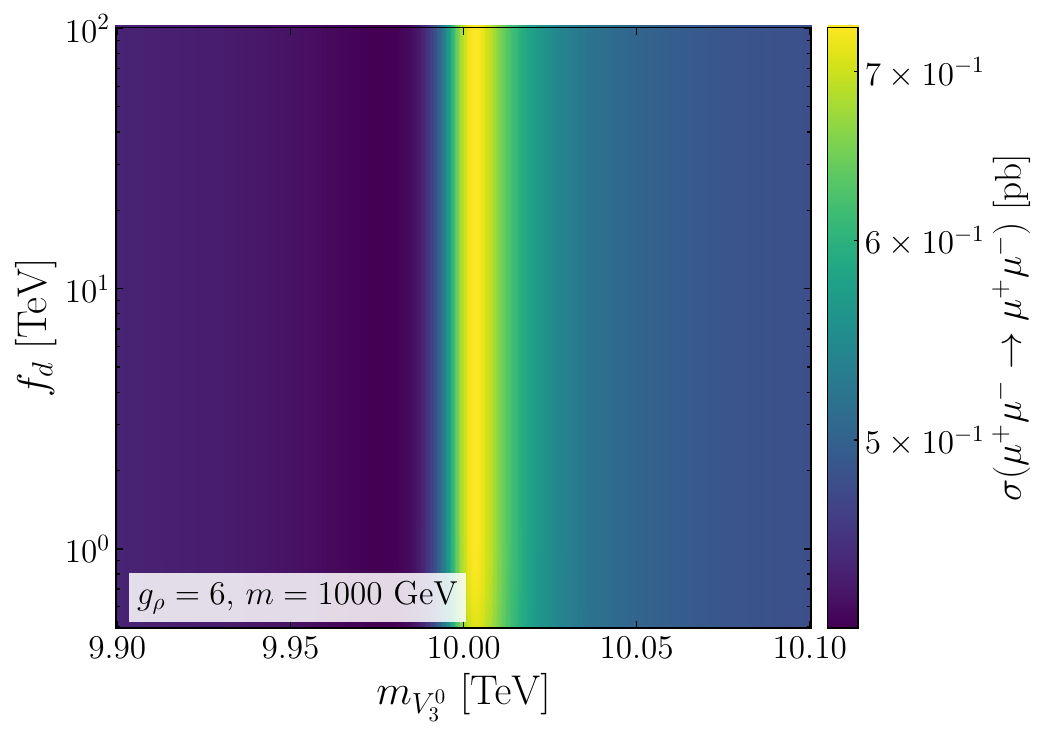}
        \caption{$m = 1000$ GeV, $g_\rho = 6$}
    \end{subfigure}
    \begin{subfigure}[!tb]{0.48\textwidth}
        \centering
        \includegraphics[width=\textwidth]{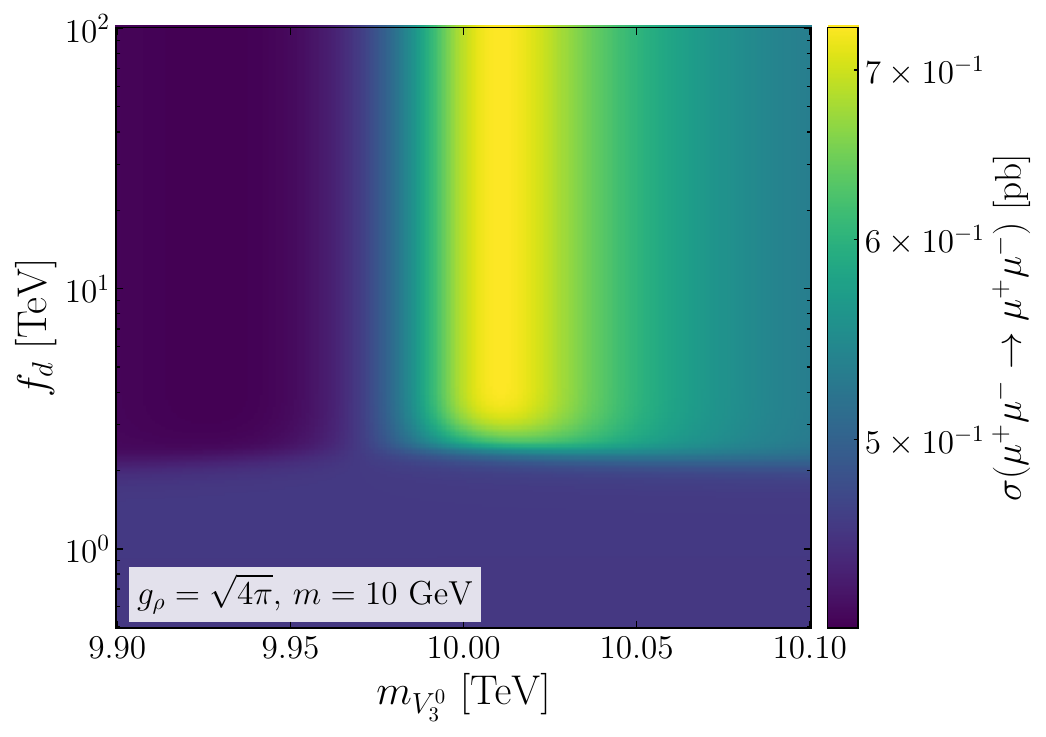}
        \caption{$m = 10$ GeV, $g_\rho = \sqrt{4\pi}$}
    \end{subfigure}
    \hfill
    \begin{subfigure}[!tb]{0.48\textwidth}
        \centering
        \includegraphics[width=\textwidth]{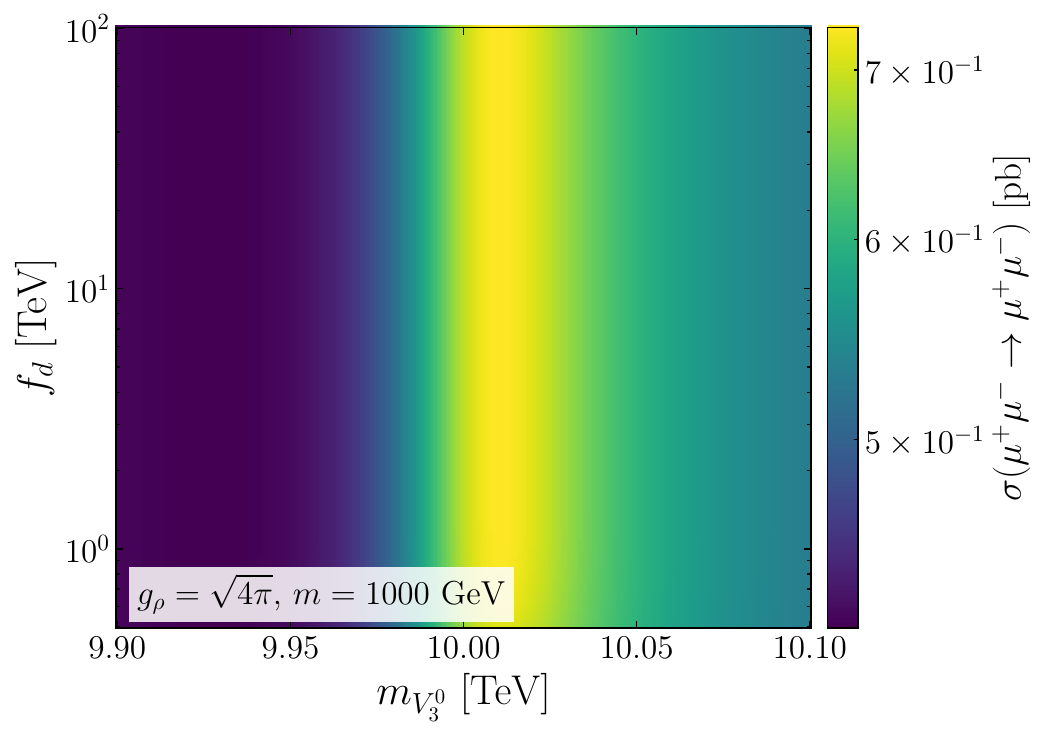}
        \caption{$m = 1000$ GeV, $g_\rho = \sqrt{4\pi}$}
    \end{subfigure}
    \caption{Total cross sections for the process $\mu^+\mu^-\to\mu^+\mu^-$ as a function of the $\rho$ triplet mass $m_{V_3^0}$ and pion decay constant $f_d$, for a center of mass energy of $\sqrt{s} = 10$ TeV and a HLS coupling of $g_\rho = 6$ and $g_\rho = \sqrt{4\pi}$.}
    \label{fig:cross_mumu}
\end{figure}

The tree-level $V_3^0$ contribution is given by the $s$- and $t$-channel diagrams shown in figure~\ref{fig:feynman_mumu}, which interfere with the SM background.  
\begin{equation}
    \mathcal{M}^{V_3^0} = \mathcal{M}^{V_3^0}_s + \mathcal{M}^{V_3^0}_t,
\end{equation}
with the $s$-channel piece given by
\begin{equation}
    \mathcal{M}^{V_3^0}_s = \frac{1}{s - m_{V_3^0}^2 + im_{V_3^0}\Gamma_{V_3^0}}
    \left[\bar{v}(p_2)\gamma^\mu\!\left(g^{\mu V}_{V_3^0} - g^{\mu A}_{V_3^0}\gamma^5\right)u(p_1)\right]
    \left[\bar{u}(p_3)\gamma_\mu\!\left(g^{\mu V}_{V_3^0} - g^{\mu A}_{V_3^0}\gamma^5\right)v(p_4)\right],
\end{equation}
and analogously for the $t$-channel piece with $s\to t$, where $p_1$ and $p_2$ are the momenta of the incoming muons and $p_3$ and $p_4$ are the momenta of the outgoing muons. The signal enhancement from $V_3^0$ is resonant near $\sqrt{s} = m_{V_3^0}$. Note that the propagator denominator contains the total width $\Gamma_{V_3^0}$, which also includes decays into $\Pi$ pairs. In other words, even though the diagrams only contain HLS parameters, we obtain an additional dependence on the dark quark mass via the width. The integrated cross plots in the vicinity of the resonance region for $\sqrt{s}=10$ TeV, in dependence on $f_d$ and $m_{V_3^0}$, are shown in figure~\ref{fig:cross_mumu}.

The figure on the left shows the cross section for $m = 10$ GeV, while the figure on the right shows the cross section for $m = 1000$ GeV. As expected, we observe a resonant enhancement of the cross section near $\sqrt{s} = m_{V_3^0}$, which is more pronounced for higher values of $f_d$, since at fixed $g_\rho$ and $m_{V_3^0}$ the value of $a$ decreases as $a\propto f_d^{-2}$, while the dark pion masses grow with $f_d$, which suppresses the pion contribution to the width and enhances the resonance. For higher dark quark mass values, the decays into dark pions are more strongly phase-space suppressed or closed, which reduces this dependence of the width. For values of $m_{V_3^0}$ far from the resonance, the cross section approaches a constant value. The width of the resonance is broader for lower values of $g_\rho$, as the coupling to muons is stronger, which also leads to a higher cross section away from the resonance.

In order to compute the $95\%$ confidence intervals, we will bin the events in the pseudorapidity $\eta$, using $10$ bins in $-2.5<\eta<0$ for $\mu^+\mu^-\to\mu^+\mu^-$ and $20$ bins in $|\eta|<2.5$ for $\mu^+\mu^-\to e^+e^-$ and $\mu^+\mu^-\to W^+W^-$. The SM background is dominated by the $t$-channel contribution, which is enhanced in the forward region, leading to a characteristic peak in the $\eta$ distribution. The $s$-channel contribution from $V_3^0$ exchange, on the other hand, produces a more isotropic distribution. Therefore, the presence of a resonance can be inferred from deviations in the $\eta$ distribution from the SM expectation, especially near the resonance mass. 

\begin{figure}
    \centering
    \includegraphics[width=0.95\textwidth]{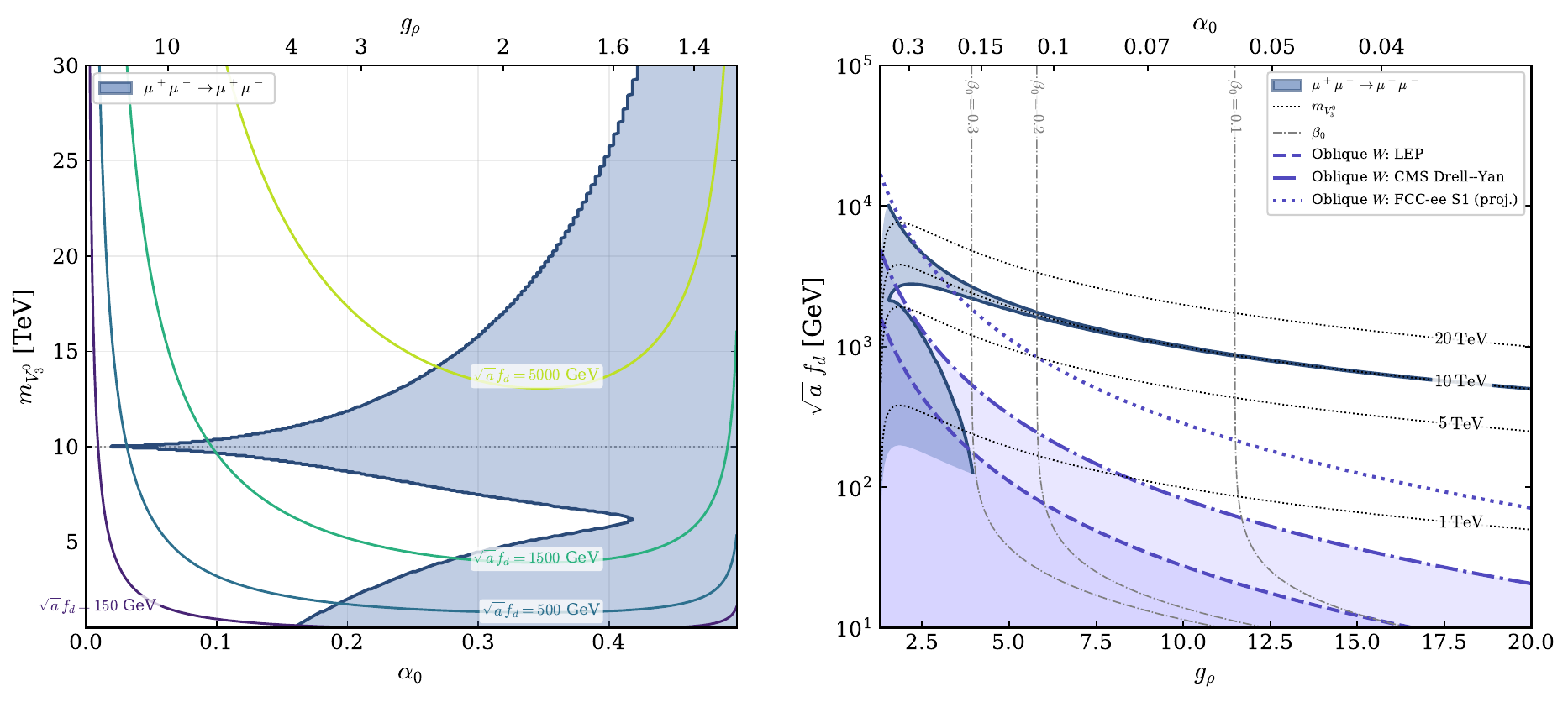}
    \caption{$95\%$ confidence exclusion regions for the $\mu^+\mu^-\to\mu^+\mu^-$ channel at $\sqrt{s}=10~\mathrm{TeV}$ with $1~\mathrm{ab}^{-1}$, assuming that the decays into dark pions are closed, shown as for the FCC-hh recast in Figs.~\ref{fig:FCClep} and~\ref{fig:FCC_grho_f}. Left: $(\alpha_0,m_{V_3^0})$ plane with $g_\rho$ on the upper axis; the coloured lines are trajectories of constant $\sqrt{a}f_d$, and the horizontal dotted line marks $m_{V_3^0}=\sqrt{s}$. Right: $(g_\rho,\sqrt{a}f_d)$ plane with $\alpha_0$ on the upper axis, $V_3^0$ mass contours (black dotted) and contours of $\beta_0$ (gray dash-dotted); the purple regions are excluded by the electroweak precision constraints on $W$ from LEP (dashed) and LHC Drell--Yan data (dash-dotted), and the dotted purple line is the projected FCC-ee bound, see appendix~\ref{app:2}.}
    \label{fig:mumuCL}
\end{figure}

The resulting $95\%$ confidence intervals are shown in figure~\ref{fig:mumuCL} in the $(\alpha_0,m_{V_3^0})$ and $(g_\rho,\sqrt{a}f_d)$ planes, as for the FCC-hh recast, assuming that the decays into dark pions are closed. As evident from the cross section color maps given in figure~\ref{fig:cross_mumu}, the exclusion region is around the mediator resonance. The line at which $m_{V_3^0}=\sqrt{s}$ is satisfied shows a small dent into the region, stemming from the anti-symmetric shape of the cross section centered on the line. The excluded range in $m_{V_3^0}$ grows with the mixing, from a narrow band around $\sqrt{s}$ that closes at $\alpha_0\simeq0.02$ ($g_\rho\simeq30$) to $9.7$--$10.4~\mathrm{TeV}$ for $g_\rho=6$ and $9.0$--$11.4~\mathrm{TeV}$ for $g_\rho=\sqrt{4\pi}$. For $\alpha_0\gtrsim0.15$, a second region appears at $m_{V_3^0}\ll\sqrt{s}$, separated from the resonant one by an allowed wedge. It stems from the off-shell $t$-channel $V_3^0$ exchange, which interferes constructively with the dominant $t$-channel photon exchange, whereas the interference of the $s$-channel $V_3^0$ with the latter is destructive below $\sqrt{s}$; the wedge marks where the two cancel. Additionally, lower values of $g_\rho$ allow for a broader region, owing to the fact that the interaction vertices with the muons become more prominent. For example, for the QCD-like case of $g_\rho = 6$ and $a=2$, the region $1.11~\mathrm{TeV}<f_d<1.
20~\mathrm{TeV}$, i.e.\ $9.64~\mathrm{TeV}<m_{V_3^0}<10.42~\mathrm{TeV}$, is excluded for $m\gtrsim0.3~\mathrm{TeV}$. For smaller $m$, the decays into dark pions open and broaden $V_3^0$, so that the excluded band shrinks. At $f_d=1.15~\mathrm{TeV}$ the exclusion persists down to $m\simeq0.25~\mathrm{TeV}$. In the $(g_\rho,\sqrt{a}f_d)$ plane, the excluded region follows $\sqrt{a}f_d\simeq m_{V_3^0}/g_\rho$ and, around $m_{V_3^0}\simeq\sqrt{s}$, lies above the electroweak precision bounds on $W$ of figure~\ref{fig:ew_constraints_combined} (for the projected FCC-ee bound, for $g_\rho\gtrsim3.1$).

\begin{figure}[!tb]
    \centering
    \begin{subfigure}[!tb]{0.48\textwidth}
        \centering
        \includegraphics[width=\textwidth]{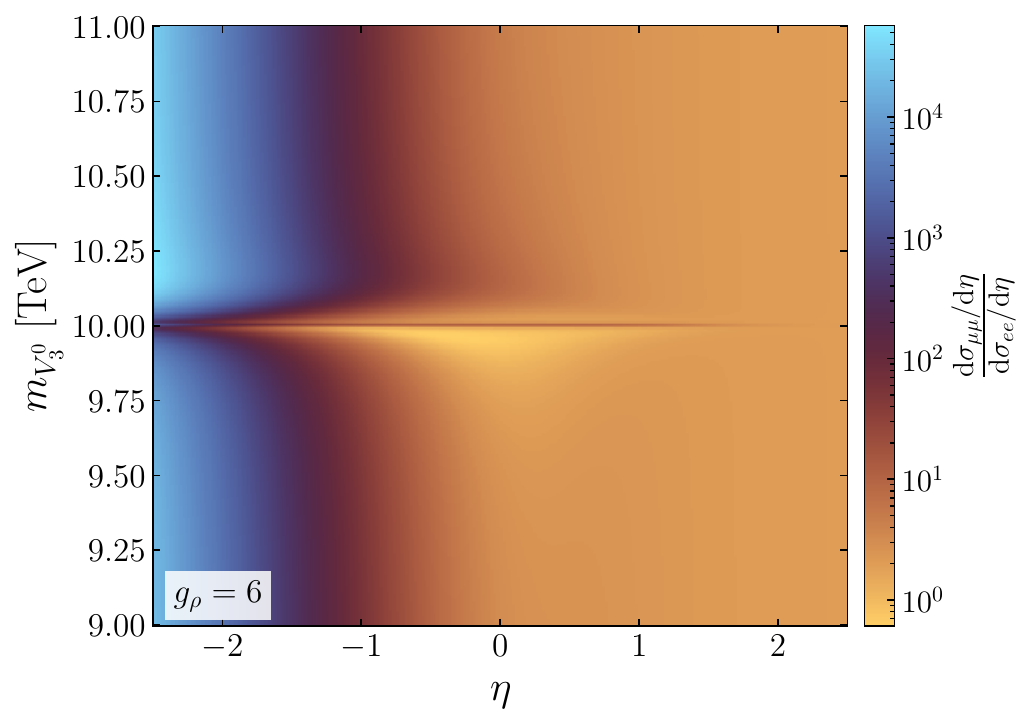}
        \caption{$g_\rho = 6$}
    \end{subfigure}
    \hfill
    \begin{subfigure}[!tb]{0.48\textwidth}
        \centering
        \includegraphics[width=\textwidth]{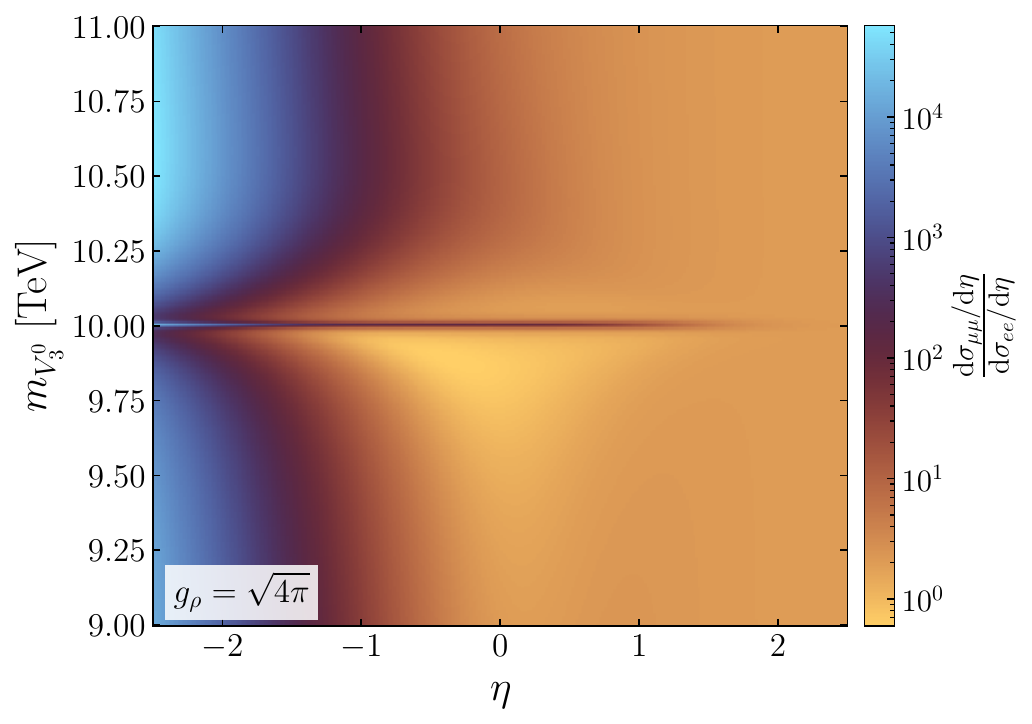}
        \caption{$g_\rho = \sqrt{4\pi}$}
    \end{subfigure}
    \caption{Ratio of the differential cross sections $\mathrm{d}\sigma/\mathrm{d}\eta$ of $\mu^+\mu^-\to\mu^+\mu^-$ and $\mu^+\mu^-\to e^+e^-$ at $\sqrt{s}=10~\mathrm{TeV}$ as a function of $\eta$ and $m_{V_3^0}$, for $m=10^{-2}~\mathrm{TeV}$, $f_d=10~\mathrm{TeV}$ and two values of $g_\rho$.}
    \label{fig:angular_mumu_ee}
\end{figure}

\paragraph{$\mu^+\mu^-\to e^+e^-$ channel}\mbox{}\\
Unlike the previous channel, lepton-flavor conservation forbids $t$-channel exchange, so only $s$-channel diagrams contribute. This eliminates the forward-scattering enhancement of the SM background and produces a cleaner angular distribution, making this channel particularly sensitive to the $V_3^0$ resonance. The $\rho$ meson contribution to the amplitude is
\begin{equation}
    \mathcal{M}^{V_3^0}_s =  \frac{
    \left[\bar{v}_\mu(p_2)\gamma^\mu\!\left(g^{\mu V}_{V_3^0} - g^{\mu A}_{V_3^0}\gamma^5\right)u_\mu(p_1)\right]
    \left[\bar{u}_e(p_3)\gamma_\mu\!\left(g^{eV}_{V_3^0} - g^{eA}_{V_3^0}\gamma^5\right)v_e(p_4)\right]}{s - m_{V_3^0}^2 + im_{V_3^0}\Gamma_{V_3^0}},
\end{equation}
Since $g^{eV/A}_\rho$ follows from Eqs.~\eqref{eq:gfV}--\eqref{eq:gfA} with $T^3_e = -1/2$ and $Q_e = -1$, the electron couplings are numerically identical to the muon ones; any difference between the two channels in the sensitivity contours thus reflects purely the absence of the $t$-channel diagrams here. Because of this, the angular distribution of the final state electrons is quite different from the muon case. As a qualitative comparison, we show the ratio of the normalized differential cross section with respect to the pseudorapidity $\eta$ for the two channels in figure~\ref{fig:angular_mumu_ee}.

\begin{figure}
    \centering
    \includegraphics[width=0.95\textwidth]{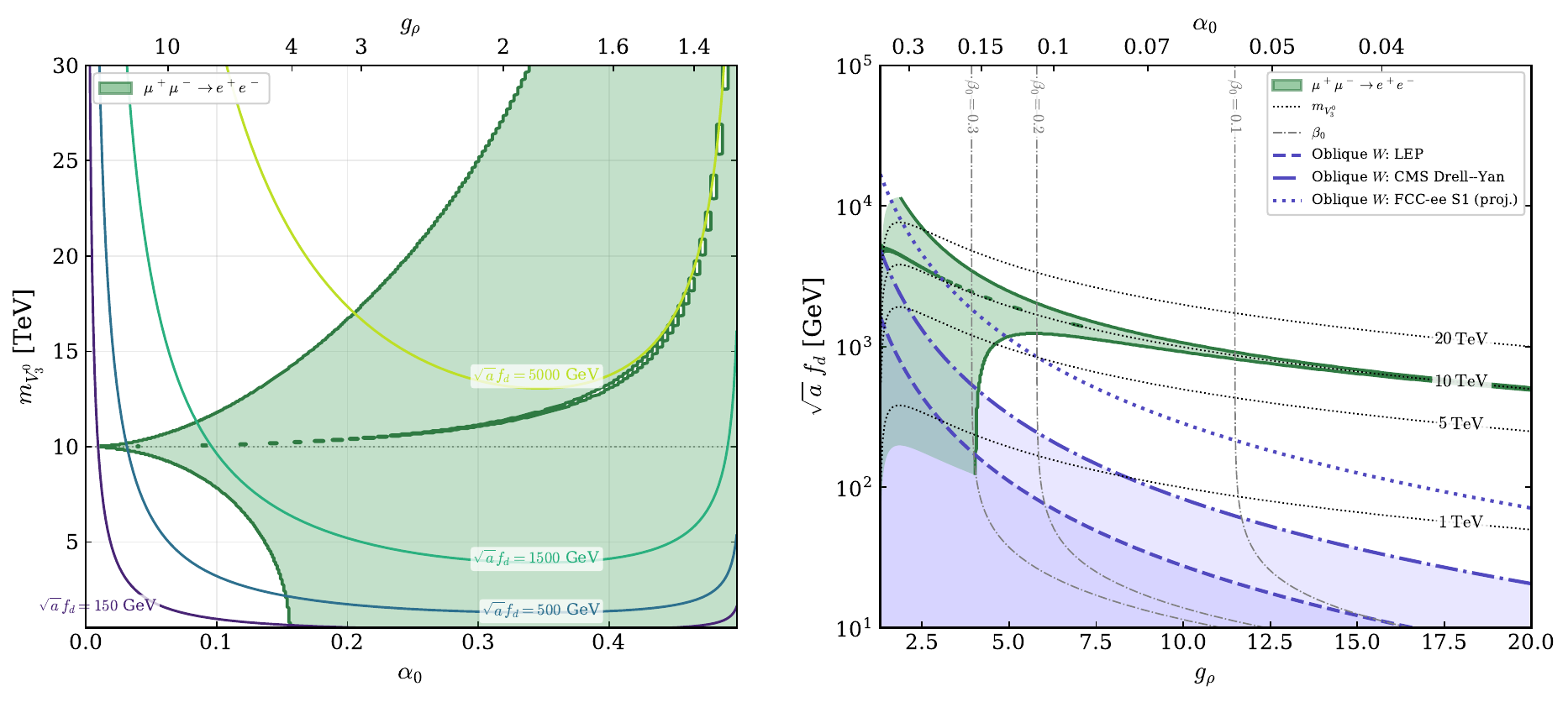}
    \caption{$95\%$ confidence exclusion regions for the $\mu^+\mu^-\to e^+e^-$ channel at $\sqrt{s}=10~\mathrm{TeV}$ with $1~\mathrm{ab}^{-1}$, assuming that the decays into dark pions are closed, shown as for the FCC-hh recast in Figs.~\ref{fig:FCClep} and~\ref{fig:FCC_grho_f}. Left: $(\alpha_0,m_{V_3^0})$ plane with $g_\rho$ on the upper axis; the coloured lines are trajectories of constant $\sqrt{a}f_d$, and the horizontal dotted line marks $m_{V_3^0}=\sqrt{s}$. Right: $(g_\rho,\sqrt{a}f_d)$ plane with $\alpha_0$ on the upper axis, $V_3^0$ mass contours (black dotted) and contours of $\beta_0$ (gray dash-dotted); the purple regions are excluded by the electroweak precision constraints on $W$ from LEP (dashed) and LHC Drell--Yan data (dash-dotted), and the dotted purple line is the projected FCC-ee bound, see appendix~\ref{app:2}.}
    \label{fig:eeCL}
\end{figure}

We have chosen a pion decay constant of $f_d=10$~TeV and a quark mass of $m=10^{-2}$~TeV for this figure, which corresponds to a point in the parameter space where the $V_3^0$ resonance is quite pronounced. For other quark masses, the ratio is not much different, however, lower decay constants lead to a uniform colormap. In this regime, we note that the $\mu^+\mu^-\to e^+e^-$ channel is more sensitive to the resonance, as the absence of the $t$-channel contribution leads to a more center peaked angular distribution. Away from the resonance, the ratio depends only on $\eta$. It reaches $10^4$ at $\eta\simeq-2.5$, where the $t$-channel exchange of $\gamma$ and $Z$ dominates $\mu^+\mu^-\to\mu^+\mu^-$, and approaches a value of order one for $\eta>0$. Near $m_{V_3^0}=\sqrt{s}$, the $s$-channel resonance dominates both channels, and the ratio drops to about one over the whole range of $\eta$. This band is broader for $g_\rho=\sqrt{4\pi}$, owing to the stronger coupling to leptons. The obtained $95\%$ confidence intervals, shown in figure~\ref{fig:eeCL}, show a similar qualitative behavior to the muon case.

Compared to the muon case, despite the shapes being very similar, the exclusion regions are much broader. For $g_\rho=6$, they span $7.
6~\mathrm{TeV}\lesssim m_{V_3^0}\lesssim12.0~\mathrm{TeV}$, and for $g_\rho=10$, they span $9.3~\mathrm{TeV}\lesssim m_{V_3^0}\lesssim10.7~\mathrm{TeV}$ for $f_d\gtrsim1$--$1.6~\mathrm{TeV}$, owing to the interference with the SM amplitude away from the resonance. Also, without $t$-channel diagrams, a destructively interfering wedge is not present in the exclusion regions. The exclusion region in the $a,f_d$ plane is therefore also thicker, as the absence of the $t$-channel contribution allows for a more pronounced resonance, and higher values of $a$ are excluded for a given value of $f_d$. Increasing the value of the quark mass leads to a shift of the exclusion region towards lower values of $f_d$, but the overall shape remains the same. For the QCD-like case of $g_\rho = 6$ and $a=2$, the excluded range is $0.
87~\mathrm{TeV}<f_d<1.38~\mathrm{TeV}$ for $m\gtrsim0.5~\mathrm{TeV}$, about six times broader than in the muon channel, and exclusion already sets in at $m\simeq0.1~\mathrm{TeV}$.
\paragraph{$\mu^+\mu^-\to W^+W^-$ channel}\mbox{}\\
This channel receives $s$-channel contributions from $\gamma$, $Z$, and $V_3^0$, together with a
single $t$-channel SM neutrino exchange. The $V_3^0 W^+W^-$ vertex inherits the standard
Lorentz structure, as confirmed by the mass-eigenstate Lagrangian. Reading off the coupling from the diagonalisation of section~\ref{sec:mass}, it receives two contributions. The first arises from the $W^3$ content of the physical $V_3^0$. The neutral-sector rotations by angles $\alpha_0$ and $\beta_0$ give $V_3^0$ a $W^3$ admixture of {$-(c_W\sin\beta_0+s_W\cos\beta_0\sin\alpha_0)=c_W\kappa_Z+s_W\kappa_\gamma$}, which inherits the $\mathrm{SU}(2)_W$ triple-gauge coupling $g_W$. The second comes from the HLS kinetic term. The physical $V_3^0$ carries a fraction $\cos\alpha_0\cos\beta_0$ of the original HLS field $\tilde{V}^0_3$, which couples to $\tilde{V}^\pm_3\tilde{V}^\mp_3$ with strength $g_\rho/2$, and the physical $W^\pm$ each carry a fraction $\sin\alpha_\pm$ of $\tilde{V}^\pm_3$. The full effective coupling is therefore
\begin{equation}
g_{V_3^0 WW} = \frac{g_\rho}{2}\cos\alpha_0\cos\beta_0\sin^2\!\alpha_\pm
    -g_W^{UV}\!\left(c_W\sin\beta_0 + s_W\cos\beta_0\sin\alpha_0\right)\cos^2\!\alpha_\pm,\label{eq:grhoWW}
\end{equation}
where $c_W = \cos\theta_W$, $s_W = \sin\theta_W$, $\alpha_\pm$ is the charged-sector mixing
angle, and $\alpha_0$, $\beta_0$ are the two successive neutral-sector rotations. 
In the decoupling limit $g_\rho \gg g_W$, all mixing angles vanish and
$g_{V_3^0 WW}\to 0$ as expected.

The $s$-channel amplitude contribution from $V_3^0$ reads
\begin{equation}
    \mathcal{M}^{V_3^0}_s = 
    \frac{g_{V_3^0 WW}}{s-m_{V_3^0}^2+im_{V_3^0}\Gamma_{V_3^0}}\,
    \bigl[\bar{v}(p_2)\gamma_\mu \left(g^{\mu V}_{V_3^0} - g^{\mu A}_{V_3^0}\gamma^5\right)u(p_1)\bigr]\,
    \mathcal{V}^{\mu\nu\rho}(p,-k_+,-k_-)\,
    \epsilon^\ast_\nu(k_+)\,\epsilon^\ast_\rho(k_-),
\end{equation}
where 
\begin{equation}
    \mathcal{V}^{\mu\nu\rho}(q_1,q_2,q_3)= g^{\mu\nu}(q_1-q_2)^\rho
+ g^{\nu\rho}(q_2-q_3)^\mu + g^{\rho\mu}(q_3-q_1)^\nu
\end{equation}
is the standard triple gauge tensor and $p=p_1+p_2$, $k_\pm$ are the $W^\pm$ momenta. 

\begin{figure}
    \centering
    \includegraphics[width=0.95\textwidth]{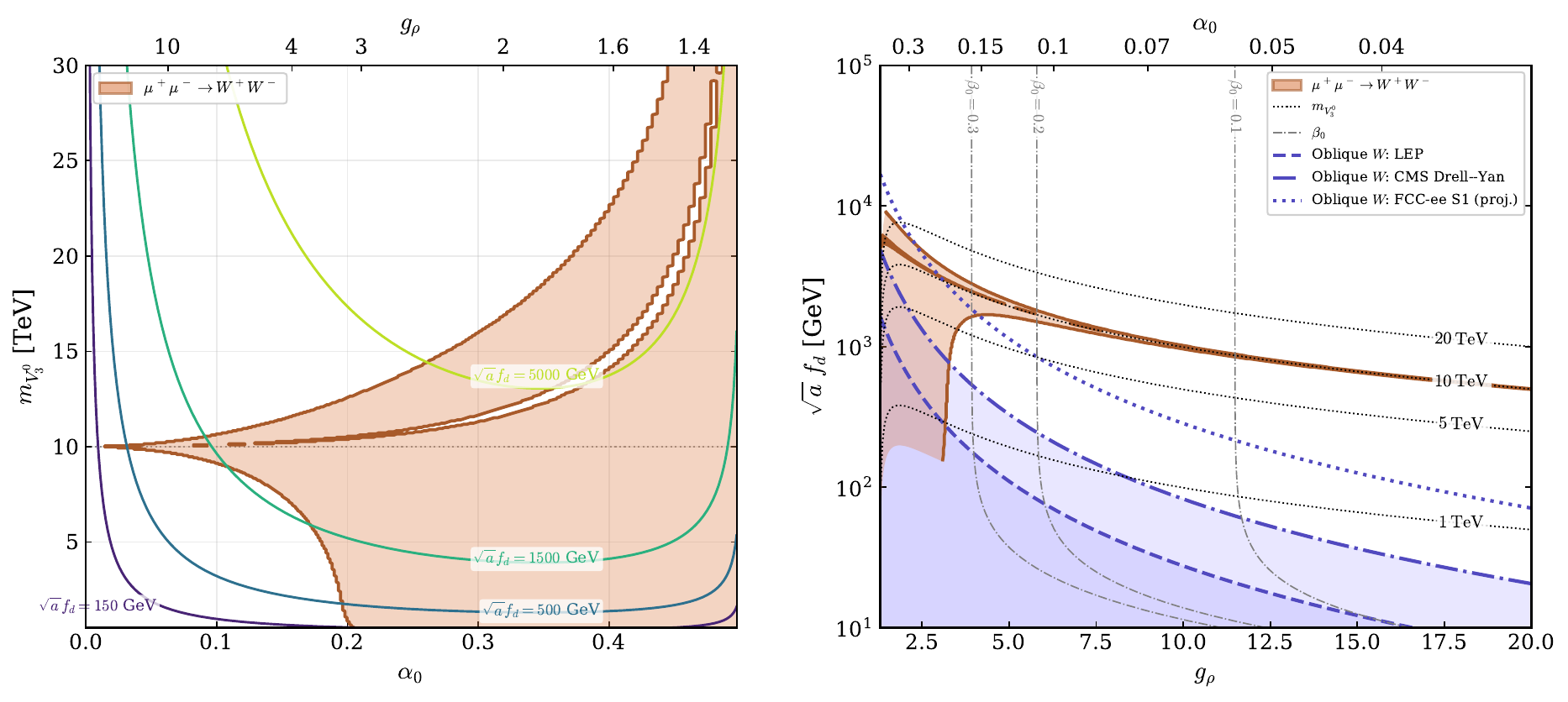}
    \caption{$95\%$ confidence exclusion regions for the $\mu^+\mu^-\to W^+W^-$ channel at $\sqrt{s}=10~\mathrm{TeV}$ with $1~\mathrm{ab}^{-1}$, assuming that the decays into dark pions are closed, shown as for the FCC-hh recast in Figs.~\ref{fig:FCClep} and~\ref{fig:FCC_grho_f}. Left: $(\alpha_0,m_{V_3^0})$ plane with $g_\rho$ on the upper axis; the coloured lines are trajectories of constant $\sqrt{a}f_d$, and the horizontal dotted line marks $m_{V_3^0}=\sqrt{s}$. Right: $(g_\rho,\sqrt{a}f_d)$ plane with $\alpha_0$ on the upper axis, $V_3^0$ mass contours (black dotted) and contours of $\beta_0$ (gray dash-dotted); the purple regions are excluded by the electroweak precision constraints on $W$ from LEP (dashed) and LHC Drell--Yan data (dash-dotted), and the dotted purple line is the projected FCC-ee bound, see appendix~\ref{app:2}.}
    \label{fig:WWCL}
\end{figure}

No BSM particle propagates in the $t$-channel, so the sensitivity to the dark sector is driven entirely by the $V_3^0$ pole in $g_{V_3^0 WW}$ of Eq.~\eqref{eq:grhoWW}. Similar to the previous cases, we use the differential cross section distribution for the $W^+W^-$ final state to compute the $95\%$ confidence intervals. Note that due to the $t$-channel neutrino diagram, the pseudorapidity distribution is slightly offset from the $\mu^+\mu^-\to e^+e^-$ case, but the overall qualitative behaviour is very similar, and thus we abstain from showing the cross section ratio for this channel. The resulting $95\%$ confidence intervals are shown in figure~\ref{fig:WWCL}.
As expected, the exclusion regions are again centered around the $V_3^0$ resonance. The shape of the exclusion region is similar to the previous two channels, and the excluded range in $m_{V_3^0}$ lies between those of the $\mu^+\mu^-$ and $e^+e^-$ channels, namely $5.35$--$12.0~\mathrm{TeV}$ for $g_\rho=\sqrt{4\pi}$, $9.1$--$10.7~\mathrm{TeV}$ for $g_\rho=6$ and $9.7$--$10.2~\mathrm{TeV}$ for $g_\rho=10$. For $g_\rho=6$ and $a=2$, $1.04~\mathrm{TeV}<f_d<1.23~\mathrm{TeV}$ is excluded for $m\gtrsim0.3~\mathrm{TeV}$.

%% file: sections/section4.tex
\section{Discussion and Outlook}\label{sec:discussion}

In this work, we have expanded the $\mathrm{SU}(2)_W$-charged composite dark matter model to include the vector mesons, and studied the phenomenology of the model for QCD-like color and flavor numbers. We have shown that under the non-Abelian charge of the dark quarks, the dark $\rho$ mesons mix with the SM gauge bosons in a way that affects both the allowed interactions of the dark pions, as well as the theoretical value of the EW gauge boson masses. We derived a lower bound for the HLS coupling $g_\rho$ with respect to $f_d$. The results in the following sections were therefore presented in a way that is easily comparable with this bound. To give a sense of the parameter space explored, table~\ref{tab:benchmarks} lists a few representative benchmark points spanning the QCD-like ($g_\rho=6$) and large-mixing ($g_\rho=\sqrt{4\pi}$) HLS couplings, together with values of $a$, $f_d$ and $m$ chosen so that each point reproduces the observed relic abundance with $c\simeq1$. The points P1--P3 have $m_\rho\simeq10~\mathrm{TeV}$, the centre-of-mass energy of the muon collider considered in section~\ref{sec:pheno}, while P4--P6 span $m_\rho\simeq6$--$30~\mathrm{TeV}$. P1 is the QCD-like point $a=2$, $g_\rho=6$. For P2, P3, P5 and P6, the decays of $V_3^0$ into dark pions are closed and $\mathrm{BR}(V_3^0\to\ell^+\ell^-)\simeq8\%$; for P1 and P4, the decays into dark pions are open and reduce it to $1.0\%$ and $2.6\%$, with $\Gamma/m_\rho=1.4\%$ and $1.7\%$, respectively. All points lie inside the projected FCC-hh exclusion region in the $\ell^+\ell^-$ channel. With $m_\rho\ge6~\mathrm{TeV}$, they lie above the current LHC limits of figure~\ref{fig:lhc_recast}, which reach at most $4.8~\mathrm{TeV}$ at the relevant mixing angles $\alpha_\pm\simeq0.22$--$0.38$; P4 lies just above the HL-LHC projection of $5.9~\mathrm{TeV}$. At a muon collider with $\sqrt{s}=10~\mathrm{TeV}$ and $1~\mathrm{ab}^{-1}$, the points P1--P3 are excluded at $95\%$~CL in all three channels of section~\ref{sec:pheno}, see table~\ref{tab:benchmarks_muon}. The strongest signal is in $\mu^+\mu^-\to e^+e^-$, where the cross section of P2 exceeds the SM one by a factor of $20$. Owing to its large mixing, P4 is excluded in the $e^+e^-$ and $W^+W^-$ channels although $m_\rho=6~\mathrm{TeV}$ lies far below $\sqrt{s}$. In terms of discovery potential, with the Asimov data set given by the BSM expectation and the SM-only hypothesis tested, P1-P3 exceed $5\sigma$ in all three channels. P4 reaches $4.4\sigma$ in $e^+e^-$ and $2.3\sigma$ in $W^+W^-$, while it is not significant in $\mu^+\mu^-$. The points P5 and P6 lie far above $\sqrt{s}$, and their cross sections differ from the SM ones by less than $1.5\%$, which is not significant.
\begin{table}[t]
	\centering
	\resizebox{\textwidth}{!}{%
	\begin{tabular}{c c c c c || c c c || c c}
		\toprule
		Benchmark & $g_\rho$      & $a$ & $f_d$            & $m$               & $m_{\Pi_3}$     & $m_{\Pi_5}$     & $m_{V_3^0}$      & $\sigma(pp\to V_3^0)$ & $\sigma\times\mathrm{BR}_{\ell\ell}$  \\
		 & & & & & & & & [fb] & [fb]\\
		\midrule
		P1 & $6$ & $2$ & ${1.15}~\text{TeV}$ & ${0.2722}~\text{TeV}$ & {4.43}~TeV & {4.78}~TeV & {9.99}~TeV & ${3.34}$ & ${3.28\times10^{-2}}$ \\
		P2 & $\sqrt{4\pi}$ & $2$ & ${1.84}~\text{TeV}$ & ${0.6070}~\text{TeV}$ & {8.27}~TeV & {8.75}~TeV & {9.90}~TeV & ${11.0}$ & ${0.885}$ \\
		P3 & $\sqrt{4\pi}$ & $3$ & ${1.54}~\text{TeV}$ & ${0.3434}~\text{TeV}$ & {5.79}~TeV & {6.26}~TeV & {10.2}~TeV & ${9.58}$ & ${0.768}$ \\
		P4 & $\sqrt{4\pi}$ & $3$ & ${0.909}~\text{TeV}$ & ${0.1017}~\text{TeV}$ & {2.52}~TeV & {2.88}~TeV & {6.00}~TeV & ${95.3}$ & ${2.43}$ \\
		P5 & $6$ & $2$ & ${2.30}~\text{TeV}$ & ${0.9654}~\text{TeV}$ & {11.6}~TeV & {12.1}~TeV & {20.0}~TeV & ${8.25\times10^{-2}}$ & ${6.57\times10^{-3}}$ \\
		P6 & $\sqrt{4\pi}$ & $7$ & ${2.97}~\text{TeV}$ & ${1.586}~\text{TeV}$ & {16.8}~TeV & {17.4}~TeV & {30.0}~TeV & ${1.40\times10^{-2}}$ & ${1.12\times10^{-3}}$ \\
		\bottomrule
	\end{tabular}}
	\caption{Benchmark points reproducing the observed dark matter abundance $\Omega_{\rm DM}h^2=0.12$ for $c=1$ and lying inside the projected FCC-hh exclusion region. The spectrum is computed with \texttt{micrOMEGAs}, and $m_\rho$ denotes the $V_3^0$ mass. The last two columns give the production cross section $\sigma(pp\to V_3^0)$ at $\sqrt{s}=100~\mathrm{TeV}$ from \textsc{MadGraph5}, the signal $\sigma\times\mathrm{BR}(V_3^0\to\ell^+\ell^-)$ ($\ell=e,\mu$) with the branching ratio evaluated at the benchmark point.}
	\label{tab:benchmarks}
\end{table}

\begin{table}[t]
	\centering
	\resizebox{\textwidth}{!}{%
	\begin{tabular}{c c || c c c}
		\toprule
		Benchmark & $\Gamma_{V_3^0}/m_{V_3^0}$ [\%] & $\sigma(\mu^+\mu^-\to\mu^+\mu^-)$ [fb] & $\sigma(\mu^+\mu^-\to e^+e^-)$ [fb] & $\sigma(\mu^+\mu^-\to W^+W^-)$ [fb] \\
		\midrule
		SM & -- & $459.6$ & $1.110$ & $12.12$ \\
		\midrule
		P1 & ${1.43}$ & ${\mathbf{459.4}}$ & ${\mathbf{4.874}}$ & ${\mathbf{14.32}}$ \\
		P2 & ${0.559}$ & ${\mathbf{421.7}}$ & ${\mathbf{21.98}}$ & ${\mathbf{26.12}}$ \\
		P3 & ${0.560}$ & ${\mathbf{492.1}}$ & ${\mathbf{2.419}}$ & ${\mathbf{13.26}}$ \\
		P4 & ${1.75}$ & ${460.6}$ & ${\mathbf{1.311}}$ & ${\mathbf{12.27}}$ \\
		P5 & ${0.177}$ & ${459.8}$ & ${1.098}$ & ${12.12}$ \\
		P6 & ${0.564}$ & ${459.9}$ & ${1.096}$ & ${12.14}$ \\
		\bottomrule
	\end{tabular}}
	\caption{Cross sections at a muon collider with $\sqrt{s}=10~\mathrm{TeV}$ for the benchmark points of table~\ref{tab:benchmarks} and in the SM, for $|\eta|<2.5$ ($-2.5<\eta<0$ for $\mu^+\mu^-\to\mu^+\mu^-$). The total width $\Gamma_{V_3^0}$ is summed over all two-body decay channels, and the $V_3^0$ masses are given in table~\ref{tab:benchmarks}. Bold values can be probed with $1~\mathrm{ab}^{-1}$; in the absence of a signal, the binned analysis of section~\ref{sec:pheno} excludes them at $95\%$~CL.}
	\label{tab:benchmarks_muon}
\end{table}

Compared to the vector mesonless case in \cite{Abe:2024mwa}, the presence of the $\rho$ mesons opens up new annihilation channels and causes resonance phenomena on the previously existing ones. For different $\rho$ triplet masses, we observe a narrow resonant bifurcation along $2m_{\Pi_3}=m_\rho$, in which the relic density contour folds back and becomes multivalued in $m$, as shown in figure~\ref{fig:relic_combined}. For very heavy $\rho$ mesons, the relic density contour approaches the one without vector mesons, as expected. The validity of the computation breaks down for dark quark masses approaching the dark scale $\Lambda_d$, or $\rho$ meson masses approaching the pion masses, whichever happens first. 

Beyond the relic density constraints, we examined collider phenomenology to provide complementary sensitivity to the model parameter space. We first considered existing constraints from ATLAS and CMS searches in section~\ref{sec:existing_constraints}, which place bounds on the $\rho$ meson interactions through resonance production with leptonic final states. Additionally, we performed a recast of an existing FCC-hh analysis~\cite{Helsens:2019bfw} to project the sensitivity at future hadron colliders, which reaches $\rho$ masses of up to $\sim40~\mathrm{TeV}$ for $g_\rho\gtrsim1.85$ if the decays into dark pions are closed. Most of the constraints come from $e^+e^-$ and $\mu^+\mu^-$ final states, and bosonic final states did not show any significant constraining power. 
We then studied the $\mu$ collider phenomenology of the model with the aim of obtaining resonant signals from mediator $\rho$ mesons. Leptonic fixed center of mass energy colliders are ideal for this purpose due to the clean resonance signals. The $\rho$ meson interacts with the SM fermions via mixing with the SM gauge bosons. Since the decay width of unstable particles becomes relevant around the region of interest, we could derive a $95\%$ exclusion zone dependent on both $\rho$ and pion variables. The $e^+e^-$ final state channel delivered the strongest results, whereas the $W^+W^-$ and $\mu^+\mu^-$ channels showcased quite narrow exclusion zones with respect to the $\rho$ mass. On the $(a,f_d)$ parameter space, this translates to a narrow band heavily dependent on the HLS coupling. Compared to the relic density results, collider analysis thus provides a complementary probe of the model still in the validity region of the theory.
The perturbative expansion of the HLS model breaks down for large values of the characteristic energy of the process, especially if $p> \Lambda_d$. Therefore, the collider results are assumed to be valid for $\Lambda_d$ values greater than $V_3$ mass. A high value of the HLS coupling $g_\rho$ does not affect the perturbativity, since it is linearly related to the $\rho$ mass, as also seen from the discussion about the lower bound. On the contrary, we obtain decoupled results in the limit of very large $g_\rho$. The ATLAS, CMS and FCC-hh results are valid for the regions where the narrow width approximation holds, as presented in figure~\ref{fig:width_ratio}. This is not the case for the $\mu$ collider results, as these are derived from a full analytical treatment of the resonance production and decay, and therefore are valid for all values of the quark mass.

We considered a general model for dark baryons in our discussion about the relic density, but we did not include them in the collider analysis. As discussed in section~\ref{sec:accidental_symmetry}, the dark baryons are protected by the dark baryon number and can be stable if they are light. Furthermore, due to the accidental $\mathcal{H}$ symmetry, the neutral dark baryons are expected to deliver weak direct detection signals, as the electromagnetic dimension 5 and 6 operators are suppressed \cite{Asadi:2024bbq}. 

Several directions remain open for future investigation. A natural extension of the present work is the explicit addition of the dark baryons to the EFT Lagrangian, considering the multiplet structure generated from the SM gauge interactions, so that baryonic contributions to the relevant observables can be computed more rigorously. Such a model would allow a consistent treatment of the dark baryons in both the relic density computation and collider phenomenology, potentially revealing new co-annihilation channels and complementary production signatures. On the collider side, on-shell dark pion production is expected to be promising as well \cite{Asadi:2025vfr}. Since the dark pion mass is sensitive to the dark quark mass, searches combining on-shell pion production with missing energy signals at fixed center-of-mass energy machines could serve as a direct probe of the dark quark spectrum, complementing the resonance-based approach pursued here. 
Finally, the HLS framework employed throughout this work lacks explicit quark mass dependence, and consequently breaks down in the regime where the $\rho$ meson mass approaches the dark pion mass. Incorporating quark mass effects into the dark spectrum calls for non-perturbative approaches such as holographic models of the dark sector~\cite{Erlich:2005qh}, building on symmetry-breaking treatments of the vector nonet~\cite{Harada:1995sj}, or direct lattice computations~\cite{CP-PACS:2007wro,Feng:2010es}, which can in principle resolve the full quark-mass dependence of the dark hadron spectrum and provide a more complete mapping between the underlying dark quark parameters and the observable meson phenomenology.

%% file: sections/acknowledgments.tex
\section*{Acknowledgments}
We thank the Studienstiftung des deutschen Volkes for financial support, and making this collaboration possible. The work of RS is supported in part by JSPS KAKENHI Grant Numbers~23K03415, 24H02236, and 24H02244. The work of TY is supported in part by JST SPRING, Grant Number JPMJSP2138.

%% file: sections/appendix1.tex
\section{$\mathcal{H}$-Symmetry}\label{app:1}
In this appendix, we briefly discuss $\mathcal{H}$ parity in the current model. This parity is introduced by ref.~\cite{Asadi:2024bbq}, and is useful for the discussion on the decay channel of the neutral particle in the model.

$\mathcal{H}$-parity is defined as
\begin{equation}
    \mathcal{H} \;\equiv\; S \otimes C_W,
\end{equation}
where $S = \exp(i\pi \tilde{T}_2)$ is the same isospin rotation as in $\mathcal{G}$-parity,
but $C_W$ is charge conjugation acting on the \emph{electroweak gauge bosons} rather
than on the dark quarks. Specifically, $C_W$ acts as
\begin{equation}
    C_W:\quad W^1_\mu \to -W^1_\mu,\quad W^2_\mu \to +W^2_\mu,
    \quad W^3_\mu \to -W^3_\mu,
\end{equation}
or equivalently $W^\pm_\mu \to -W^\mp_\mu$ in the charged basis. In the dark quark
sector, $S$ compensates the sign change from $C_W$ in the $\mathrm{SU}(2)_W$ covariant
derivative, leaving the Lagrangian of Eq.~\eqref{eq:lag_quark} exactly
$\mathcal{H}$-invariant.

The key structural difference from $\mathcal{G}$-parity is that $\mathcal{H}$ does not involve
dark charge conjugation. On dark mesons, $\mathcal{H}$ therefore acts \emph{only}
through the $S$ rotation. The $\mathcal{H}$-parity eigenvalue of the neutral ($m = 0$)
component of a $J$-multiplet under $S = \exp(i\pi \tilde{T}_2)$ is \cite{Asadi:2024bbq}
\begin{equation}
    S\,\bigl|J,\,m{=}0\bigr\rangle \;=\; (-1)^J\,\bigl|J,\,m{=}0\bigr\rangle.
\end{equation}
The resulting
$\mathcal{H}$-parity eigenvalues for the neutral dark sector states are:
\begin{itemize}
    \item $\Pi_3^0$ ($J=1$): $\mathcal{H} = -1$ ($\mathcal{H}$-parity odd).
    \item $\Pi_5^0$ ($J=2$): $\mathcal{H} = +1$ ($\mathcal{H}$-parity even).
    \item $V_3^0$ ($J=1$): $\mathcal{H} = -1$ ($\mathcal{H}$-parity odd).
    \item $V_5^0$ ($J=2$): $\mathcal{H} = +1$ ($\mathcal{H}$-parity even).
\end{itemize}
For the charged components, $\mathcal{H}$ is not diagonal but rather maps states to
their charge conjugates: $V_3^\pm \mapsto -V_3^\mp$, $\Pi_3^\pm \mapsto
+\Pi_3^\mp$, and $\Pi_5^\pm \mapsto -\Pi_5^\mp$ (the sign depends on the
representation). 
Note that the dark baryons are not eigenstates of $\mathcal{G}$-parity, as they are mapped to their antiparticle counterparts. 

$\mathcal{H}$-parity provides an important constraint on the decay channels of the neutral triplet
dark $\rho$ meson $V_3^0$. Since $\mathcal{H}(V_3^0) = -1$, any
$\mathcal{H}$-parity-conserving final state must also satisfy $\mathcal{H} = -1$. It follows immediately that \emph{any final state with an even number of neutral dark pions
has $\mathcal{H} = +1$}, which is incompatible with $\mathcal{H}(V_3^0) = -1$.
Hence
\begin{equation}
    V_3^0 \;\to\; 2\,\Pi^0_3 \quad\text{is $\mathcal{H}$-parity forbidden}
    \qquad (\mathcal{H}: {-1} \to {+1}).
\end{equation}
A final state with an odd number of $\Pi_3^0$ mesons would have
$\mathcal{H} = -1$ and is thus $\mathcal{H}$-parity allowed, but this channel is $\mathcal{G}$-parity
forbidden: $\mathcal{G}(V_3^0) = +1$ while $\mathcal{G}(3\Pi_3) = (-1)^3 = -1$.

This pattern holds for every number of neutral pions: even $n$ is $\mathcal{H}$-parity forbidden, odd
$n$ is $\mathcal{G}$-parity forbidden. Since both $\mathcal{G}$-parity and $\mathcal{H}$-parity are exact discrete
symmetries of the dark sector, the decay $V_3^0 \to n\,\Pi^0_3$ is forbidden for \emph{all} $n$. This is a remarkable consequence of the two symmetries, absent if only one were present. Although the $\mathcal{G}$ parity argument also works for charged pions, the $\mathcal{H}$-parity argument does not apply to the charged components, since they are not $\mathcal{H}$-parity eigenstates. Therefore, the triplet dark $\rho$ mesons can decay into charged dark pions as well.

$V_3^0$ can also decay into the Standard Model particles. Such decays are not $\mathcal{G}$-parity forbidden, since SM particles are $\mathcal{G}$-parity even. They are not $\mathcal{H}$-parity forbidden either: since $C_W$ maps $W^3_\mu\to-W^3_\mu$, the neutral $\mathrm{SU}(2)_W$ gauge boson is $\mathcal{H}$-odd, exactly like $V_3^0$, so the mass mixing between $V_3^0$ and $W^3$ derived in Section~\ref{subseq:dark_rho} conserves $\mathcal{H}$ and is present even for $g'\to0$. The decay therefore does not rely on the breaking of $\mathcal{H}$ by the hypercharge interactions~\cite{Asadi:2024bbq}. Through this mixing, the physical mass eigenstate $V_3^0$ inherits a small admixture of $\tilde{Z}_\mu$ and $\tilde{A}_\mu$ through the mixing angles $\alpha_0$ and $\beta_0$, which at leading order combine into a pure $W^3$ admixture (see section~\ref{sec:mass}), so that the partial widths into SM fermions scale as $({g_W^{UV}}/{g_\rho})^2$. In the limit $g_\rho \gg g_W$, these mixing angles are suppressed, so $V_3^0$ is parametrically narrow and can be long-lived for large $g_\rho$.

For the neutral quintet $V_5^0$, the symmetry constraints are similar, but caused by different means. Compared to the triplet, the quintet has the opposite $\mathcal{H}$ and $\mathcal{G}$ assignments, preventing its decay into neutral dark pions of any number. Decays into charged dark pions are only allowed for mixtures of triplet and quintet pions, as other combinations would violate the $\mathcal{G}$-parity. Furthermore, $V_5$ does not mix with the SM gauge bosons (as shown in the mass matrix of section~\ref{subseq:dark_rho}), and any direct decay into SM particles would require violating $\mathcal{G}$-parity. Its stability therefore depends entirely on the mass hierarchy between $V_5$ and its decay products.

We would like to note that unlike $\mathcal{G}$-parity, neutral baryons are in fact eigenstates of $\mathcal{H}$-parity. Their eigenvalues depend on the corresponding $\mathrm{SU}(2)_L$ representation. This has very important consequences for the collider phenomenology of the neutral baryons, as the related electromagnetic dimension-5 and 6 operators are suppressed \cite{Asadi:2024bbq}. 

%% file: sections/appendix2.tex
\section{Electroweak Oblique Parameters}\label{app:2}

In this section, we give a detailed explanation on the electroweak constraints.
The new SU(2)$_L$ triplet vector bosons affect the electroweak observables.  
It is effective to constrain the new physics effect by $\hat{S}$, $\hat T$, $W$, and $Y$~\cite{Barbieri:2004qk}.
To obtain these parameters, we focus on the kinetic terms for the gauge bosons which are non-canonically normalized. {Here $V$ denotes the neutral triplet field in the normalization of $\tilde W$, which couples with $\hat g_\rho=g_\rho/2$, cf.\ section~\ref{sec:mass}. }
The kinetic terms for the electrically neutral sector are given by
\begin{align}
-
\frac{1}{2}
g_{\mu \nu}
\mqty( W_{0}^{\mu} & B^{\mu} & V^{\mu} )
\mqty(
 \Pi_{00} & \Pi_{02} & \Pi_{01}\\
 \Pi_{20} & \Pi_{22} & \Pi_{21}\\
 \Pi_{10} & \Pi_{12} & \Pi_{11}\\
)
\mqty( W_{0}^{\nu}\\  B^{\nu} \\  V^{\nu} )
+
(q_{\mu} q_{\nu} \text{ terms}),
\end{align}
where
{\begin{align}
 \Pi_{00} =& \frac{1}{g_W^2} q^2 - \frac{v^2}{4} - 4a f_d^2, \\ 
 \Pi_{22} =& \frac{1}{g'^2} q^2 - \frac{v^2}{4}, \\ 
 \Pi_{11} =& \frac{1}{\hat g_\rho^2} q^2 - 4a f_d^2, \\ 
 \Pi_{01} =& 4a f_d^2, \\
 \Pi_{02} =& \frac{1}{4} v^2, \\
 \Pi_{12} =& 0,
\end{align}}
and $\Pi_{xy} = \Pi_{yx}$.
We consider only $g_{\mu \nu}$ terms.
Since $V$ is decoupled from the fermion sector, we integrate it out. Then the quadratic terms become
\begin{align}
 &
-
\frac{1}{2}
g_{\mu \nu}
\left(
\begin{array}{cc}
W_{0}^{\mu} & B^{\mu} 
\end{array}
\right)
\left(
\begin{array}{cc}
 \Pi_{W_3 W_3 } & \Pi_{W_3 B} \\
 \Pi_{W_3 B} & \Pi_{BB} \\
\end{array}
\right)
\left(
\begin{array}{c}
 W_{0}^{\nu}\\
 B^{\nu}
\end{array}
\right)
,
\end{align} 
where
{\begin{align}
 \Pi_{W_3 W_3}
=&
 \Pi_{00} - \Pi_{01} (\Pi_{11})^{-1} \Pi_{10}
 \simeq - \frac{v^2}{4} + \qty(\frac{1}{g_W^2} + \frac{1}{\hat g_\rho^2}) q^2 + \frac{q^4}{\hat g_\rho^4 4a f_d^2}
, \\ 
 \Pi_{W_3 B}
=&
 \Pi_{02} - \Pi_{01} (\Pi_{11})^{-1} \Pi_{12}
= \frac{v^2}{4}
 , \\ 
 \Pi_{B B}
=&
 \Pi_{22} - \Pi_{21} (\Pi_{11})^{-1} \Pi_{12}
 = - \frac{v^2}{4} + \frac{1}{g'^2} q^2
.
\end{align}}
In a similar manner, we can calculate the charged sector and find that 
$\Pi_{W_1 W_1}(q^2)= \Pi_{W_3 W_3}(q^2)$. This is due to the custodial symmetry.
In Ref.~\cite{Barbieri:2004qk}, 
the SU(2)$_L$ and U(1)$_Y$ couplings and the Fermi constant are defined by
\begin{align}
  \frac{1}{g^2} = \Pi'_{W^+W^-}(0), \quad \frac{1}{g'^2} = \Pi'_{BB}(0), \quad \frac{1}{\sqrt{2} G_F} = - 4 \Pi_{W^+W^-}(0),
\end{align}
which leads to
{\begin{align}
  \frac{1}{g^2} = \frac{1}{g_W^2} + \frac{1}{\hat g_\rho^2},\\
  \frac{1}{\sqrt{2} G_F} = v^2. 
\end{align}}
The definitions of the $\hat{S}$ and other parameters, and their tree level predictions are given by
{\begin{align}
 g^{-2} \hat{S}
=&
 \Pi'_{W^3 B}(0) = 0
, \\ 
g^{-2} m_W^2 \hat{T}
=&
 \Pi_{W^3W^3}(0) - \Pi_{W^+W^-}(0)
=0
, \\ 
-g^{-2} \hat{U}
=&
 \Pi'_{W^3W^3}(0) - \Pi'_{W^+W^-}(0)
=0
, \\ 
2 g^{-2} m_W^{-2} V
=&
 \Pi''_{W^3W^3}(0) - \Pi''_{W^+W^-}(0)
=0
, \\ 
2 g^{-1} g'^{-1} m_W^{-2} X
=&
 \Pi''_{W^3B}(0) 
=0
, \\ 
2 g'^{-2} m_W^{-2} Y
=&
 \Pi''_{BB}(0)
=0
, \\ 
2 g^{-2} m_W^{-2} W
=&
 \Pi''_{W^3W^3}(0) 
=
\frac{2}{\hat g_\rho^4 4a f_d^2}
.
\end{align}}
We find only $W$ is non-zero at the tree level,
{\begin{align}
  W = \frac{g^2 m_W^2}{4a\,\hat g_\rho^4 f_d^2} \simeq \frac{m_W^4}{a\,\hat g_\rho^4 f_d^2 v^2}{=\frac{16\, m_W^4}{a\,g_\rho^4 f_d^2 v^2}}.
\end{align}}

The parameter $W$ is the Wilson coefficient of the following mass-dimension-six operator,
\begin{align}
\mathcal{L}
\supset - \frac{W}{4 m_W^2} (D_\rho W_{\mu \nu}^a)^2  
= - \frac{g^2 W}{2 m_W^2} j_{L \mu}^a J_{L}^{a \mu}
\end{align}
with canonical normalization of the gauge field, 
where $j_L^{a \mu}$ is the SU(2)$_L$ current consists of the SM fermions~\cite{Farina:2016rws}. 
The second equality is justified by using the equation of motion for the SU(2)$_L$ gauge boson.  

\subsection{Constraint on $W$ from LEP}
The central values, standard deviations, and correlations of these
parameters are given in Tab.~4 in the Ref.~\cite{Barbieri:2004qk}.
\begin{align}
 10^3 \hat{S}
=&
 0 \pm 1.3
\equiv 
 10^{3} \left( \hat{S}_0 \pm \sigma_{\hat{S}}\right)
, \\ 
 10^3 \hat{T}
=&
 0.1 \pm 0.9
\equiv 
 10^{3} \left( \hat{T}_0 \pm \sigma_{\hat{T}}\right)
, \\ 
 10^3 Y
=&
 0.1 \pm 1.2
\equiv 
 10^{3} \left( \hat{Y}_0 \pm \sigma_{\hat{Y}}\right)
, \\ 
 10^3 W
=&
 -0.4 \pm 0.8
\equiv 
 10^{3} \left( \hat{W}_0 \pm \sigma_{\hat{W}}\right)
, \\ 
\rho
=&
\left(
\begin{matrix}
 1    & 0.68 & 0.65 & -0.12 \\
 0.68 &    1 & 0.11 & 0.19 \\
 0.65 & 0.11 &   1  & -0.59 \\
-0.12 & 0.19 & -0.59& 1
\end{matrix}
\right)
.
\end{align}
The confidence ellipse is given as
\begin{align}
 \vec{v}^{T} V^{-1} \vec{v} = \sigma_{\text{CL}}^2
,
\end{align}
where
\begin{align}
 \vec{v}^{T}
=&
10^{3}
\left(
\begin{matrix}
\hat{S} - \hat{S}_0
&
\hat{T} - \hat{T}_0
&
Y- Y_0
&
W - W_0
\end{matrix}
\right)
, \\ 
V
=&
(10^{3})^2
\left(
\begin{matrix}
 \sigma_{\hat{S}} & 0 & 0 & 0 \\ 
 0 & \sigma_{\hat{T}} & 0 & 0 \\ 
 0 & 0 & \sigma_Y & 0 \\
 0 & 0 & 0 & \sigma_W \\
\end{matrix}
\right)
\rho
\left(
\begin{matrix}
 \sigma_{\hat{S}} & 0 & 0 & 0 \\ 
 0 & \sigma_{\hat{T}} & 0 & 0 \\ 
 0 & 0 & \sigma_Y & 0 \\
 0 & 0 & 0 & \sigma_W \\
\end{matrix}
\right)
,
\end{align}
and where
\begin{align}
 \sigma_{\text{CL}}^2 
=&
\begin{cases}
 & 4.71957 \quad (68.27 \text{\% CL})  \\
 & 7.77944 \quad (90 \text{\% CL})  \\
 & 9.48773 \quad (95 \text{\% CL})  \\
 & 13.2767 \quad (99 \text{\% CL})  \\
\end{cases}
.
\end{align}
The set of parameters should be in this ellipse.
Since $\hat{S}=\hat{T}=Y=0$ in our model, we can constrain $W$. We find
\begin{align}
  -2.12\times 10^{-3} \leq W \leq 1.45 \times 10^{-3} \qq{at 95\% CL}.
\end{align}
Using $v \simeq 246.22$~GeV and $m_W \simeq 80.4$~GeV, this leads
\begin{align}
  -{(0.439)^2} g_\rho^4 \leq \frac{\text{TeV}^2}{a f_d^2} \leq {(0.363)^2} g_\rho^4.
\end{align}
If we ignore the negative value, we obtain
\begin{align}
\frac{\sqrt{a} f_d}{\text{TeV}} \gtrsim \frac{1}{{0.363} g_\rho^2}.    
\end{align}

\subsection{Constraint on $W$ from LHC}
The left panel in figure~2 in 
The CMS high-mass dilepton search~\cite{CMS:2026dilepton} gives the observed $95\%$~CL bound

\begin{align}
  -2.1 \times 10^{-4} \lesssim W \lesssim 1.6 \times 10^{-4},
\end{align}
which leads to
\begin{align}
 \frac{1}{{0.120} g_\rho^2} \lesssim \frac{\sqrt{a} f_d}{\text{TeV}}.
\end{align}

\subsection{Prospect for FCC-ee}
Three scenarios for the FCC-ee are discussed in Ref.~\cite{Greljo:2025ggc}. We take the S1 scenario, which is the most conservative one. 
\begin{align}
 \hat{S}
=&
 \pm 2.42 \times 10^{-5}
, \\ 
 \hat{T}
=&
 \pm 1.29 \times 10^{-5}
, \\ 
 W
=&
 \pm 0.48 \times 10^{-5}
, \\ 
 Y
=&
 \pm 1.55 \times 10^{-5}
, \\ 
\rho
=&
\left(
\begin{matrix}
 1    & 0.856 & 0.315 & 0.672 \\
 0.856 &    1 & 0.154 & 0.352 \\
 0.315 & 0.154 &   1  & 0.221 \\
0.672 & 0.352 & 0.221& 1
\end{matrix}
\right)
.
\end{align}
For $\hat{S}= \hat{T} = Y=0$, we find

\begin{align}
  -1.35 \times 10^{-5}  \leq W \leq 1.35 \times 10^{-5} \qq{(95\% CL)},
\end{align}
which leads
\begin{align}
  \frac{\sqrt{a} f_d}{\text{TeV}}\gtrsim \frac{1}{{0.0350} g_\rho^2}.
\end{align}

\subsection{Combined constraints}
Any upper bound $W < W^X_{95}$ translates into
\begin{equation}\label{eq:oblique_bound}
    \sqrt{a}f_d \;>\; \frac{C_X}{g_\rho^{2}}\,,
    \qquad
    C_X = \frac{{2}\,g\, m_W}{\sqrt{W^X_{95}}}\,,
\end{equation}
\begin{figure}[!tb]
    \centering
    \includegraphics[width=0.85\textwidth]{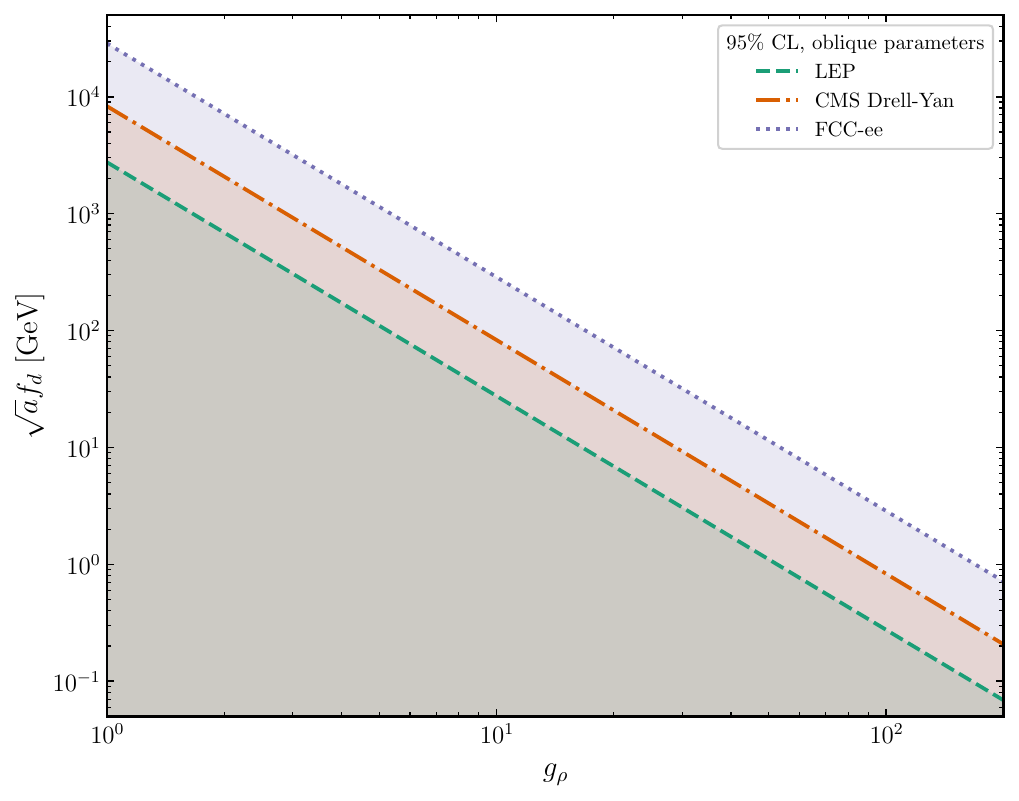}
    \caption{Electroweak precision constraints on the HLS coupling $g_\rho$ and the scale $\sqrt{a}f_d$. The dashed, dash-dotted and dotted lines show the $95\%$~CL lower bounds on $\sqrt{a}f_d$ from the oblique-parameter fits to LEP data~\cite{Barbieri:2004qk}, LHC Drell--Yan measurements~\cite{CMS:2026dilepton}, and FCC-ee projections (Scenario~S1)~\cite{Greljo:2025ggc}; the region below each line is excluded. Since the model generates only the oblique parameter $W$, with $\hat S=\hat T=\hat U=Y=0$ at tree level, all bounds scale as $\sqrt{a}f_d \propto g_\rho^{-2}$.}
    \label{fig:ew_constraints_combined}
\end{figure}

%% file: sections/appendix3.tex
\newpage
\section{Decay Width of the Dark $\rho$-Meson}\label{app:3}
\begin{figure}[!tb]
    \centering
    \includegraphics[width=\textwidth]{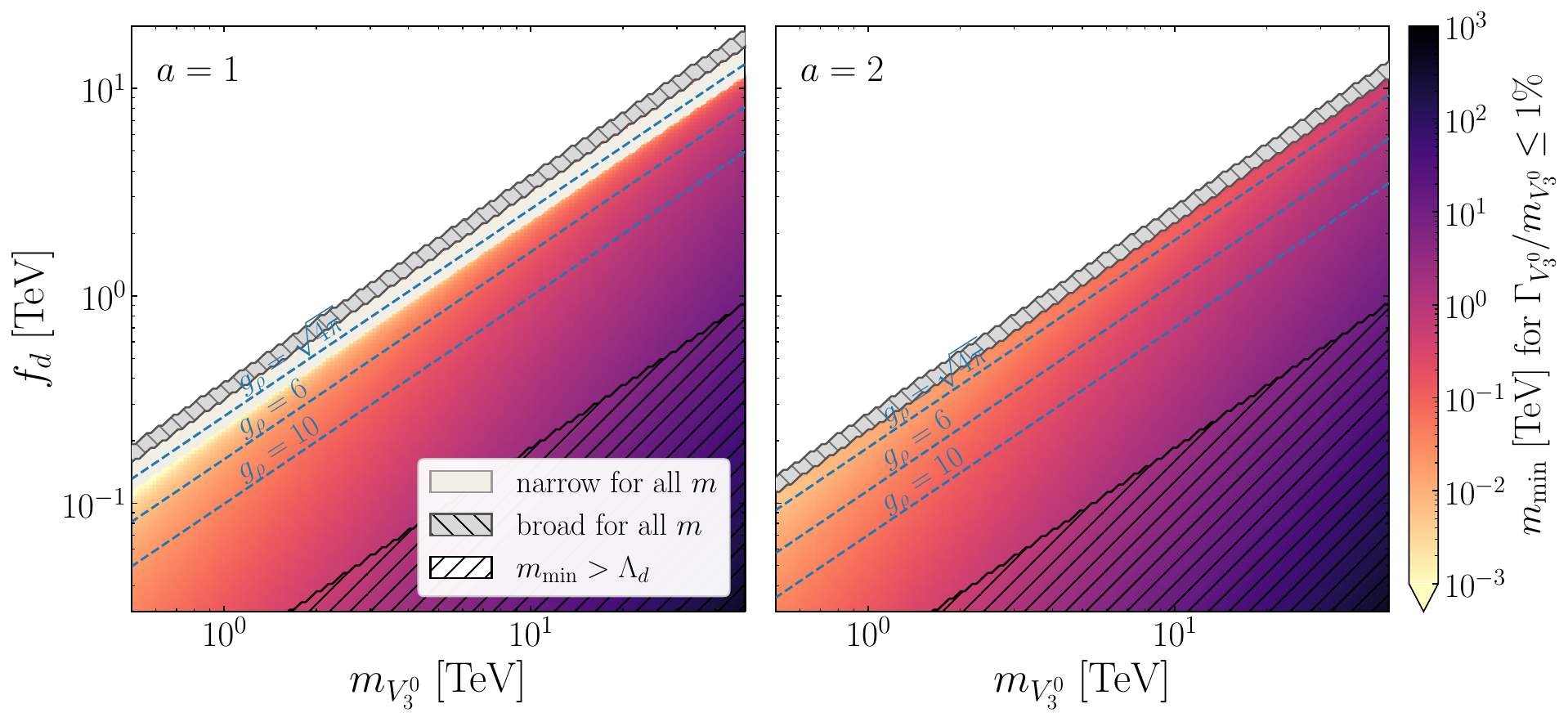}
    \caption{Minimum quark mass $m_{\min}$ for which $\Gamma_{V_3^0}/m_{V_3^0}\le1\%$ in the $(m_{V_3^0},f_d)$ plane, for $a=1$ (left) and $a=2$ (right). In the light area, the width ratio stays below $1\%$ for every $m$, while in the grey hatched area it exceeds $1\%$ for every $m$. Both statements hold for arbitrary $m$, not only for a scanned range: the width decreases monotonically with $m$, approaches a finite value for $m\to0$, where the dark pion masses remain of order $f_d$, and no longer depends on $m$ once the decays into dark pions are closed. In the black hatched area, $m_{\min}$ exceeds $\Lambda_d=4\pi f_d$. The dashed lines are contours of $g_\rho$.}
    \label{fig:width_ratio}
\end{figure}

Most of the generalized studies assume a narrow width approximation in order to factorize the production and decay of the vector bosons. In our model, the width of the dark $\rho$ can be quite large and has a non-trivial dependence on all of the model parameters, including quark mass. The latter is especially important, as the dark $\rho$ can decay to dark pions, which can be kinematically forbidden for certain parameter choices, causing steep changes in the width ratio. The parameter region corresponding to a ratio of $\Gamma/M\approx0.01$ is shown in figure~\ref{fig:width_ratio} for $a$ values of $1$ and $2$.
We compute the total width of $V_3^0$ from all two-body decay channels with the exact mixing angles of section~\ref{sec:mass}. At fixed $m_{V_3^0}$, $f_d$ and $a$, $g_\rho$ follows from the exact $V_3^0$ mass on the branch connected to the decoupling limit $g_\rho\to\infty$, and the width decreases with the quark mass $m$ as the decays into dark pions close. The colour in figure~\ref{fig:width_ratio} shows the smallest $m$ for which $\Gamma_{V_3^0}/m_{V_3^0}\le1\%$. For $g_\rho\lesssim{2.9}$, the mixing with the SM gauge bosons alone makes $V_3^0$ broader than $1\%$. {For $a=1$ and $2.9\lesssim g_\rho\lesssim4.1$, the coupling of $V_3^0$ to dark pions is small enough that the width ratio stays below $1\%$ even if the decays into dark pions are open; for $a=2$, where the $\rho\Pi\Pi$ coupling $a g_\rho/2$ is twice as large, no such region exists. Above it, a narrow $V_3^0$ requires $m$ to grow with $m_{V_3^0}$; for the QCD-like point $a=2$, $g_\rho=6$, from $m\simeq0.28~\mathrm{TeV}$ at $m_{V_3^0}=10~\mathrm{TeV}$ to $1.3~\mathrm{TeV}$ at $m_{V_3^0}=45~\mathrm{TeV}$, and up to $2.5~\mathrm{TeV}$ for $m_{V_3^0}\simeq45~\mathrm{TeV}$ and $g_\rho=10$.} For large $m_{V_3^0}$ and small $f_d$, the required $m$ exceeds $\Lambda_d$.